\documentclass[a4paper,12pt,twoside]{article}
\usepackage[pdftex,colorlinks,bookmarksnumbered,bookmarksopen,citecolor=red,urlcolor=red]{hyperref}
\usepackage{graphicx}\usepackage{a4wide}
\usepackage{amssymb,amsmath,amsthm,mathrsfs}
\usepackage{siunitx}
\usepackage{color,colordvi,graphicx,xcolor}
\usepackage[only,llbracket,rrbracket,interleave]{stmaryrd}
\usepackage{gensymb}

\usepackage{enumitem}

\usepackage{float}
\usepackage{sidecap}
\usepackage[font=small,labelfont=bf]{caption}
\usepackage[export]{adjustbox}
\usepackage{multirow}
\usepackage[caption = false]{subfig}

\usepackage{booktabs,array,dcolumn}
\newcommand{\PreserveBackslash}[1]{\let\temp=\\#1\let\\=\temp}
\newcolumntype{C}[1]{>{\PreserveBackslash\centering}p{#1}}
\newcolumntype{R}[1]{>{\PreserveBackslash\raggedleft}p{#1}}
\newcolumntype{L}[1]{>{\PreserveBackslash\raggedright}p{#1}}

\makeatletter
\newcommand*\wt[2][0.2ex]{%
        \begingroup
        \mathchoice{\wt@helper{#1}{#2}{\displaystyle}{\textfont}}
                   {\wt@helper{#1}{#2}{\textstyle}{\textfont}}
                   {\wt@helper{#1}{#2}{\scriptstyle}{\scriptfont}}
                   {\wt@helper{#1}{#2}{\scriptscriptstyle}{\scriptscriptfont}}%
        \endgroup
        #2%
}
\newcommand*\wt@helper[4]{%
        \def\currentfont{\the#41}%
        \def\currentskewchar{\char\the\skewchar\currentfont}%
        \setbox\tw@\hbox{\currentfont#2\currentskewchar}%
        \dimen@ii\wd\tw@
        \setbox\tw@\hbox{\currentfont#2{}\currentskewchar}%
        \advance\dimen@ii-\wd\tw@
        \rlap{\raisebox{-#1}{$\m@th#3\kern\dimen@ii\widetilde{\phantom{#2}}$}}%
}
\makeatother

\newcommand{\bm}[1]{\text{\boldmath $#1$\unboldmath}}
\newcommand{\abs}[1]{\lvert#1\rvert}
\newcommand{\norm}[1]{\lVert#1\rVert}
\DeclareMathOperator{\sgn}{sgn}
\newcommand{\sign}[1]{{\sgn \left( #1 \right)}}

\newcommand{\bcdot}{\operatorname{\bm{\cdot}}}
\newcommand{\vect}[1]{\bm{#1}}
\newcommand{\mat}[1]{\mathbf{#1}}
\newcommand{\bvect}[1]{\mathbf{#1}}
\newcommand{\bmat}[1]{\mathbf{#1}}

\newcommand{\Div}{{\bm{\nabla}\bcdot\,}}
\newcommand{\Grad}{\bm{\nabla}}
\newcommand{\GradS}{\bm{\nabla^{\texttt{S}}}}

\newcommand{\pd}[2]{\frac{\partial{#1}}{\partial{#2}}}

\newcommand{\RR}{\mathbb{R}}

\newcommand{\eltwo}{\ensuremath{\mathcal{L}_2}}

\newcommand{\nsd}{\ensuremath{\texttt{n}_{\texttt{sd}}}}
\newcommand{\msd}{\ensuremath{\texttt{m}_{\texttt{sd}}}}
\newcommand{\nrr}{\ensuremath{\texttt{n}_{\texttt{rr}}}}
\newcommand{\numel}{\ensuremath{\texttt{n}_{\texttt{el}}}}
\newcommand{\nface}{\ensuremath{\texttt{n}_{\texttt{fa}}}}
\newcommand{\nfaceE}{\ensuremath{\nface^e}}

\newcommand{\tras}{^{{T}}}
\newcommand{\tr}{\operatorname{tr}}
\newcommand{\bu}{\bm{U}}
\newcommand{\bv}{\bm{v}}

\newcommand{\bn}{\bm{n}}
\newcommand{\bx}{\bm{x}}
\newcommand{\bbAll}{\bm{B}}
\newcommand{\bb}{\widehat{\bbAll}}

\newcommand{\bhu}{\widehat{\bu}}
\newcommand{\bhv}{\widehat{\bv}}
\newcommand{\bhT}{\widehat{T}}

\newcommand{\bF}{\bm{F}}
\newcommand{\bG}{\bm{G}}

\newcommand{\btau}{\boldsymbol{\tau}}
\newcommand{\bq}{\bm{q}}
\newcommand{\bsigma}{\boldsymbol{\sigma}^d}
\newcommand{\beps}{\boldsymbol{\varepsilon}^{\! d}}
\newcommand{\bphi}{\boldsymbol{\phi}}

\newcommand{\buE}{\bu_{\! e}}

\newcommand{\bepsE}{\beps_{\! e}}
\newcommand{\bphiE}{\bphi_{\! e}}

\newcommand{\bUinf}{\bm{U}_{\! \infty}}
\newcommand{\Tend} {\textrm{T}_{\texttt{end}}}

\newcommand{\hlamax} {\widehat{\lambda}_{\texttt{max}}}
\newcommand{\Id}[1]{\bmat{I}_{#1}}
\newcommand{\Jd}[1]{\bmat{J}_{#1}}
\newcommand{\dev} {\ensuremath{\mathcal{D}}}
\newcommand{\Dt}{\Delta t}

\newcommand{\hu}{\widehat{\bvect{U}}}
\newcommand{\Dhu}{\widehat{\Delta \! \bvect{U}}}
\newcommand{\uE}{\bvect{U}_{\! e}}
\newcommand{\qE}{\bvect{Q}_{e}}
\newcommand{\bPsi}{\bm{\Psi}}
\newcommand{\AmatE}[2]{\bmat{A}_{{#1}{#2}}^e}
\newcommand{\FvectE}[1]{\bvect{F}_{\!{#1}}^e}
\newcommand{\Kmat}{\bmat{K}}
\newcommand{\Fvect}{\bvect{F}}
\newcommand{\jump}[1]{\llbracket #1\rrbracket}

\newcommand{\Rey}{\small{\textsl{Re}}}
\newcommand{\Pra}{\small{\textsl{Pr}}}
\newcommand{\Ma}{\small{\textsl{M}}}
\newcommand{\vinf}{v_{\! \infty}}
\newcommand{\rhoinf}{\rho_{\! \infty}}
\newcommand{\muinf}{\mu_{\! \infty}}
\newcommand{\Minf}{\Ma_{\! \infty}}

\newcommand{\cp}{c_p}
\newcommand{\cv}{c_v}
\newcommand{\Tref}{\ensuremath{T_{\texttt{ref}}}}

\newcommand{\setAe}{\ensuremath{\mathcal{A}_e}}
\newcommand{\setEe}{\ensuremath{\mathcal{E}_e}}
\newcommand{\setIe}{\ensuremath{\mathcal{I}_e}}

\newcommand{\bqE}{\bm{Q}_{\! e}}
\newcommand{\re}{\bvect{R}_{\! e}}
\newcommand{\hre}{\widehat{\bvect{R}}_{\! e}}

\newcommand{\bepsV}{\boldsymbol{\varepsilon}^{\! d}_{\texttt{V}}}
\newcommand{\bDv}{\mat{D}_{\texttt{V}}}
\newcommand{\bNv}{\mat{N}_{\texttt{V}}}
\newcommand{\epsD}{\varepsilon^{\! d}}
\newcommand{\Gsym}{\bm{\nabla}_{\!\texttt{S}}}

\newcommand{\HU}{\widehat{U}}
\newcommand{\Ga}[1]{\Gamma_{\! { #1}}}

\newenvironment{keywords}{\begin{quote}\emph{\textbf{Keywords:}}}{\end{quote}}

\theoremstyle{definition}

\newtheorem{remark}{Remark}

\usepackage{scalerel}
\usepackage{stackengine}
\stackMath
\def\hatgap{0pt}
\def\subdown{-2pt}
\newcommand\reallywidehat[2][]{
	\renewcommand\stackalignment{l}
	\stackon[\hatgap]{#2}{
		\stretchto{\scalerel*[\widthof{$#2$}]{\kern-.6pt\bigwedge\kern-.6pt}{\rule[-\textheight/2]{1ex}{\textheight}}}{0.5ex}_{\smash{\belowbaseline[\subdown]{\scriptstyle#1}}}}}

\RequirePackage{algorithm}[0.1] 
\usepackage{algorithmic} 

\graphicspath{{figsPDF/}{figsPNG/}{figs/}}
\usepackage{fancyhdr}

\newcommand{\hl}[1]{{\color{black} #1\color{black}}}

\usepackage{hyphenat}

\begin{document}

\title{A non-oscillatory face-centred finite volume method for compressible flows}

\author{
\renewcommand{\thefootnote}{\arabic{footnote}}
			  Jordi Vila-P\'erez\footnotemark[1]\textsuperscript{ \ ,}\footnotemark[2] \ ,
			  Matteo Giacomini\footnotemark[1]\textsuperscript{ \ ,}\footnotemark[3]\textsuperscript{ \ ,}*  ,
			  Ruben Sevilla\footnotemark[4], \ and \\
\renewcommand{\thefootnote}{\arabic{footnote}}
             Antonio Huerta\footnotemark[1]\textsuperscript{ \ ,}\footnotemark[3]
}

\date{}
\maketitle

\renewcommand{\thefootnote}{\arabic{footnote}}

\footnotetext[1]{Laboratori de C\`alcul Num\`eric (LaC\`aN), ETS de Ingenieros de Caminos, Canales y Puertos, Universitat Polit\`ecnica de Catalunya, Barcelona, Spain}
\footnotetext[2]{Barcelona Graduate School of Mathematics (BGSMath), Barcelona, Spain}
\footnotetext[3]{Centre Internacional de M\`etodes Num\`erics en Enginyeria (CIMNE), Barcelona, Spain.}
\footnotetext[4]{Zienkiewicz Centre for Computational Engineering, Faculty of Science and Engineering, Swansea University, Wales, UK.
\vspace{5pt}\\
* Corresponding author: Matteo Giacomini. \textit{E-mail:} \texttt{matteo.giacomini@upc.edu}
}

\begin{abstract}
This work presents the face-centred finite volume (FCFV) paradigm for the simulation of compressible flows. The FCFV method defines the unknowns at the face barycentre and uses a hybridisation procedure to eliminate all the degrees of freedom inside the cells. In addition, Riemann solvers are defined implicitly within the expressions of the numerical fluxes.
The resulting methodology provides first-order accurate approximations of the conservative quantities, i.e. density, momentum and energy, as well as of the viscous stress tensor and of the heat flux, without the need of any gradient reconstruction procedure. Hence, the FCFV solver preserves the accuracy of the approximation in presence of distorted and highly stretched cells, providing a solver insensitive to mesh quality.
\hl{In addition, FCFV is capable of constructing non-oscillatory approximations of sharp discontinuities without resorting to shock capturing or limiting techniques.}
For flows at low Mach number, the method is robust and is capable of computing accurate solutions in the incompressible limit without the need of introducing specific pressure correction strategies.
A set of 2D and 3D benchmarks of external flows is presented to validate the methodology in different flow regimes, from inviscid to viscous laminar flows, from transonic to subsonic incompressible flows, demonstrating its potential to handle compressible flows in realistic scenarios.
\end{abstract}


\section{Introduction}
\label{sc:Introduction}

Finite volume (FV) solvers are the most widespread technology within the aerospace community for the simulation of steady-state compressible flows~\cite{Versteeg2007,Leveque2013}. The success of such methodologies is mainly due to their capability of providing results for large-scale flow problems involving complex geometries by means of overnight simulations. For this reason, FV implementations are accessible in many computational fluid dynamics (CFD) platforms, from open-source to commercial and industrial software~\cite{FluentManual,CFL3D:06,FUN3D:19,Morgan1991,sorensen2003a,sorensen2003b,jasak2009openfoam,gerhold2005overview,chalot2004industrial}. Nonetheless, these methods require delicate mesh generation procedures, limiting unstructured regions and distorted cells~\cite{Diskin2010,Diskin2011}, in order to construct high-quality meshes suitable for computation. 

Traditional FV solvers, see e.g.~\cite{Leveque2013,Morton2007,Ohlberger-BO-04,Eymard2000}, rely on two paradigms, the so-called cell-centred finite volume (CCFV) method~\cite{Maire2007} and the vertex-centred finite volume (VCFV) method~\cite{Asouti2010}. The former defines the unknown solution at the centroid of the cells, whereas the latter at the mesh nodes. Recently, the face-centred finite volume (FCFV) paradigm was proposed for a series of linear elliptic partial differential equations (PDEs)~\cite{RS-SGH:2018_FCFV1,RS-SGH:2019_FCFV2,MG-RS-20,RS-VGSH:20}. The FCFV method utilises a mixed FV formulation and eliminates the degrees of freedom within each cell via a hybridisation step, leading to a problem defined in terms of the unknowns at the face barycentres only. Finally, the variables inside each cell are retrieved via a computationally inexpensive postprocessing step performed independently cell-by-cell. 

The present work discusses the first FCFV formulation for nonlinear hyperbolic PDEs modelling steady-state compressible flows, spanning from viscous compressible Navier-Stokes to inviscid Euler equations. The proposed approach introduces the flow variables, i.e. density, momentum and energy, the deviatoric strain rate tensor and the gradient of temperature as unknowns inside each cell, whereas the conservative quantities on the cell faces define the so-called \emph{hybrid} unknowns. \hl{The resulting method yields first-order accuracy for all above mentioned variables. Contrary to traditional first-order CCFV and VCFV paradigms, the FCFV method provides first-order convergence of the deviatoric strain rate tensor and of the gradient of temperature. In addition, this is achieved without resorting to any flux reconstruction strategy required by traditional second-order CCFV and VCFV approaches. This is especially important as it allows the FCFV method to preserve optimal accuracy in the computation of the stress tensor and the heat flux even in presence of distorted and stretched cells~\cite{RS-SGH:2018_FCFV1,RS-SGH:2019_FCFV2,MG-RS-20,RS-VGSH:20}, where traditional second-order CCFV and VCFV approaches are known to suffer from a loss of accuracy~\cite{Diskin2010,Diskin2011,Svaerd2008}}. In addition, the FCFV method showed its versatility in devising approximations based on meshes of different cell types and on hybrid meshes~\cite{RS-SGH:2018_FCFV1,MG-RS-20}, proving to be a robust methodology insensitive to element type and mesh quality. Although the FCFV method is known to introduce a larger number of unknowns than CCFV and VCFV approaches~\cite{RS-SGH:2018_FCFV1}, the simplified procedure required to generate meshes suitable for computation and its capability to avoid gradient reconstruction make the FCFV scheme a competitive alternative to traditional FV solvers from the computational viewpoint.

\hl{In the approximation of conservation laws, a critical aspect is represented by the strategy employed to handle the nonlinear convective fluxes. In this context, Riemann solvers have been extensively studied~\cite{Toro2009,Hesthaven2017,Cockburn-CS:1998,Qiu-QKS:2006}.} The present work revisits the traditional Lax-Friedrichs, Roe, HLL and HLLEM Riemann solvers in the context of the FCFV method. An extensive set of numerical examples is employed to validate and compare these approaches. Special attention is devoted to the development of positivity-preserving approximations in order to devise a robust methodology able to compute physically-admissible solutions across a wide range of flow regimes. \hl{More precisely, numerical experiments are reported to showcase the capability of the FCFV method to construct non-oscillatory approximations of shock waves and sharp fronts, without the need of shock capturing techniques based on artificial viscosity~\cite{VonNeumann1950,Donea2003,Persson-PP:2006,Fernandez-FNP:2018} or flux limiters~\cite{vanLeer1979,Sweby1984,Toro2000,Krivodonova2007}.}

It is worth mentioning that FV methods have been reinterpreted in recent years as particular cases of finite element discretisations~\cite{Morton2007}: the CCFV method can be seen as a discontinuous Galerkin (DG) approach with piecewise constant approximations in each cell~\cite{Eymard2000,Cockburn-CKS:2000}, whereas a VCFV discretisation on a simplicial mesh is equivalent to a continuous Galerkin finite element method with piecewise linear approximations~\cite{Selmin1993,Bank1987}. Concerning the FCFV paradigm, it can be interpreted as the lowest-order version of the hybridisable discontinuous Galerkin (HDG) method~\cite{Jay-CGL:09,Cockburn2016,HDGlab-GSH-20}, in which constant approximations are selected for all the variables. More precisely, the FCFV formulation presented in this work is inspired by the high-order HDG discretisations for compressible flows discussed in~\cite{Nguyen-NP:2012,Peraire-PNC:10}. The proposed FCFV method thus inherits its properties from HDG, including its stability in the incompressible limit~\cite{RS-SGH:2018_FCFV1,RS-SGH:2019_FCFV2,MG-RS-20,RS-VGSH:20}, circumventing the Ladyzhenskaya-Babu\v ska-Brezzi (LBB) condition~\cite{Donea2003,Jay-CG:2009}. Hence, the FCFV method provides accurate approximations of incompressible flows without the need of introducing specific pressure corrections like the well-known SIMPLE algorithm~\cite{Patankar-PS:72} implemented by commercial and open-source software, see e.g.~\cite{FluentManual,jasak2009openfoam}. 

The remainder of this paper is organised as follows. In section~\ref{sc:Equations}, the compressible Navier-Stokes equations are recalled. Section~\ref{sc:FCFV} introduces the corresponding FCFV discretisation and the integration of the different Riemann solvers in the definitions of the numerical fluxes. A set of convergence tests to validate the optimal accuracy properties of the method for inviscid and viscous laminar flows is presented in section~\ref{sc:Convergence}, with special emphasis on the robustness to cell stretching and distortion. Two- and three-dimensional benchmarks of external flows of aerodynamic interest are reported in section~\ref{sc:Benchmarks} to demonstrate the capabilities of the method in different flow regimes, from inviscid to viscous laminar flows, from transonic to subsonic incompressible flows. Finally, section~\ref{sc:Conclusions} reviews the main results of this work and three appendices provide technical details on the definition of the boundary conditions (\ref{app:BoundaryConditions}), on the construction of the mixed variable (\ref{app:Voigt}) and on the hybridisation procedure (\ref{app:FCFVsolver}) in the FCFV solver.


\section{Governing equations for compressible flows}
\label{sc:Equations}

Consider the compressible Navier-Stokes equations, written in non-dimensional conservative form as
\begin{equation}\label{eq:NScompact}
\left\{
\begin{aligned}
\pd{\bu}{t} + \Div \left( \bF (\bu)- \bG (\bu, \Grad \bu) \right) &= \vect{0} &  \text{in } &\Omega \times (0,\Tend], \\
\bu &= \bu^0  &  \text{in } & \Omega \times \{0\}, \\
\bbAll (\bu, \Grad \bu) &= \bm{0}  & \text{on } &\partial \Omega \times (0,\Tend], 
\end{aligned}
\right.
\end{equation}
where $\Omega \subset \RR^{\nsd}$ denotes an $\nsd$-dimensional open bounded domain with boundary $\partial \Omega$, $\Tend>0$ is the final time of interest, $\bu^0$ represents the initial state and $\bbAll$ is an operator imposing inlet, outlet or wall boundary conditions as detailed in~\ref{app:BoundaryConditions}.

\begin{remark}[Pseudo-time in steady-state flows]
	The present work focuses on the development of a novel FV spatial discretisation for steady-state compressible flows. In this context, $t$ represents an artificial pseudo-time and time marching algorithms are introduced to speed-up the convergence of the nonlinear solver, as detailed in section~\ref{ssc:FCFVdiscretisation}.
\end{remark}

The system~\eqref{eq:NScompact} is written in terms of the vector of conserved quantities $\bu \in \RR^{\nsd + 2}$ and the advection, $\bF$, and diffusion, $\bG$, flux tensors, namely
\begin{equation}\label{eq:NSterms}
\bu = \begin{Bmatrix} \rho\\ \rho\bv\\ \rho E \end{Bmatrix}\!,
\quad
\bF (\bu) = \begin{bmatrix} \rho \bv\tras\\
\rho\bv \otimes\bv + p \Id{\nsd}\\
(\rho E + p) \bv\tras \end{bmatrix} \!,
\quad
\bG (\bu, \Grad \bu)= \begin{bmatrix} \vect{0}\\
\bsigma \\
(\bsigma \bv + \bq )\tras \end{bmatrix} ,
\end{equation}
where $\rho$ is the density, $\bv$ the velocity vector, $E$ the total specific energy and $p$ denotes the pressure field. The viscous stress tensor $\bsigma$ and the heat flux vector $\bq$ are given by
\begin{equation}\label{eq:StressTensorHeatFlux}
\bsigma = \frac{\mu}{\Rey} \left(2 \GradS \bv - \frac{2}{3} (\Div \bv) \Id{\nsd} \right), \quad \bq = \frac{\mu}{\Pra \Rey} \Grad T,
\end{equation}
where $\GradS := (\Grad + \Grad^T)/2$ is the symmetric part of the gradient operator and $T$ denotes the temperature. The non-dimensional dynamic viscosity is defined according to the Sutherland's law, i.e. $\mu = \left(T/T_\infty\right)^{3/2} (T_\infty + S)/(T + S)$, $T_\infty = 1/\left((\gamma - 1) \Minf^2\right)$ and $S = S_0/\left((\gamma - 1) \Tref \Minf^2\right)$ being the non-dimensional free-stream temperature and the Sutherland constant, respectively, with $S_0 = 110K$ and $\Tref = 273K$.

In addition, following the assumption of a calorically perfect gas, the pressure is given by $p = (\gamma - 1) \rho \left( E -  \norm{\bv}^2/2 \right)$, where $\gamma = \cp / \cv$ (equal to 1.4 for air) is the ratio of specific heats at constant pressure, $\cp$, and constant volume, $\cv$. Finally, from the ideal gas law it follows that $\gamma p = (\gamma - 1) \rho T$.

\begin{remark}[Non-dimensional quantities]
	The problem is described in terms of the Mach, Reynolds and Prandtl numbers, defined respectively as
	\begin{equation} \label{eq:non-dimensionalQuantities}
	\Minf = \frac{\vinf}{c_\infty}, \qquad \Rey = \frac{\rhoinf \vinf L}{\muinf}, \qquad \Pra = \frac{\cp \muinf}{\kappa} \text{ (= 0.71 for air)},
	\end{equation}
	$c = \sqrt{\gamma p/\rho}$ being the speed of sound, $L$ a characteristic length and $\kappa$ the thermal conductivity. The subscript $_\infty$ denotes free-stream reference values.
\end{remark}


\section{FCFV formulation for compressible flows}
\label{sc:FCFV}

The domain $\Omega$ is partitioned in a set of $\numel$ disjoint cells $\Omega_e$ satisfying $\Omega = \bigcup_{e=1}^{\numel} \Omega_e$. In addition, the boundary of the cell $\Omega_e$, denoted by $\partial \Omega_e$, is obtained as the union of its faces, namely
\begin{equation}
\partial \Omega_e :=\bigcup_{j=1}^{\nfaceE} \Gamma_{e\!, j} ,
\end{equation}
where $\nfaceE$ is the total number of faces of the cell $\Omega_e$. Finally, the internal interface $\Gamma$ is defined as
\begin{equation}
\Gamma:=\left[ \bigcup_{e=1}^{\numel}\partial\Omega_e \right] \setminus \partial \Omega.
\end{equation}

\subsection{Introducing a set of mixed variables}
\label{ssc:mixedVar}

Following the HDG~\cite{Jay-CGL:09,Jay-CG:2009,Nguyen-NP:2012} and FCFV~\cite{RS-SGH:2018_FCFV1,RS-SGH:2019_FCFV2} rationales, the second-order problem~\eqref{eq:NScompact} is written as a system of first-order PDEs via the introduction of a set of so-called \emph{mixed variables}. In the context of compressible flows, the most common approach, see~\cite{Peraire-PNC:10,Nguyen-NP:2012,May-WBMS-14, Fernandez-FNP:2017}, relies on defining the mixed variable as the gradient of the so-called \emph{primal variable} $\bu$, namely $\bPsi = \Grad \bu$. Nonetheless, this choice leads to the introduction of a mixed variable, associated with the gradient of density, which is redundant since the mass conservation equation is a first-order PDE. In addition, several nonlinearities appear in the resulting expressions to compute the viscous stress and the heat flux starting from $\bu$ and $\bPsi$. Following~\cite{JVP_HDG-VGSH:20}, in this work only two mixed variables are considered, namely the deviatoric strain rate tensor $\beps$ and the gradient of temperature $\bphi$, given by
\begin{equation}\label{eq:mixedVariables}
\beps = 2 \GradS \bv - \frac{2}{3} (\Div \bv) \Id{\nsd}, \qquad \bphi =  \Grad T .
\end{equation}

\begin{remark}[Deviatoric strain rate]\label{rmrk:operatorVoigt}
	It is worth noticing that the deviatoric strain rate tensor can be expressed as a function of the gradient of velocity as $\beps = \dev  \GradS \bv $, where the linear operator $\dev$ is defined as
	\begin{equation} \label{eq:deviatoric}
	\dev \bm{W}  = \left( \bm{W} + \bm{W}\tras \right) - \frac{2}{3} \tr (\bm{W}) \Id{}.
	\end{equation}
	Interested readers are referred to~\ref{app:Voigt} for the details concerning the construction of the operator $\dev$ and its implementation.
\end{remark}

Besides reducing the number of mixed variables involved, from~\eqref{eq:mixedVariables} it also follows that the viscous stress tensor and the heat flux vector in equation~\eqref{eq:StressTensorHeatFlux} can be obtained using the linear expressions
\begin{equation} \label{eq:ViscousTermsNewMixed} 
\bsigma = \frac{\mu}{\Rey} \beps, \qquad
\bq = \frac{\mu}{\Rey \Pra} \bphi.
\end{equation}

\subsection{A mixed hybrid finite volume framework}

The FCFV method solves the compressible Navier-Stokes equations in two stages. First, an independent hybrid variable $\bhu$, representing the vector of conservative variables on the mesh faces $\Gamma \cup \partial \Omega$, is introduced. Equation~\eqref{eq:NScompact} is thus rewritten in each cell $\Omega_e$, $e = 1,\dotsc,\numel$ as
\begin{equation} \label{eq:NSstrongLocal} 
\left\{\begin{aligned}
\beps - \dev  \GradS \bv  &= \bm{0} & \text{in } &\Omega_e \times (0,\Tend],  \\
\bphi - \Grad T &= \bm{0} & \text{in } &\Omega_e \times (0,\Tend],  \\
\pd{\bu}{t} +\Div \left( \bF (\bu) - \bG (\bu, \beps, \bphi) \right) &= \bm{0}  & \text{in } &\Omega_e \times (0,\Tend],  \\
\bu &= \bu^0  &  \text{in } & \Omega_e \times \{0\}, \\
\bu &= \bhu  \qquad & \text{on } &\partial \Omega_e \times (0,\Tend] . \\
\end{aligned}\right.
\end{equation}
Equation~\eqref{eq:NSstrongLocal} represents the $\numel$ FCFV local problems. They define the vector of conservative variables and the mixed variables $(\bu,\beps,\bphi)$ in each cell as functions of the hybrid vector $\bhu$ on the cell faces, in order to reduce the global number of unknowns of the problem.

Second, the vector $\bhu$ of conservative variables on the faces is computed by solving the FCFV global problem, which prescribes the continuity of the conservative variables and of the normal fluxes on $\Gamma$ and the boundary conditions on $\partial \Omega$, namely
\begin{equation} \label{eq:NSstrongGlobal} 
\left\{\begin{aligned}
\jump{\bu\otimes \bn} &= \bm{0} &\text{on } &\Gamma \times (0,\Tend],\\
\jump{\left( \bF(\bu) - \bG(\bu,\beps,\bphi) \right) \bn} &= \bm{0} &\text{on } &\Gamma \times (0,\Tend], \\
\bb(\bu,\bhu,\beps,\bphi) &= \bm{0},  & \text{on } &\partial \Omega \times (0,\Tend] ,
\end{aligned}\right.
\end{equation}
where $\bn$ stands for the outward normal vector to the cell face and $\jump{\circledcirc} = \circledcirc^+ + \circledcirc^-$ denotes the \emph{jump} operator defined on an internal face as the sum of the values in the neighbouring elements $\Omega^+$ and $\Omega^-$, respectively~\cite{AdM-MFH:08}. The trace boundary operator $\bb(\bu,\bhu,\beps,\bphi)$ imposes the boundary conditions on $\partial \Omega$ exploiting the hybrid variable, as detailed in~\ref{app:BoundaryConditions}.

It is worth noticing that the first condition in equation~\eqref{eq:NSstrongGlobal} is automatically satisfied due to the Dirichlet boundary conditions imposed in the local problems~\eqref{eq:NSstrongLocal} and because of the unique definition of the hybrid variable $\bhu$ on each face.

\subsection{Integral form of the FCFV local and global problems}
\label{ssc:IntegralForm}

For each cell $\Omega_e, \ e = 1,\dotsc,\numel$, the integral form of the FCFV local problem is obtained by applying the divergence theorem to equation~\eqref{eq:NSstrongLocal}. Given $\bu = \bu^0$ at time $t=0$, it holds that 
\begin{subequations} \label{eq:IntegralLocal}
	\begin{align}
	\int_{\Omega_e} \beps \, d\Omega- \int_{\partial \Omega_e} \dev \bhv \otimes \bn \, d\Gamma &= \vect{0}, \label{eq:IntegralLocal_eps}\\
	\int_{\Omega_e} \bphi \, d\Omega - \int_{\partial \Omega_e} \bhT \bn \, d\Gamma &= \vect{0}, \label{eq:IntegralLocal_phi}\\
	\int_{\Omega_e} \pd{\bu}{t}  \, d\Omega  + \int_{\partial \Omega_e}  \left( \reallywidehat{ \bF(\bu) \bn} - \reallywidehat{\bG(\bu, \beps, \bphi) \bn} \right) \, d\Gamma &= \vect{0}, \label{eq:IntegralLocal_U}
	\end{align}
\end{subequations}
where $\bhv$ and $\bhT$ denote the velocity and temperature fields on the cell faces $\partial \Omega_e$, respectively, and they are defined using the hybrid vector $\bhu$ of conservative variables. This problem corresponds to the hybridisation step of the FCFV method: the goal is to eliminate the unknowns $(\bu,\beps,\bphi)$ within each cell by expressing them in terms of the hybrid variable $\bhu$.

The unknown $\bhu$ is thus computed by means of the global problem~\eqref{eq:NSstrongGlobal} whose integral form is
\begin{equation} \label{eq:IntegralGlobal}
\begin{aligned} 
\sum_{e = 1}^{\numel} \left\lbrace \int_{\partial \Omega_e \setminus \partial \Omega}  \left( \reallywidehat{ \bF(\bu) \bn} - \reallywidehat{\bG(\bu, \beps, \bphi) \bn}  \right) \, d\Gamma \right. \hspace{70pt} &\\ 
\left.+ \int_{\partial \Omega_e \cap \partial \Omega} \bb(\bu,\bhu,\beps,\bphi) \, d\Gamma \right\rbrace = \vect{0} & .
\end{aligned}
\end{equation}

The terms $\reallywidehat{ \bF(\bu) \bn}$ and $\reallywidehat{\bG(\bu, \beps, \bphi) \bn}$ appearing in equations~\eqref{eq:IntegralLocal_U} and~\eqref{eq:IntegralGlobal} stand for the convection and diffusion numerical fluxes of the conservation equations, respectively. It is worth recalling that the approximation of the numerical fluxes in the FCFV method plays a crucial role in the accuracy and stability of the computed solution, see~\cite{RS-SGH:2018_FCFV1,RS-SGH:2019_FCFV2,RS-VGSH:20,MG-RS-20}. Following the rationale discussed for high-order HDG discretisations in~\cite{JVP_HDG-VGSH:20,Peraire-PNC:10,Peraire-PNC:11,Nguyen-NP:2012,Fernandez-FNP:2017}, the traces of the numerical fluxes on the cell faces are defined as
\begin{subequations} \label{eq:NumericalFluxes}
	\begin{align}
	\reallywidehat{ \bF(\bu) \bn} &=  \bF(\bhu) \bn + \btau^a(\bhu) \, (\bu - \bhu), \label{eq:fluxF}\\
	\reallywidehat{\bG(\bu, \beps, \bphi) \bn} &= \bG(\bhu, \beps, \bphi) \bn - \btau^d \, (\bu - \bhu) . \label{eq:fluxG}
	\end{align}
\end{subequations}
On the one hand, the stabilisation tensor $\btau^a$ is associated with convection phenomena. Different expressions of $\btau^a$ are derived from the theory of Riemann solvers for nonlinear hyperbolic PDEs, as described in section~\ref{ssc:RiemannSolvers}. On the other hand, the term $\btau^d$ stands for the stabilisation tensor related to viscous effects and  is defined by means of the diagonal matrix
\begin{equation} \label{eq:diffStabilisation}
\btau^d = \frac{1}{\Rey}  \, \text{diag} \left(  0,  \vect{1}_{\nsd}, \frac{1}{(\gamma - 1) \Minf^2 \Pra} \right) ,
\end{equation}
$\bmat{1}_{\nsd}$ being an $\nsd$-dimensional vector of ones.

\begin{remark}[FCFV method for inviscid Euler equations]
	The inviscid Euler equations are obtained as the limit of the compressible Navier-Stokes equations when $\Rey \to \infty$. Setting $\bG = \bm{0}$, equation~\eqref{eq:NScompact} thus reduces to the well-known system of first-order hyperbolic PDEs modelling inviscid compressible flows. For each cell $\Omega_e, \ e = 1,\dotsc,\numel$, the corresponding FCFV local problem for the Euler equations is obtained from equation~\eqref{eq:IntegralLocal} by neglecting the mixed variables and the viscous term, namely
	\begin{equation} \label{eq:IntegralLocalEuler}
	\int_{\Omega_e} \pd{\bu}{t}  \, d\Omega  + \int_{\partial \Omega_e}  \reallywidehat{ \bF(\bu) \bn} \, d\Gamma = \vect{0} .
	\end{equation}
	Similarly, the global problem follows from the simplification of equation~\eqref{eq:IntegralGlobal} as
	\begin{equation} \label{eq:IntegralGlobalEuler}
	\sum_{e = 1}^{\numel} \left\lbrace \int_{\partial \Omega_e \setminus \partial \Omega}  \reallywidehat{ \bF(\bu) \bn}  \, d\Gamma + \int_{\partial \Omega_e \cap \partial \Omega} \bb(\bu,\bhu) \, d\Gamma \right\rbrace = \vect{0} .
	\end{equation}
\end{remark}

\subsection{Riemann solvers for the FCFV method}
\label{ssc:RiemannSolvers}

In the context of the FCFV method, Riemann solvers are defined implicitly within the convection fluxes, see equation~\eqref{eq:fluxF}, by means of appropriate expressions of the stabilisation term $\btau^a$. In this section, the stabilisation terms leading to the formulation of the Lax-Friedrichs, Roe, HLL and HLLEM numerical fluxes are presented. 

Let $\bm{A}_n(\bhu):=[\partial \bF(\bhu) / \partial \bhu] \cdot \bn$ be the Jacobian of the convective fluxes along the normal direction to a cell face. Moreover, denote by $\hlamax := \abs{\bhv \cdot \bn} + \widehat{c}$ the maximum eigenvalue in absolute value of the matrix $\bm{A}_n(\bhu)$, $\bhv$ and $\widehat{c}$ being the velocity and the speed of sound evaluated from $\bhu$, respectively. Following the unified formulation in~\cite{JVP_HDG-VGSH:20}, the Riemann solvers for the FCFV method are detailed below.

\subsubsection{Lax-Friedrichs Riemann solver}

The FCFV stabilisation tensor inspired by the Lax-Friedriches Riemann solver, see~\cite{Toro2009}, is defined as
\begin{subequations}
	\begin{equation}
	\btau^a = \hlamax \bmat{I}_{\nsd + 2} .
	\end{equation}

	\subsubsection{Roe Riemann solver} 
	
	Consider the spectral decomposition $\bm{A}_n(\bhu) = \mat{R} \mat{\Lambda} \mat{R}^{-1}$, where $\mat{\Lambda}$ is a diagonal matrix containing the eigenvalues $\lambda_i, \ i=1,\ldots,\nsd+2$ of $\bm{A}_n(\bhu)$ and $\mat{R}$ is the corresponding matrix of right eigenvectors. In addition, the diagonal matrix $\mat{\Phi}$ is given by $\mat{\Phi} = \text{diag}\left( \varphi_1, \ldots, \varphi_{\nsd+2} \right)$ where $\varphi_{i} = \max (\abs{\lambda_{i}}, \delta)$, $\delta \geq 0$ being a user-defined parameter. The Roe Riemann solver with Harten-Hyman \emph{entropy fix}~\cite{Harten-HH:1983} is obtained for the FCFV method by setting
	\begin{equation}
	\btau^a = \abs{\bm{A}_n^\delta(\bhu)} = \mat{R} \mat{\Phi} \mat{R}^{-1} .
	\end{equation}
	The parameter $\delta$ represents the threshold value of the aforementioned Harten-Hyman entropy fix. Such correction aims to remedy the failure of entropy conditions of the Roe solver, which may produce nonphysical solutions in transonic and supersonic cases. It is worth noticing that for $\delta = 0$, $\mat{\Phi} = \abs{\mat{\Lambda}}$ and the traditional Roe Riemann solver is retrieved, namely $\btau^a = \abs{\bm{A}_n(\bhu)}$.

	\subsubsection{HLL Riemann solver}
	
	The HLL Riemann solver is devised to recover the Rankine-Hugoniot condition, for a simplified scenario in which contact discontinuities are neglected, without the need of any user-defined parameter~\cite{Harten-HLL:1983}. The resulting positivity-preserving Riemann solver for the FCFV method is given by
	\begin{equation}\label{eq:RS-HLL}
	\btau^a = s^+ \bmat{I}_{\nsd + 2},
	\end{equation}
	where $s^+ := \max (0,\bhv\cdot \bn + \widehat{c})$ is an estimate of the largest wave speed of the Riemann problem.

	\subsubsection{HLLEM Riemann solver}
	
	In order to exploit both the positivity-preserving properties of the HLL Riemann solver and the capability of Roe's method to capture shear layers, the HLLEM Riemann solver~\cite{Einfeldt1988,Einfeldt1991} is employed to construct the stabilisation tensor
	\begin{equation}\label{eq:RS-HLLEM}
	\btau^a = s^+ \bm{\theta} (\bhu),
	\end{equation}
\end{subequations}
where $s^+$ is the HLL wave speed estimate and $\bm{\theta}(\bhu) = \mat{R} \bm{\Theta} \mat{R}^{-1}$ replaces the identity $\bmat{I}_{\nsd + 2}$ in equation~\eqref{eq:RS-HLL}. Note that the definition of  $\bm{\theta}(\bhu)$ exploits the matrix of right eigenvectors arising from the spectral decomposition of $\bm{A}_n(\bhu)$, whereas the diagonal matrix $\bm{\Theta}$ is given by $\bm{\Theta} = \text{diag}\left( 1, \widehat{\theta} \bmat{1}_{\nsd} , 1 \right)$, where $\widehat{\theta} =  \abs{\bhv \cdot \bn}/\hlamax$~\cite{Rohde2001}.

\subsection{FCFV discrete problem}
\label{ssc:FCFVdiscretisation}

The discrete form of the FCFV method is obtained by introducing the definition~\eqref{eq:NumericalFluxes} of the numerical fluxes into the local~\eqref{eq:IntegralLocal} and global~\eqref{eq:IntegralGlobal} problems. In addition, the vector of conservative variables $\bu$ and the mixed variables $\beps$ and $\bphi$ are discretised using a constant value at the centroid of each cell, whereas a constant approximation at the barycentre of the faces is employed for the hybrid vector $\bhu$. Finally, a quadrature rule based on a single integration point is utilised to evaluate the integral quantities on cells and faces.

For each cell $\Omega_e$, the sets of all, $\setAe$, internal, $\setIe$, and boundary, $\setEe$, faces are introduced
\begin{equation}
\setAe := \lbrace 1,\dotsc,\nfaceE \rbrace, \qquad \setIe := \lbrace j \in \setAe \mid \Gamma_{e\!, j} \cap \Gamma \neq \emptyset \rbrace , \qquad \setEe := \setAe \setminus \setIe.
\end{equation}
Moreover, $\chi_{\setIe}$ and $\chi_{\setEe}$ are defined to represent the indicator functions associated with the sets $\setIe$ and $\setEe$, respectively.

The semi-discrete form of the FCFV local problem~\eqref{eq:IntegralLocal} is: for $e = 1,\dotsc,\numel$, given the initial state $\buE = \buE^0$ at $t=0$ and the hybrid vector $\bhu_{\! j}$ on the faces $\Gamma_{e\!, j}, \ j=1,\ldots,\nfaceE$, compute $(\buE,\bepsE,\bphiE)$ that satisfy
\begin{subequations} \label{eq:FCFVLocal}
	\begin{align}
	&\abs{\Omega_e} \bepsE- \sum_{j \in \setAe} \abs{\Gamma_{e\!, j}} \dev \widehat{\bv}_{\! j} \otimes \bn_j= \vect{0}, \label{eq:FCFVLocal_eps}\\
	&\abs{\Omega_e} \bphiE - \sum_{j \in \setAe} \abs{\Gamma_{e\!, j}} \bhT_{\! j} \bn_j = \vect{0}, \label{eq:FCFVLocal_phi}\\
	& \begin{aligned} \int_{\Omega_e} \pd{\buE}{t}  \, d\Omega +  \sum_{j \in \setAe} \abs{\Gamma_{e\!, j}} & \left\lbrace \bF(\bhu_{\! j}) \bn_j - \bG(\bhu_{\! j}, \bepsE, \bphiE) \bn_{\! j} \right. \\
	& \left. + \left( \btau^a(\bhu_{\! j}) + \btau^d\right) \, (\buE - \bhu_{\! j}) \right\rbrace = \vect{0} .
	\end{aligned} \label{eq:FCFVLocal_U}
	\end{align}
\end{subequations}

\begin{remark}[Symmetry of the mixed variable]
	The mixed variable $\beps$ is a second-order symmetric tensor commonly represented using a matrix of dimension $\nsd \times \nsd$. In order to exploit the symmetry property in its discretisation, Voigt notation is employed to store only its $\msd = \nsd(\nsd + 1)/2$ non-redundant components. This approach, detailed in~\ref{app:Voigt}, was first proposed in the context of hybrid discretisation methods for high-order HDG formulations, see~\cite{Sevilla-SGKH:2018,Giacomini-GKSH:2018,MG-GS-19,MG-SGH-20,Tutorial-GSH:2020}, and later exploited also for FCFV approaches in~\cite{RS-SGH:2019_FCFV2}. For the simulation of compressible and weakly-compressible flows, this approximation of the deviatoric strain rate tensor was employed in~\cite{JVP_HDG-VGSH:20,AlS-SKGWH:20} in the context of high-order HDG methods.
\end{remark}

\begin{remark}[Time integration scheme]
	In order to obtain the fully-discrete form of the local problem~\eqref{eq:FCFVLocal_U}, an appropriate time integration scheme needs to be introduced. As previously mentioned, the present work proposes a novel FV spatial discretisations and the numerical examples in sections~\ref{sc:Convergence} and~\ref{sc:Benchmarks} focus on steady-state flows, whence this term is neglected. Nonetheless, it is worth mentioning that time marching based on an artificial time is a common relaxation approach to speed-up the convergence of a nonlinear solver. In this context, time derivative may be discretised using a backward Euler scheme, that is
	\begin{equation}\label{eq:backwardEuler}
	\int_{\Omega_e} \pd{\buE}{t}  \, d\Omega\simeq \frac{\abs{\Omega_e}}{\Dt} \left(\buE^{n+1} - \buE^{n}\right) ,
	\end{equation}
	where $\Dt$ is the artificial time step. Note that similar FCFV discrete problems are obtained using other implicit time integration schemes, e.g. higher-order backward difference formulae (BDF)~\cite{Jaust-JS:2014,Sanjay2020,Nguyen-NP:2012}, providing additional accuracy in the simulation of transient phenomena. Of course, in the case of transient simulations, Newton-Raphson iterations are performed at each time step to solve the nonlinear problem.
\end{remark}

In a similar fashion, the discrete form of the FCFV global problem~\eqref{eq:IntegralGlobal} is: compute the hybrid vector $\bhu$ such that
\begin{equation} \label{eq:FCFVGlobal}
\begin{aligned}
\sum_{e = 1}^{\numel} \abs{\Gamma_{e\!, i}} \left\lbrace \left[ \bF(\bhu_{\! i}) \bn_i - \bG(\bhu_{\! i}, \bepsE, \bphiE) \bn_{\! i} \right. \right. \hspace{110pt} & \\
+ \left. \left. \left( \btau^a(\bhu_{\! i}) + \btau^d\right) \, (\buE - \bhu_{\! i}) \right] \chi_{\setIe}(i) \right. \hspace{60pt} &\\
\left. + \, \bb(\bu,\bhu,\beps,\bphi) \chi_{\setEe}(i) \right\rbrace = \vect{0} &,
\end{aligned}
\end{equation}
for all $i \in \setAe$.

It is worth noticing that both the local~\eqref{eq:FCFVLocal} and global~\eqref{eq:FCFVGlobal} problems are nonlinear. More precisely, let $\bqE = (\bepsE,\bphiE)$ be the set of mixed variables introduced by the FCFV formulation in the cell $\Omega_e$. The resulting system of algebraic-differential equations arising from the local problem~\eqref{eq:FCFVLocal} is
\begin{subequations} \label{eq:LocalSemiDiscreteCompact} 
	\begin{align} 
	\qE  &= \qE(\hu) , \label{eq:LocalSemiDiscreteCompact_Q}  \\
	\abs{\Omega_e} \frac{d \uE}{d t} + \re(\uE,\qE,\hu) &= \bmat{0} \label{eq:LocalSemiDiscreteCompact_U}  ,
	\end{align}
\end{subequations}
where $\uE$ and $\qE$ are the vectors containing the values of the local and mixed variables, respectively, at the centroid of the cell, whereas the vector $\hu$ collects the values of the hybrid variable at the barycentres of its faces. On the one hand, equations~\eqref{eq:FCFVLocal_eps} and~\eqref{eq:FCFVLocal_phi} provide analytical expressions of the mixed variables in terms of the hybrid unknown, see equation~\eqref{eq:LocalSemiDiscreteCompact_Q}. On the other hand, equation~\eqref{eq:FCFVLocal_U} is nonlinear and the residual vector obtained from its spatial discretisation is denoted by $\re$. Upon linearisation of equation~\eqref{eq:LocalSemiDiscreteCompact_U} via a Newton-Raphson procedure, the $\numel$ local problems allow to express $\uE$ and $\qE$ in each cell $\Omega_e, \ e=1,\ldots,\numel$ in terms of the unknown $\hu$ on its faces. The resulting expressions are plugged into equation~\eqref{eq:FCFVGlobal} and all the degrees of freedom inside the cells are eliminated from the global problem, leading to
\begin{equation} \label{eq:GlobalSemiDiscreteCompact} 
\sum_{e = 1}^{\numel} \hre(\hu) = \bmat{0},
\end{equation}
where $\hre$ is the nonlinear residual vector related to the unknowns $\hu$ associated with cell $\Omega_e$. The global problem~\eqref{eq:GlobalSemiDiscreteCompact}, whose structure is detailed in~\ref{app:FCFVsolver}, is thus solved by means of a Newton-Raphson linearisation.

\section{Numerical convergence studies}
\label{sc:Convergence}

In this section, the optimal convergence of the FCFV method is examined for different compressible flows, namely inviscid and viscous laminar flows. The accuracy of the method is evaluated using different types of meshes, employing both triangular and quadrilateral elements, with special attention to the robustness of the methodology to cell distortion and stretching.

\subsection{Inviscid Ringleb flow}
\label{ssc:Ringleb}

The convergence properties of the FCFV method in the inviscid limit are examined through the Ringleb flow problem~\cite{Sanjay2020,Nguyen-NP:2012}. This 2D example describes a smooth transonic flow, for which an analytical expression of the solution can be computed via the hodograph method~\cite{Chiocchia1985}. At a given point $(x,y)$, the solution is obtained as result of the nonlinear implicit equation
\begin{equation}\label{eq:RinglebNLeq}
\left( x + \frac{J}{2} \right)^2 + y^2 = \frac{1}{4 \rho^2 V^4},
\end{equation}
where $c$ is the speed of sound, whereas density $\rho$, velocity magnitude $V$, pressure $p$ and $J$ are determined as
\begin{equation}
\begin{aligned}
\rho &= c^{2/(\gamma - 1)}, & V &= \sqrt{\frac{2(1 - c^2)}{\gamma - 1}}, \\
p &= \frac{1}{\gamma} c^{2\gamma/(\gamma - 1)}, & J &= \frac{1}{c} + \frac{1}{3c^3} + \frac{1}{5c^5} - \frac{1}{2}\log\left( \frac{1+c}{1-c}\right).
\end{aligned}
\end{equation}
Finally, the velocity vectorfield is given by
\begin{equation}\label{eq:RinglebV}
\bv = \begin{Bmatrix} -  \sign{y} V \sin \theta\\ V \cos \theta 	\end{Bmatrix} ,
\end{equation}
$\sign{\cdot}$ being the \emph{sign} operator and $\sin (2 \theta) = 2 \rho V^2 y$.

\begin{figure}[!ht]
	\subfloat[Mesh M2 \label{fig:Ringleb_mesh_QUA_H3}]{\includegraphics[width=0.24\textwidth]{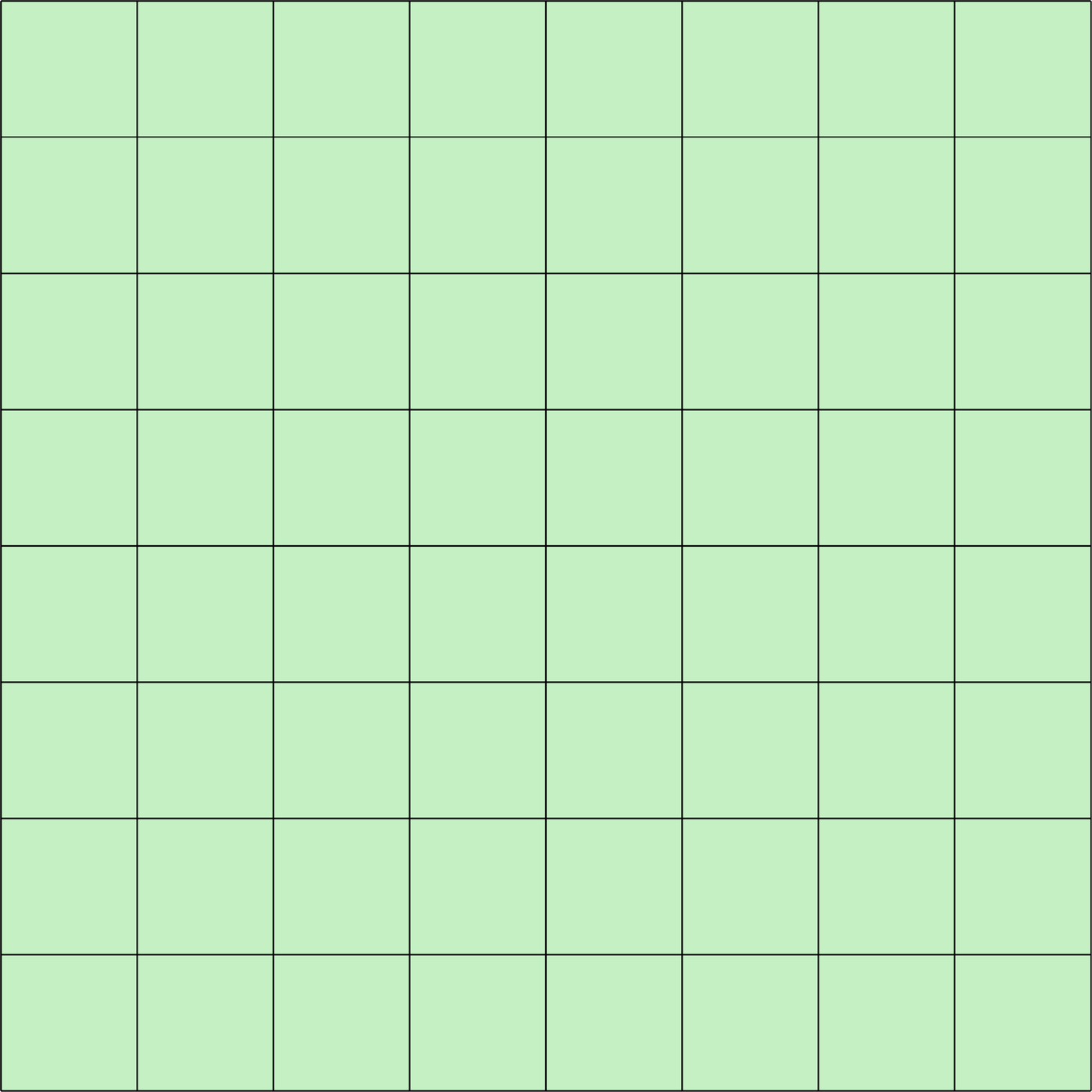}} \hfill
	\subfloat[Mesh M4 \label{fig:Ringleb_mesh_QUA_H5}]{\includegraphics[width=0.24\textwidth]{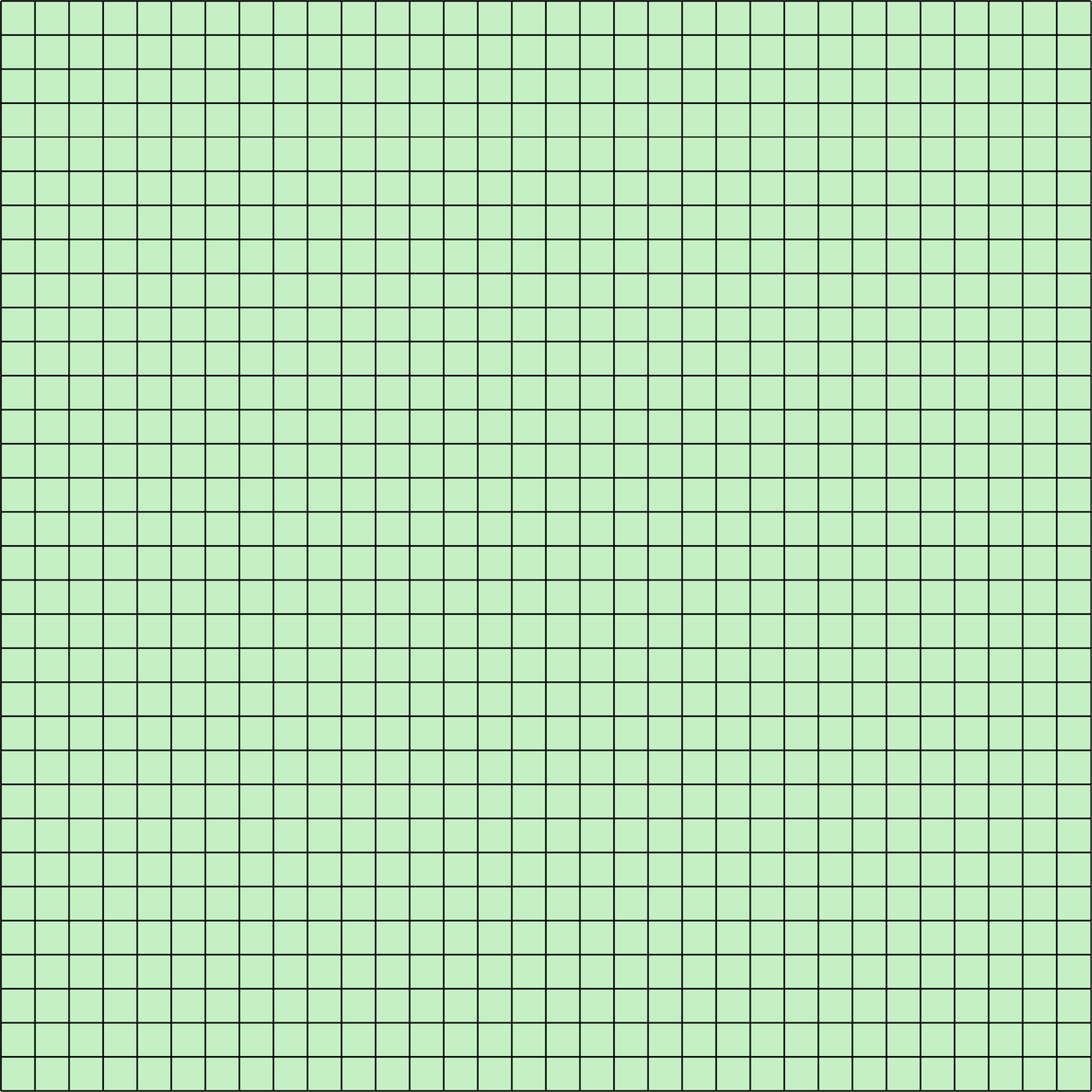}} \hfill
	\subfloat[Mesh M2 \label{fig:Ringleb_mesh_TRI_H3}]{\includegraphics[width=0.24\textwidth]{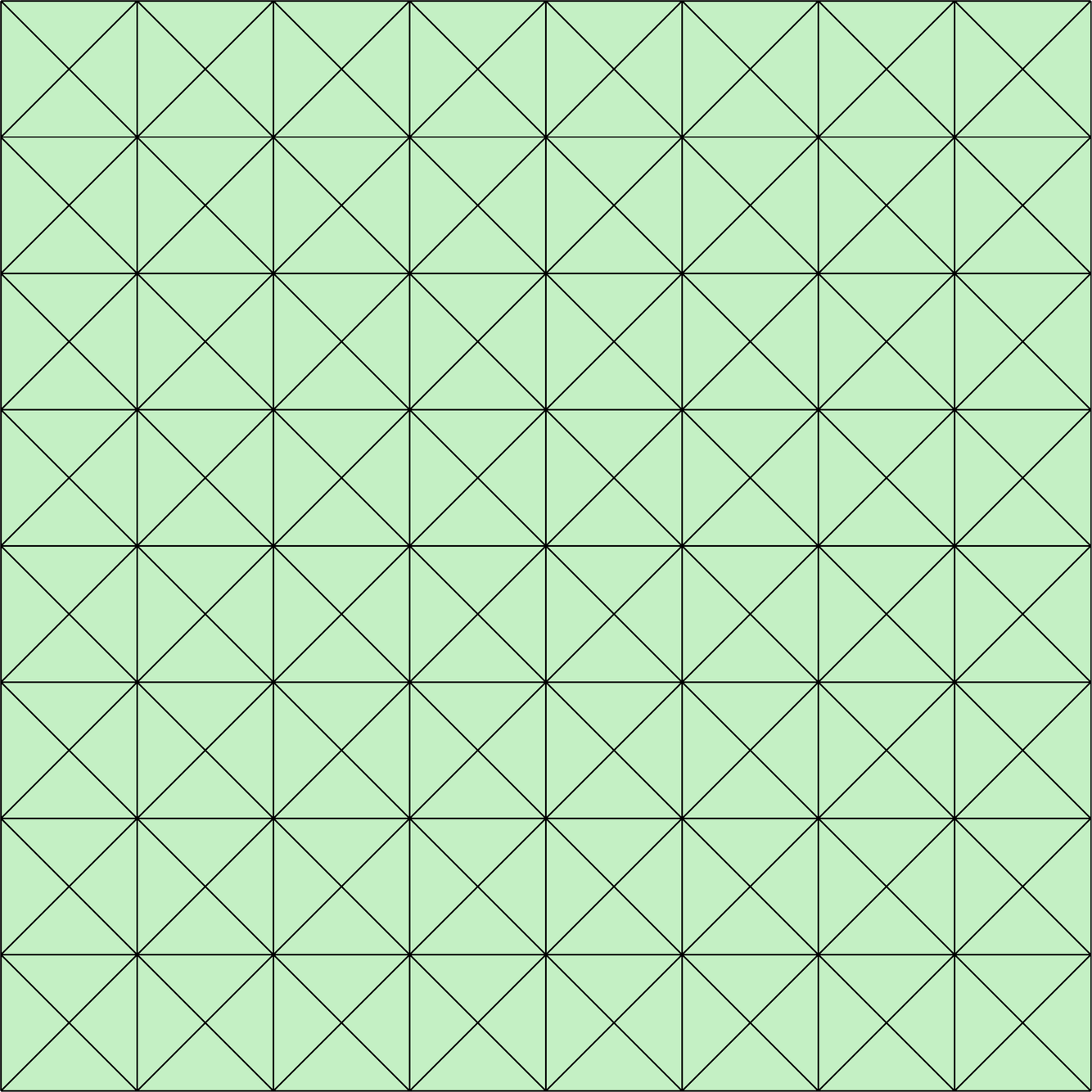}} \hfill
	\subfloat[Mesh M4 \label{fig:Ringleb_mesh_TRI_H5}]{\includegraphics[width=0.24\textwidth]{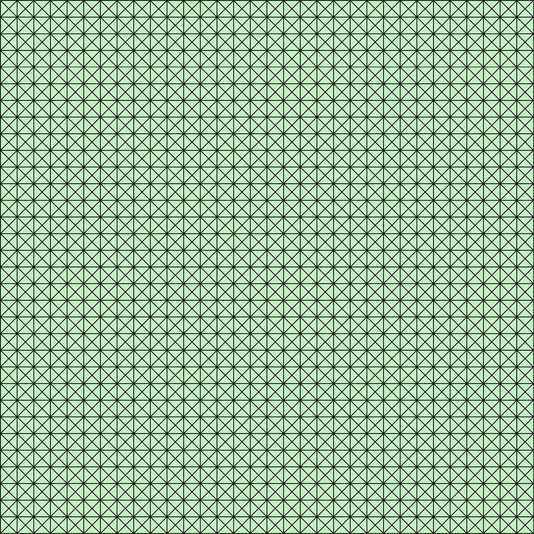}}
	\caption{Uniform (a-b) quadrilateral and (c-d) triangular meshes of $\Omega = [0,1]^2$.}
	\label{fig:Ringleb_meshes}
\end{figure}

\begin{figure}[!ht]
	\subfloat[Quad. M2 \label{fig:Ringleb_Ma_QUA_H3}]{\includegraphics[width=0.24\textwidth]{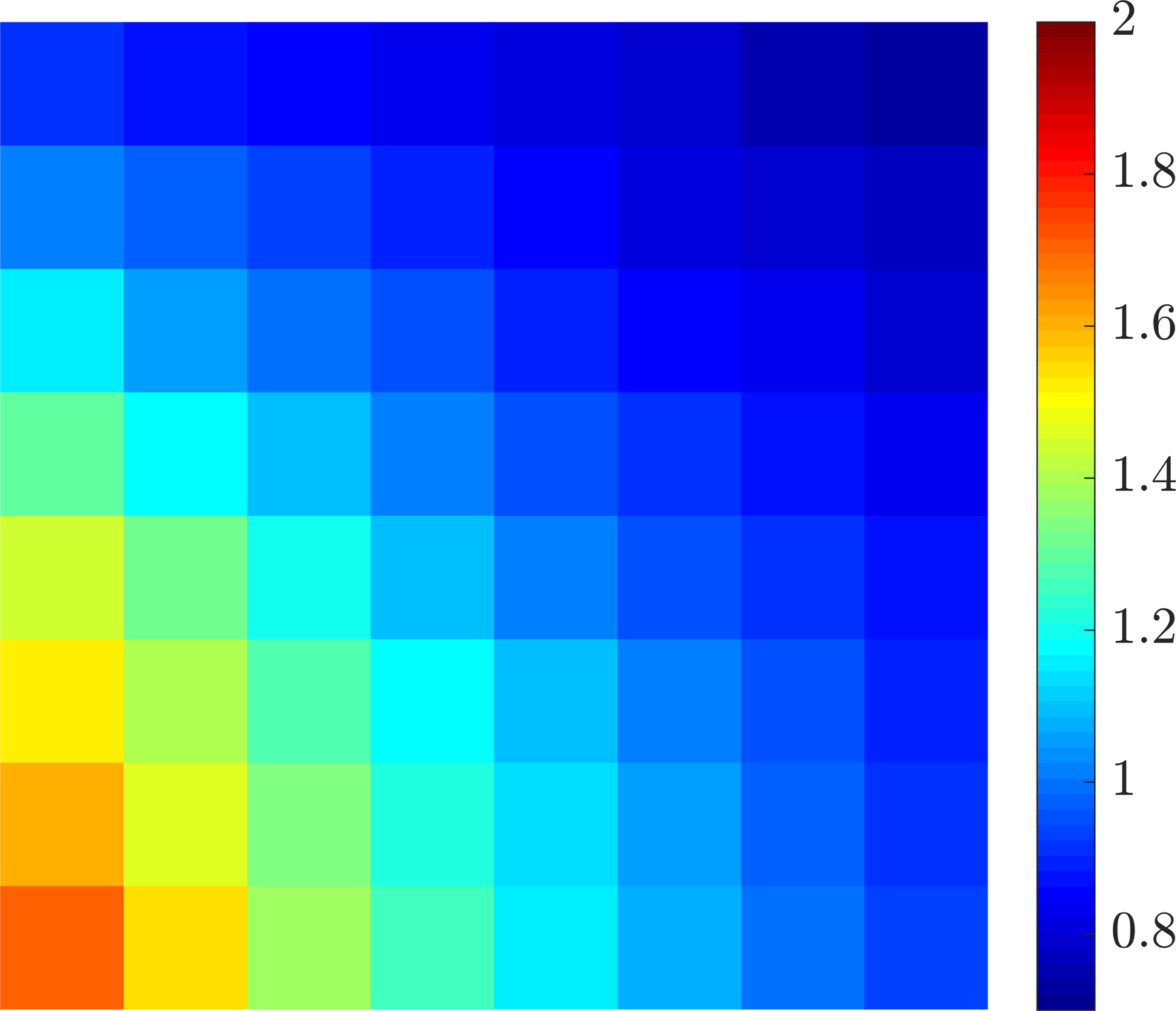}} \hfill
	\subfloat[Quad. M4 \label{fig:Ringleb_Ma_QUA_H5}]{\includegraphics[width=0.24\textwidth]{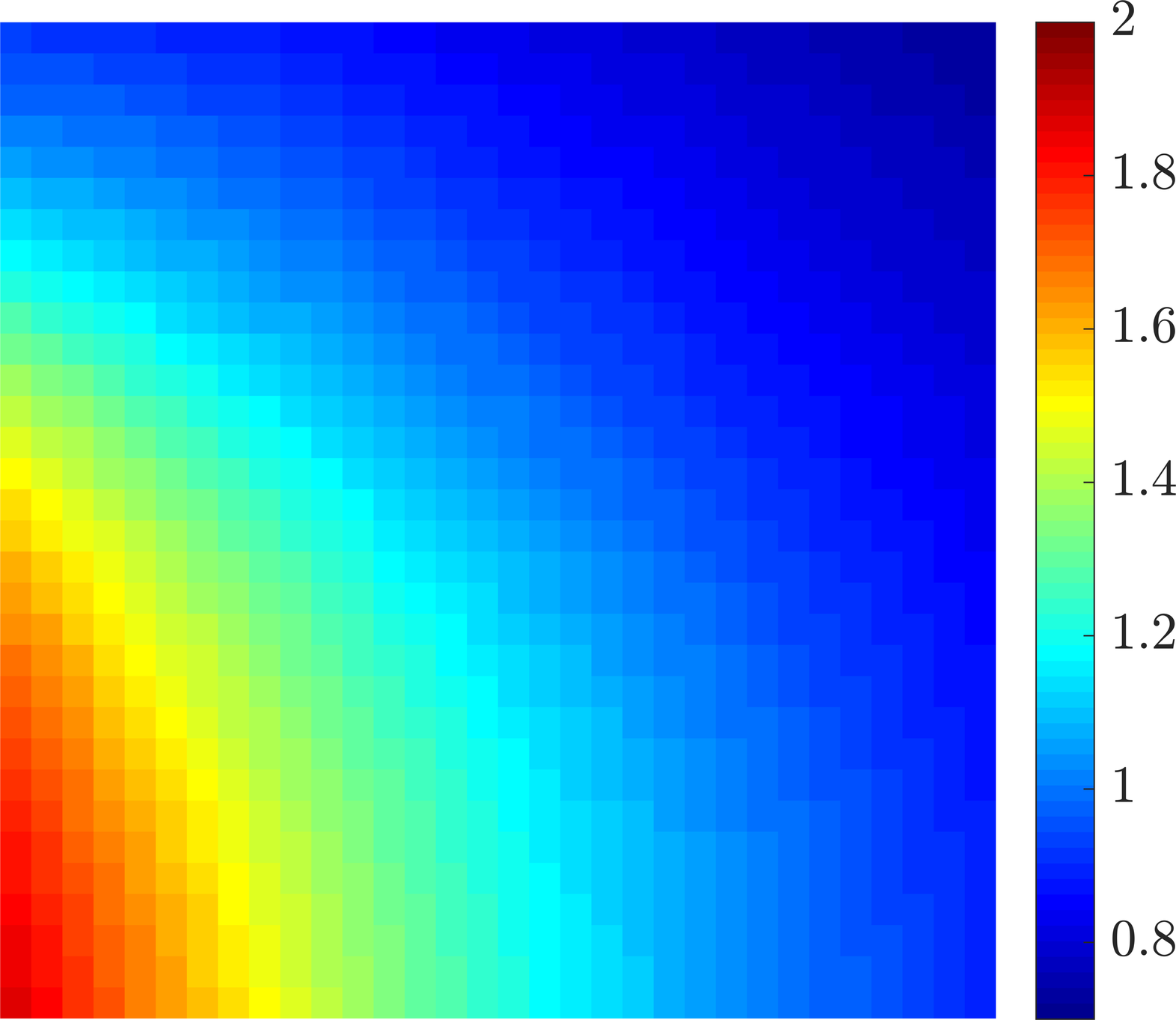}} \hfill
	\subfloat[Triangles M2 \label{fig:Ringleb_Ma_TRI_H3}]{\includegraphics[width=0.24\textwidth]{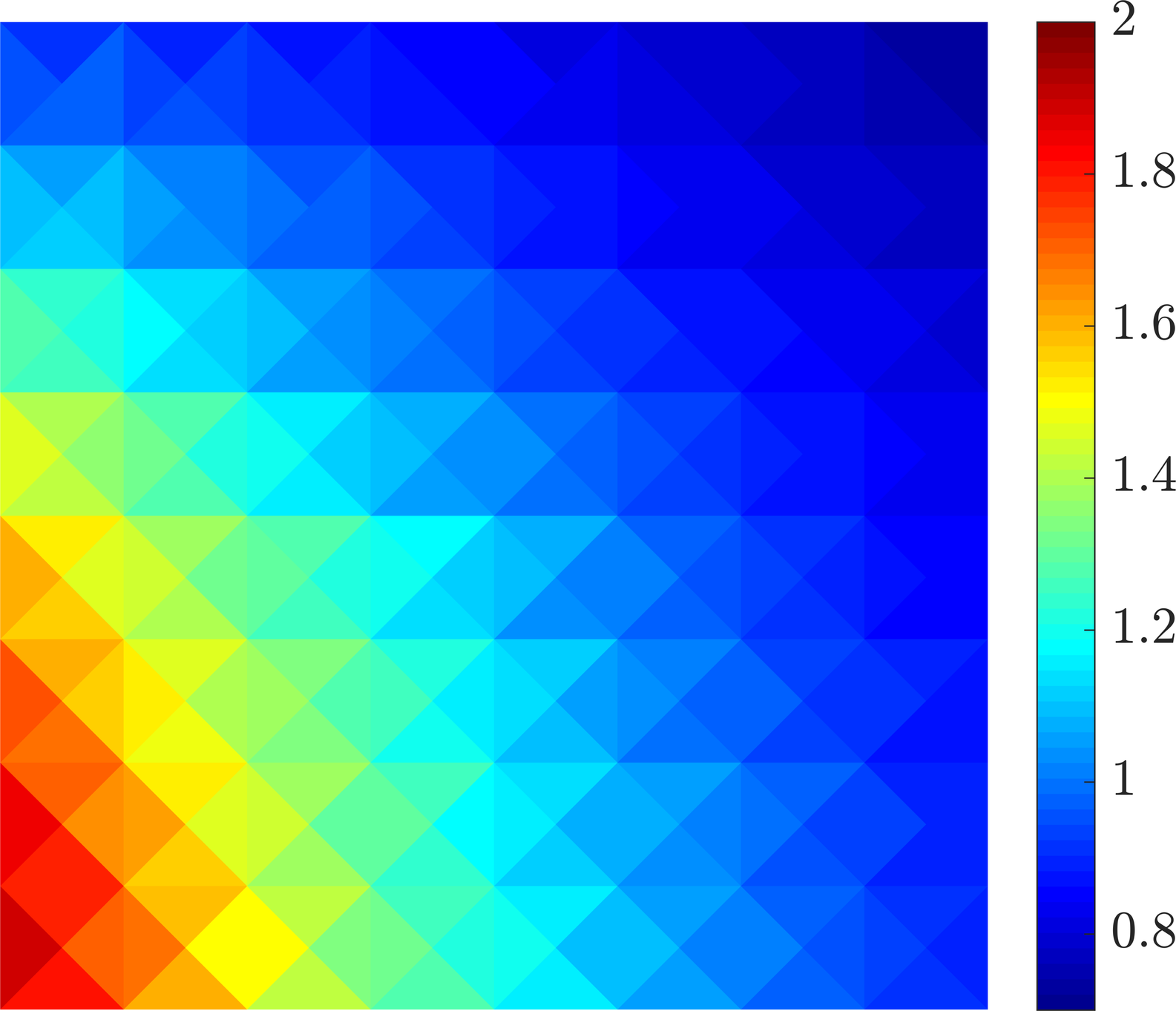}} \hfill
	\subfloat[Triangles M4 \label{fig:Ringleb_Ma_TRI_H5}]{\includegraphics[width=0.24\textwidth]{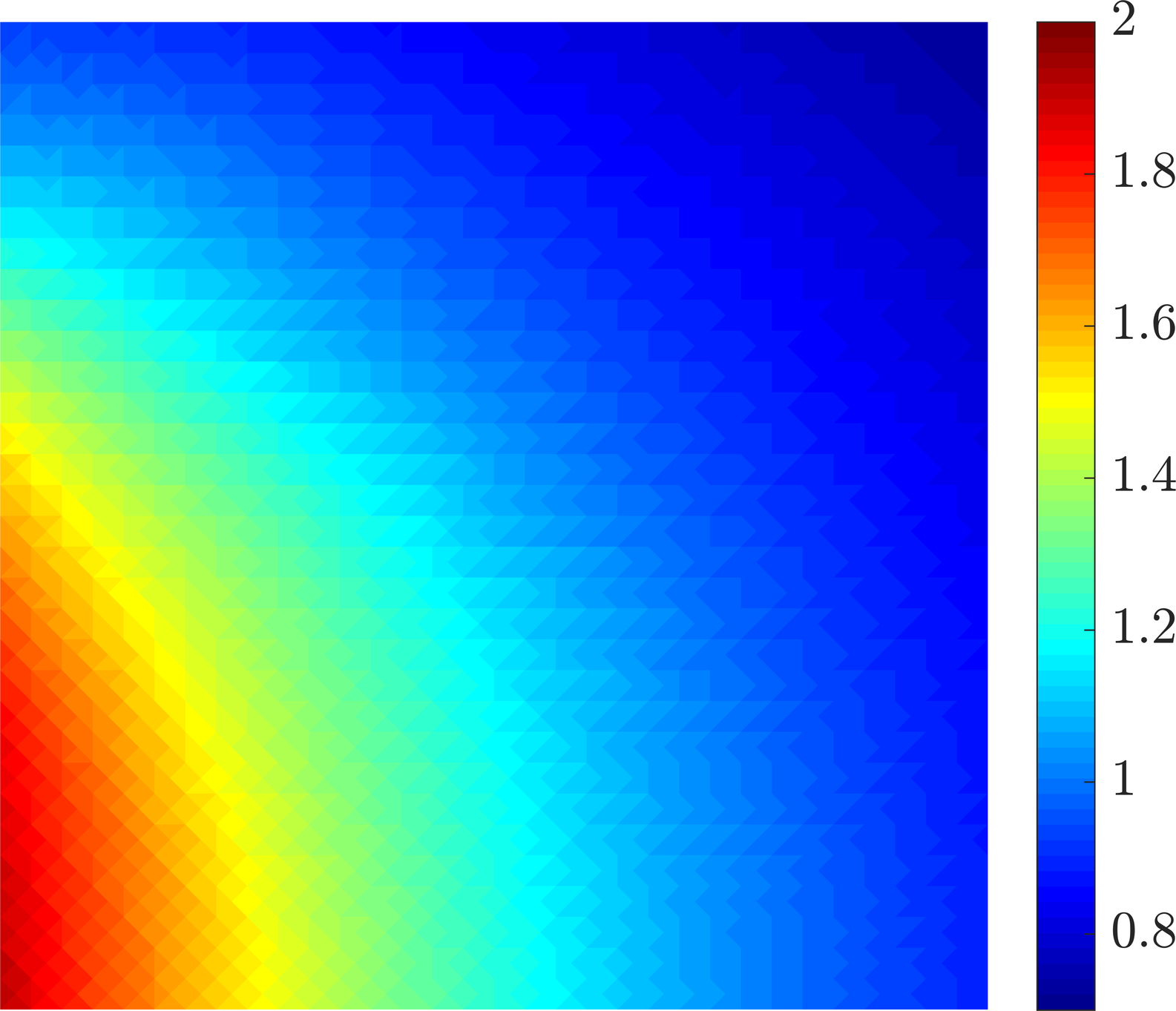}}
	\caption{Ringleb flow - Mach number distribution using the (a-b) quadrilateral and (c-d) triangular meshes in Figure \ref{fig:Ringleb_meshes} employing the HLL Riemann solver.}
	\label{fig:Ringleb_quantities}
\end{figure}

The computational domain is defined as $\Omega = [0,1]^2$ and far-field boundary conditions are imposed on $\partial \Omega$. Figure~\ref{fig:Ringleb_meshes} displays two levels of refinement of the domain using uniform meshes of triangular and quadrilateral cells.
The corresponding approximation of the Mach number distribution on these meshes is reported in figure~\ref{fig:Ringleb_quantities}.

\begin{figure}[!ht]
	\centering
	\subfloat[Density, $\rho$ \label{fig:Ringleb_convergence_density}]{\includegraphics[width=0.33\textwidth]{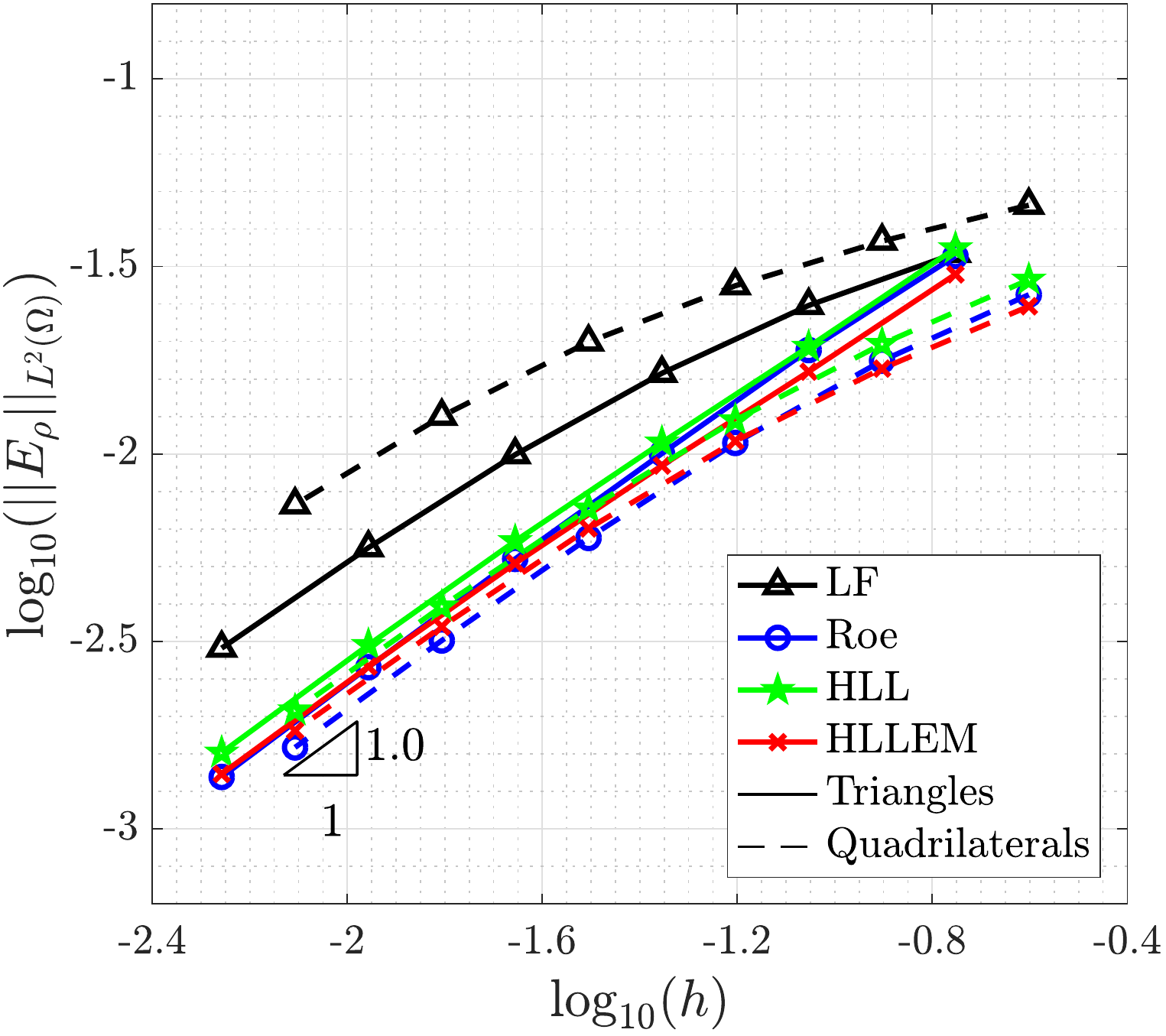}} \hfill
	\subfloat[Momentum, $\rho \bm{v}$ \label{fig:Ringleb_convergence_momentum}]{\includegraphics[width=0.33\textwidth]{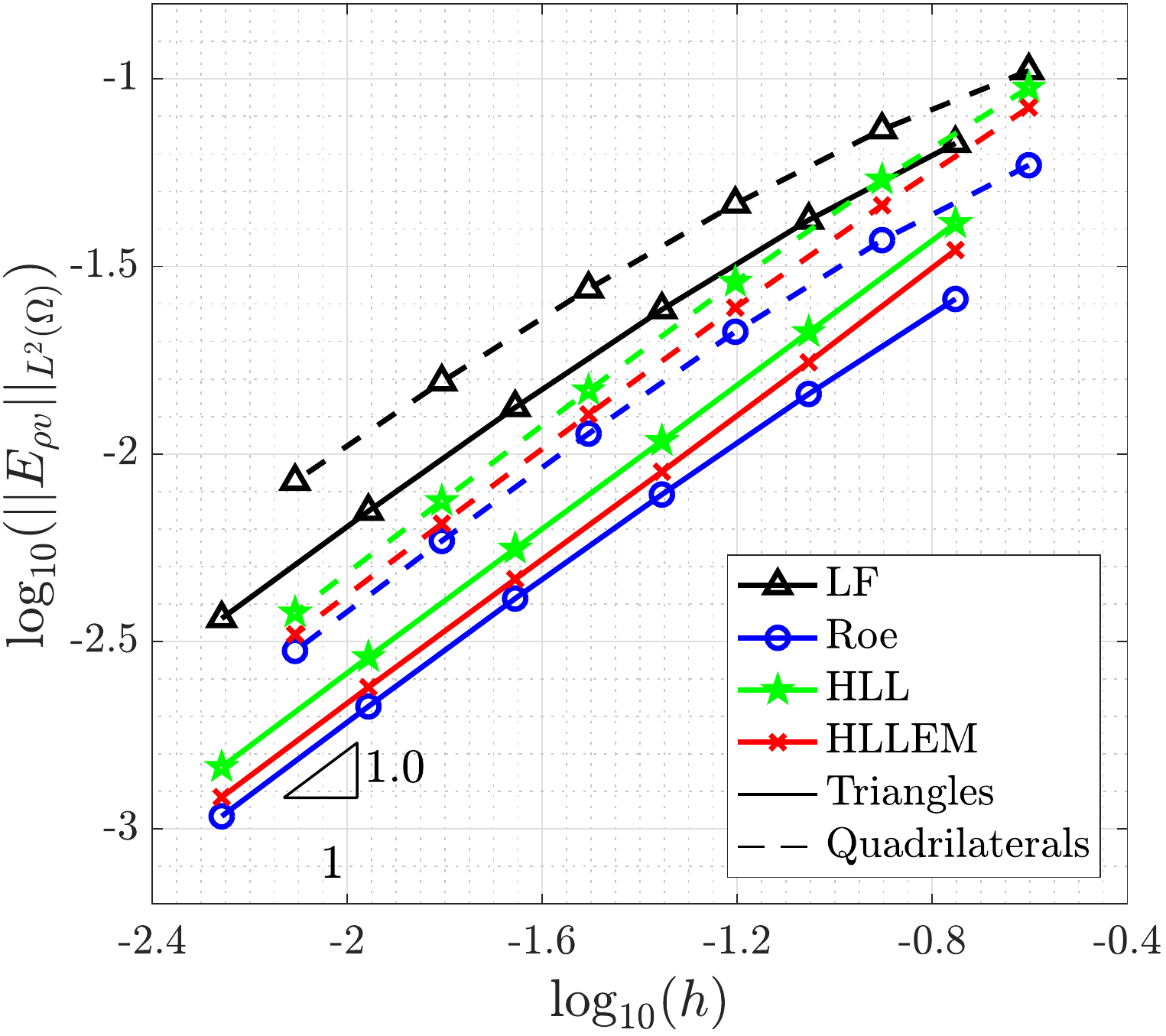}} \hfill
	\subfloat[Energy, $\rho E$ \label{fig:Ringleb_convergence_energy}]{\includegraphics[width=0.33\textwidth]{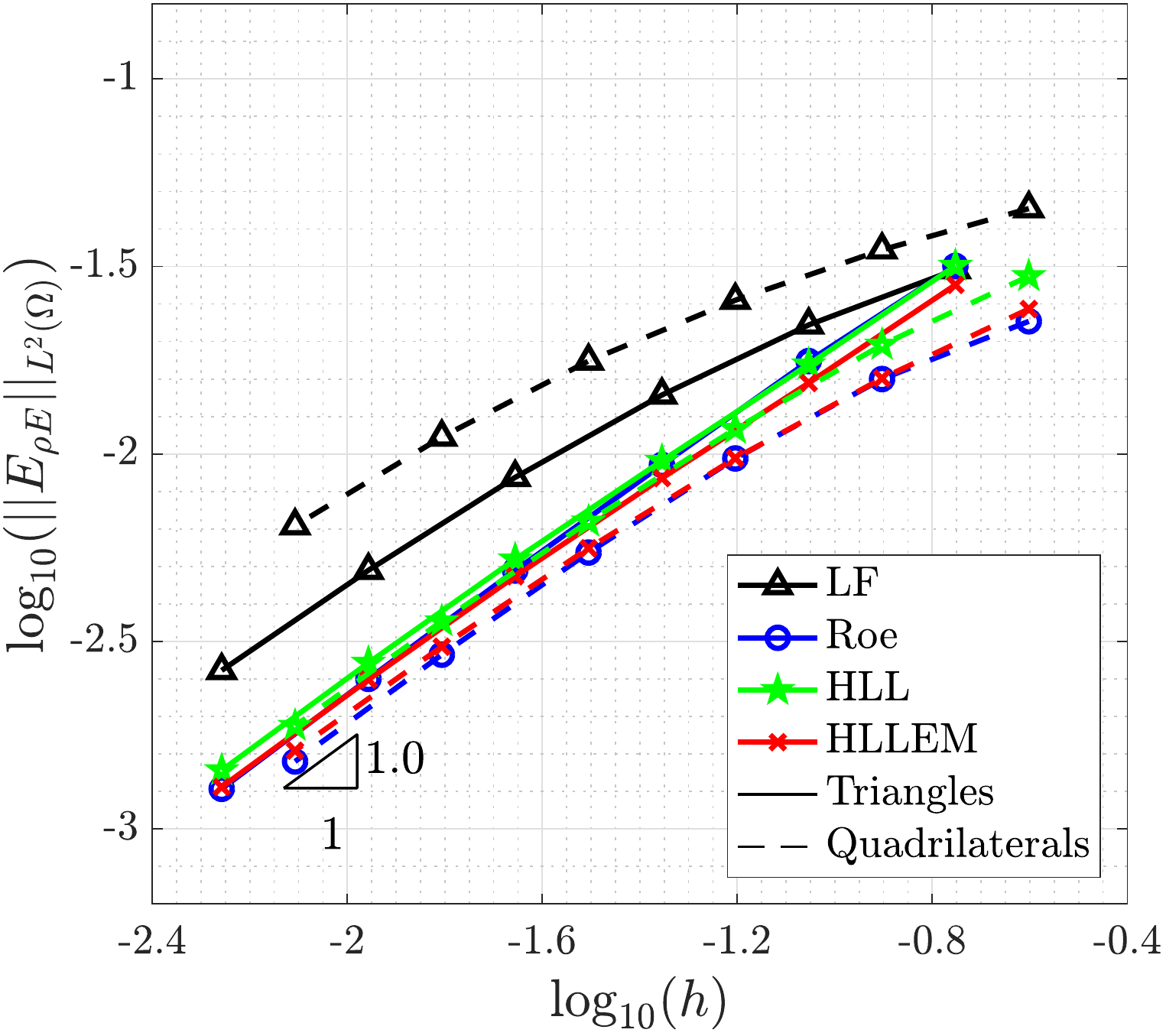}}
	\caption{Ringleb flow - $h$-convergence of the error of (a) density, (b) momentum and (c) energy in the $\eltwo(\Omega)$ norm, using Lax-Friedrichs (LF), Roe, HLL and HLLEM Riemann solvers and uniform meshes of triangles and quadrilaterals.}
	\label{fig:Ringleb_Error}
\end{figure}

The relative error of the numerical approximation, measured in the $\eltwo(\Omega)$ norm as function of the characteristic mesh size $h$, is examined for the four Riemann solvers discussed in section~\ref{ssc:RiemannSolvers}. The $h$-convergence study is performed using the sets of meshes introduced above and the results are displayed in figure~\ref{fig:Ringleb_Error}.
Optimal convergence of order $1$ is obtained for the approximation of the conservative variables regardless of the Riemann solver utilised, showing the robustness of the FCFV approach in the inviscid case. The method displays optimal convergence properties using both triangular and quadrilateral cells. Errors of the order of $10^{-3}$ are achieved, independently of the type of cells, by Roe, HLL and HLLEM Riemann solvers in the approximation of density and energy. Concerning the momentum, similar levels of accuracy are obtained using the three Riemann solvers on triangular meshes, whereas the errors are slightly higher using quadrilateral cells. The Lax-Friedrichs (LF) numerical flux displays the worst performance among the analysed Riemann solvers, showing errors almost half an order of magnitude and almost one order of magnitude higher using triangular and \hl{quadrilateral meshes}, respectively.

\subsection{Viscous laminar Couette flow}
\label{ssc:Couette}

A Couette flow with source term~\cite{Schutz-SWM:2012,Nguyen-NP:2012} is defined in the domain $\Omega = [0,1]^2$ to examine the convergence properties of the FCFV method in the viscous laminar regime. The analytical expressions of  velocity, pressure and temperature, are
\begin{equation}\label{eq:Couette_analytical}
\begin{aligned}
\bv &= \begin{Bmatrix} y \log (1 + y) \\ 0  \end{Bmatrix}, \qquad p =\frac{ 1}{\gamma \Minf^2} \\
T &= \frac{1}{(\gamma - 1) \Minf^2} \left[ \alpha_c + y(\beta_c - \alpha_c) + \frac{(\gamma - 1)\Minf^2 \Pra}{2} y(1-y)\right],
\end{aligned}
\end{equation}
where $\alpha_c = 0.8$, $\beta_c = 0.85$ and the free-stream Mach number is set to $\Minf=0.15$. Assuming constant viscosity, the source term is determined from the exact solution and is given by
\begin{equation}\label{eq:Couette_source}
\begin{aligned}
\bm{S} =  {-} \frac{1}{\Rey} \Bigg\lbrace \Bigg. & 0, \,  \frac{2 + {y}}{(1+{y})^2}, \, 0, \,  \\
& \Bigg. \log^2(1 + {y}) + \frac{{y}\log(1+{y})}{1+{y}} + \frac{{y}(3+2{y})\log(1+{y})-2{y}-1}{(1+{y})^2} \Bigg\rbrace  \tras.
\end{aligned}
\end{equation}
Finally, the boundary conditions are prescribed on $\partial \Omega$ employing the expression of the analytical solution.

\begin{figure}[!ht]
	\centering
	\subfloat[Density, $\rho$ \label{fig:Couette_convergence_density}]{\includegraphics[width=0.33\textwidth]{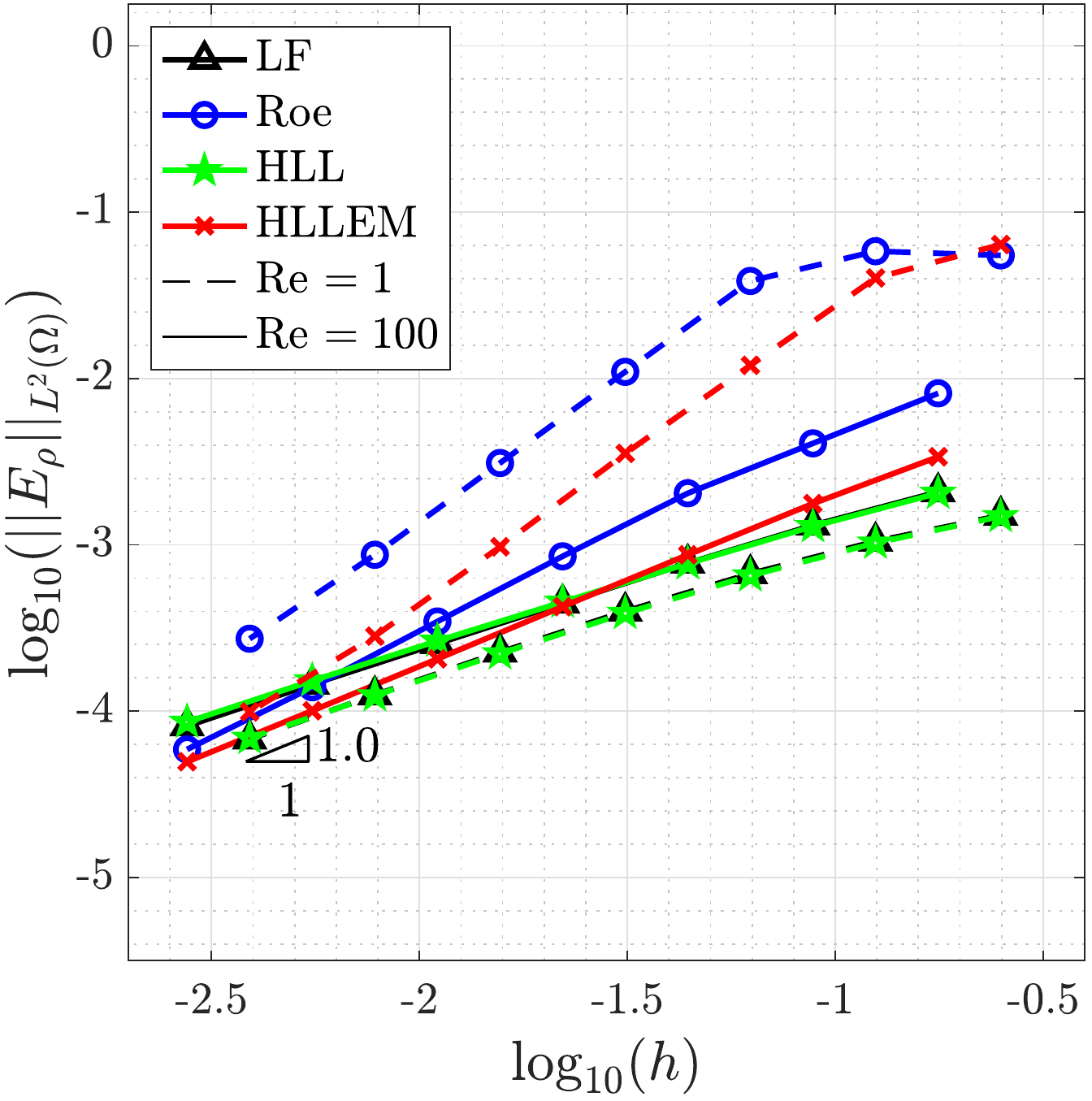}} \hfill
	\subfloat[Momentum, $\rho \bm{v}$ \label{fig:Couette_convergence_momentum}]{\includegraphics[width=0.33\textwidth]{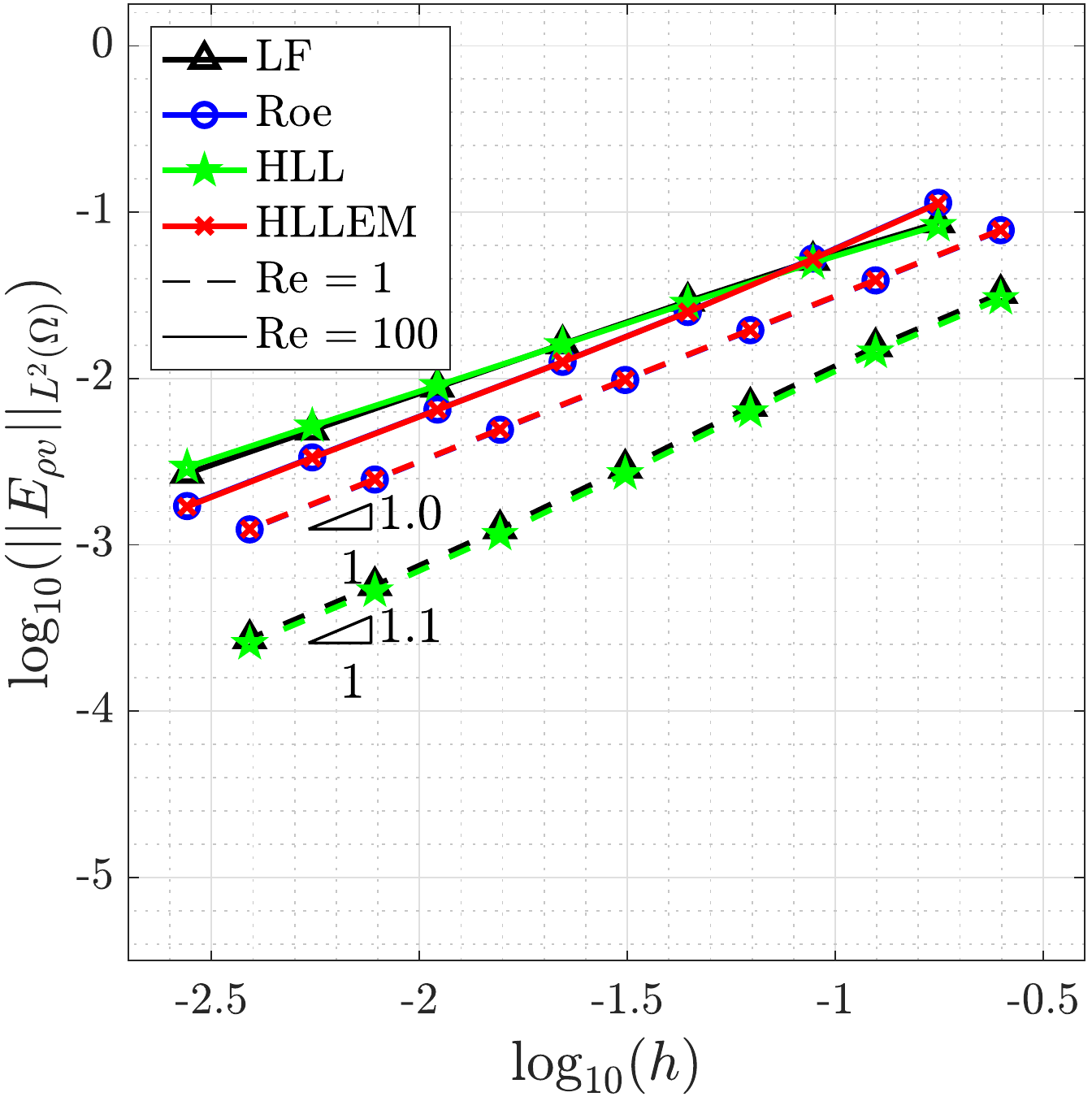}} \hfill
	\subfloat[Energy, $\rho E$ \label{fig:Couette_convergence_energy}]{\includegraphics[width=0.33\textwidth]{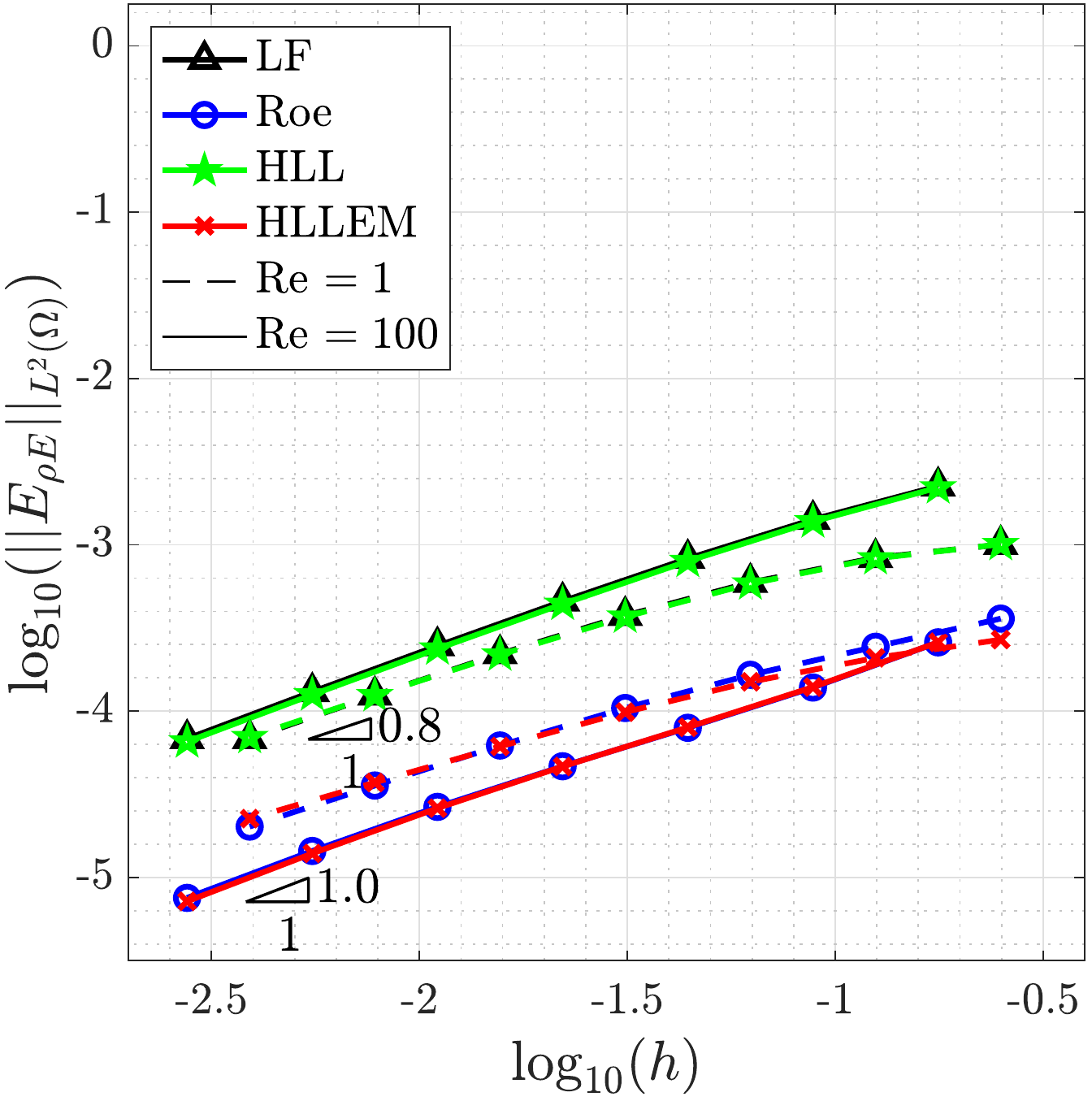}} \\
	\subfloat[Viscous stress, $\bsigma$ \label{fig:Couette_convergence_stress}]{\includegraphics[width=0.33\textwidth]{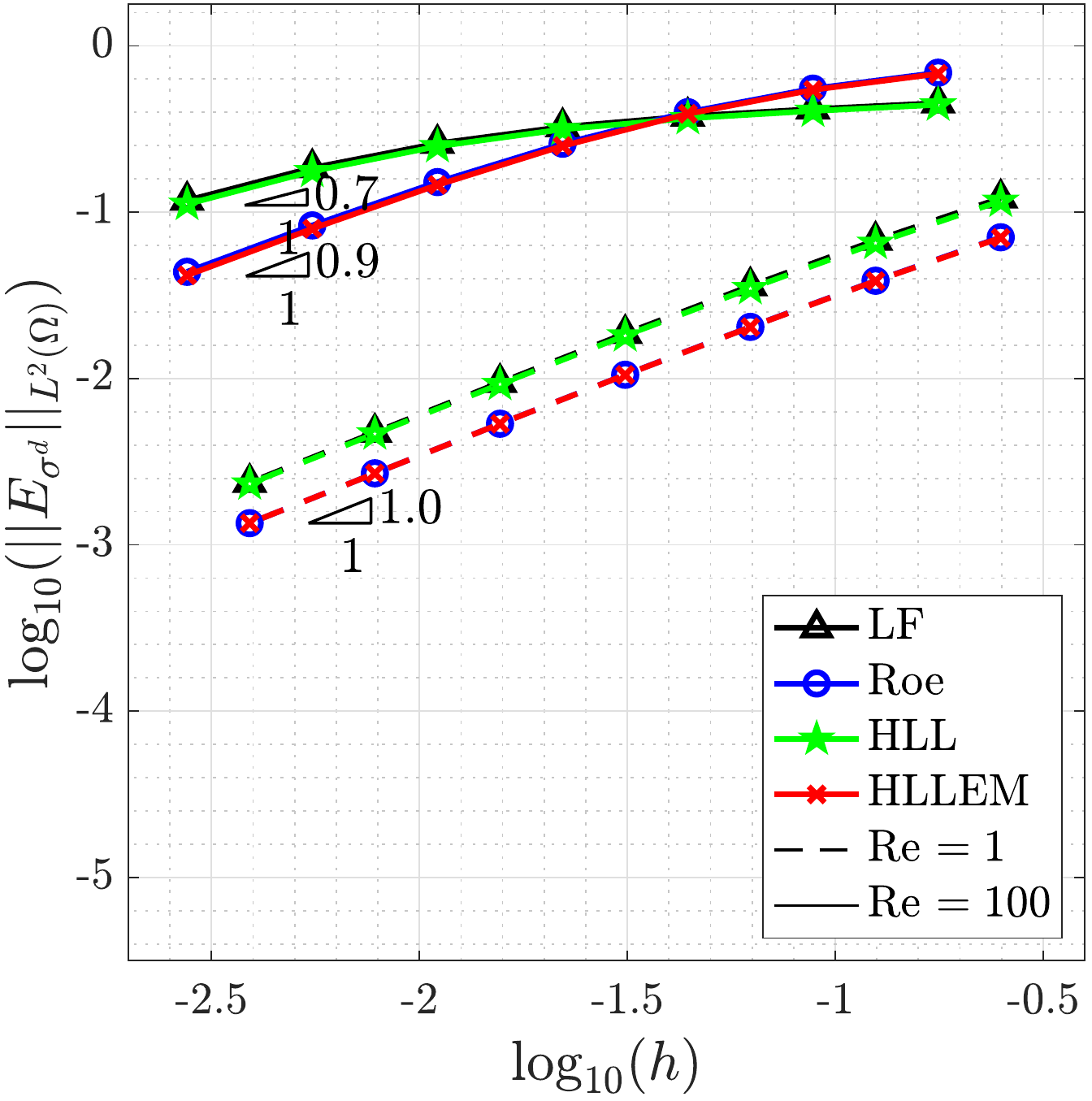}} \quad
	\subfloat[Heat flux, $\bq$ \label{fig:Couette_convergence_heatFlux}]{\includegraphics[width=0.33\textwidth]{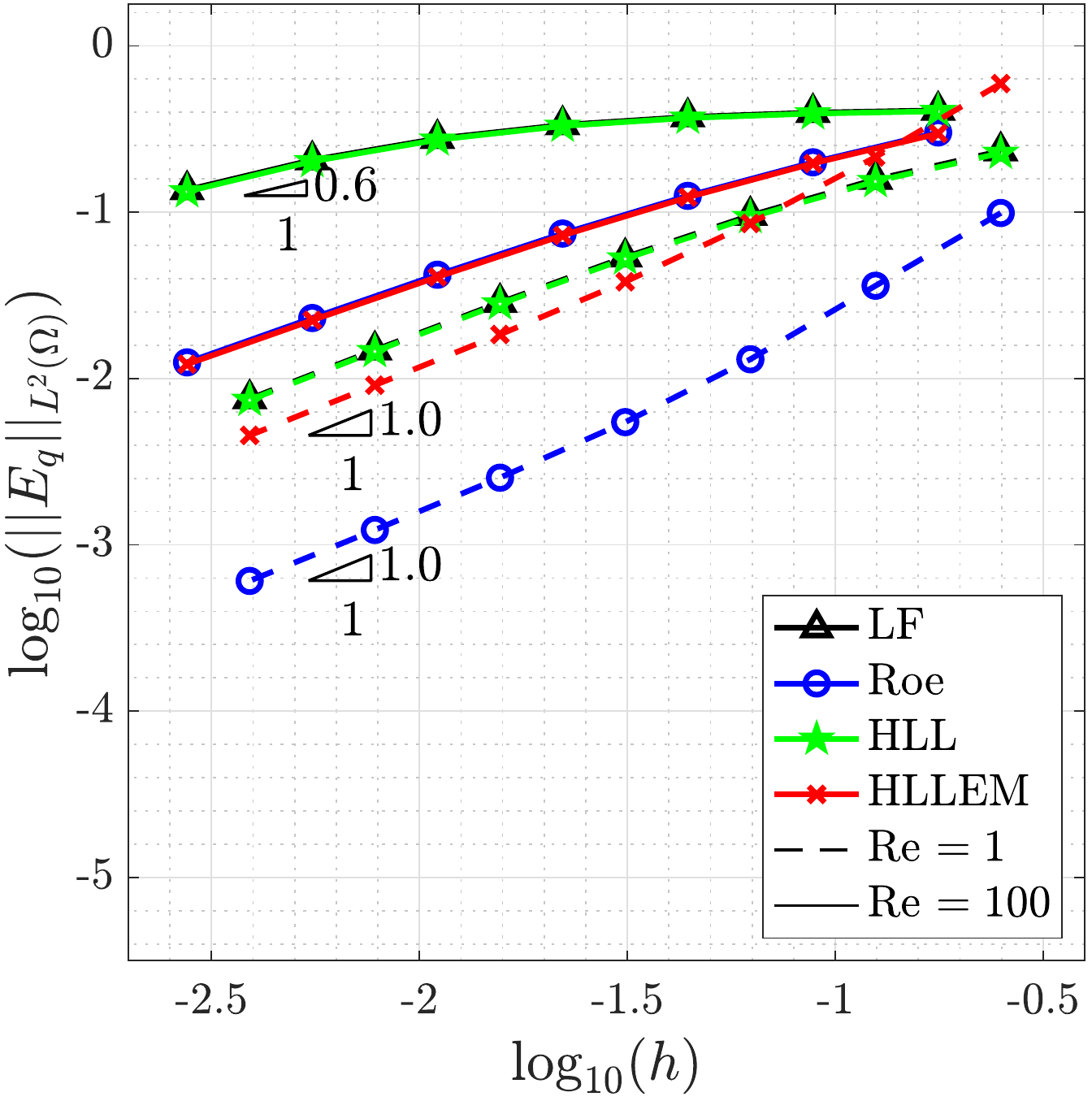}} 
	\caption{Couette flow - $h$-convergence of the error of (a) density, (b) momentum (c) energy, (d) viscous stress tensor and (e) heat flux in the $\eltwo(\Omega)$ norm, using Lax-Friedrichs (LF), Roe, HLL and HLLEM Riemann solvers and Reynolds number $\Rey = 1$ and $\Rey = 100$.}
	\label{fig:Couette_ErrorRe}
\end{figure}

The $h$-convergence study is performed using the meshes of triangular cells in figure~\ref{fig:Ringleb_meshes}. Figure~\ref{fig:Couette_ErrorRe} reports the results using different Riemann solvers and for Reynolds number $\Rey = 1$ and $\Rey = 100$.
The approximation of the conservative variables displays optimal convergence regardless of the Reynolds number and of the employed Riemann solver. The proposed method is thus able to provide optimal accuracy in the conservative quantities also in case viscous phenomena are considered, with errors between $10^{-3}$ and $10^{-5}$. Regarding the viscous stress tensor and the heat flux, optimal accuracy is achieved using HLLEM and Roe Riemann solvers independently of the Reynolds number, whereas the Lax-Friedrichs and HLL fluxes appear to be more sensitive to increasing values of the Reynolds number. 

Finally, it is worth mentioning that the test case under analysis features an incompressible flow ($\nabla \cdot \bv = 0$, $\Minf=0.15$). Despite this additional difficulty, the FCFV method is capable of computing an accurate approximation without the need of introducing specific pressure corrections like the well-known SIMPLE algorithm~\cite{Patankar-PS:72}. Thus, an important advantage of the proposed methodology is its robustness in the incompressible limit as futher detailed in section~\ref{ssc:lowMach}.

\subsection{Influence of cell distortion and stretching}
\label{ssc:StretchingDistortion}

In this section, the sensitivity of the FCFV method to cell distortion and stretching is investigated. For the sake of brevity, this study focuses on the viscous case in order to analyse the effect of mesh regularity on both conservative and mixed variables. To this end, two sets of meshes are generated for the domain $\Omega = [0,1]^2$ by modifying the regular ones employed in the previous examples. 

\begin{figure}[!ht]
	\subfloat[Mesh M3D \label{fig:Mesh_QUA_Distortion_H4}]{\includegraphics[width=0.24\textwidth]{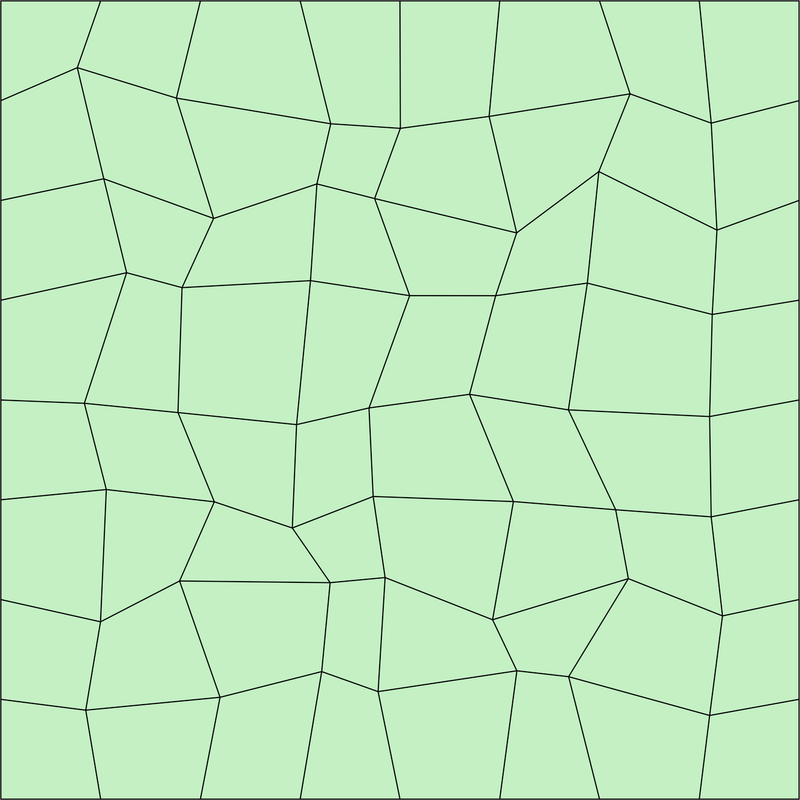}} \hfill
	\subfloat[Mesh M5D \label{fig:Mesh_QUA_Distortion_H6}]{\includegraphics[width=0.24\textwidth]{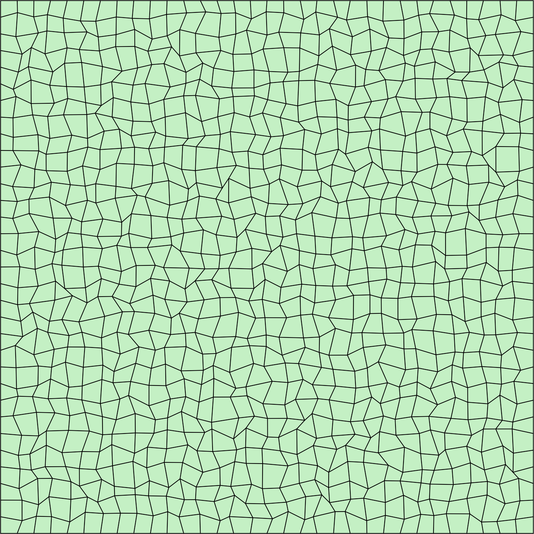}} \hfill
	\subfloat[Mesh M3D \label{fig:Mesh_TRI_Distortion_H4}]{\includegraphics[width=0.24\textwidth]{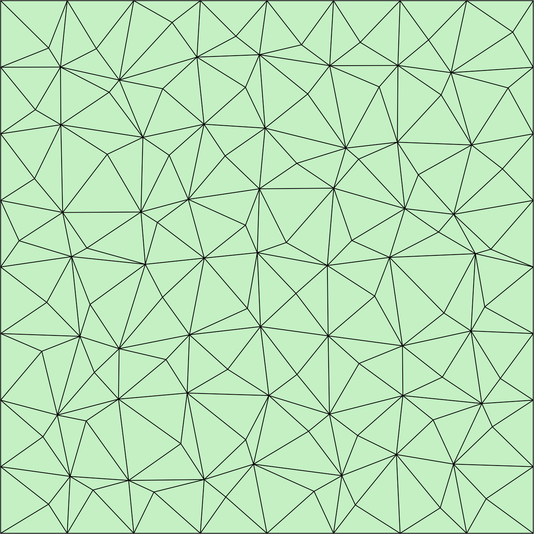}} \hfill
	\subfloat[Mesh M5D \label{fig:Mesh_TRI_Distortion_H6}]{\includegraphics[width=0.24\textwidth]{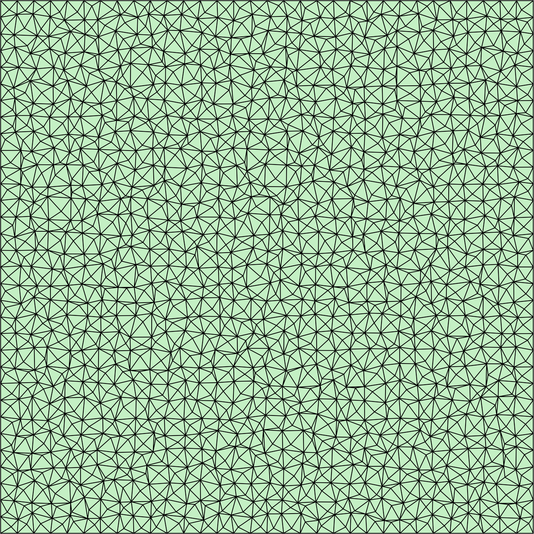}} 
	\caption{Distorted meshes of $\Omega = [0,1]^2$ featuring (a-b) quadrilateral and (c-d) triangular cells.}
	\label{fig:Couette_DistortedMeshes}
\end{figure}

First, a set of highly distorted meshes is generated by introducing a perturbation on the position of the interior nodes of the regular meshes illustrated in figure~\ref{fig:Ringleb_meshes}.
In particular, for a given node $i$, its new position is defined as $\tilde{\bx_i} = \bx_i + \bm{r}_i$, $\bm{r}_i$ being an $\nsd$-dimensional vector whose components are randomly generated within the interval $[-\ell_{\min}/3,\ell_{\min}/3]$, where $\ell_{\min}$ denotes the minimum edge length of the regular mesh. The third and fifth level of refinement of the meshes featuring distorted quadrilateral and triangular cells are illustrated in figure~\ref{fig:Couette_DistortedMeshes}.

\begin{figure}[!ht]
	\centering
	\subfloat[Quad. M3D \label{fig:Couette_Mach_QUA_distorted_H4}]{\includegraphics[width=0.24\textwidth]{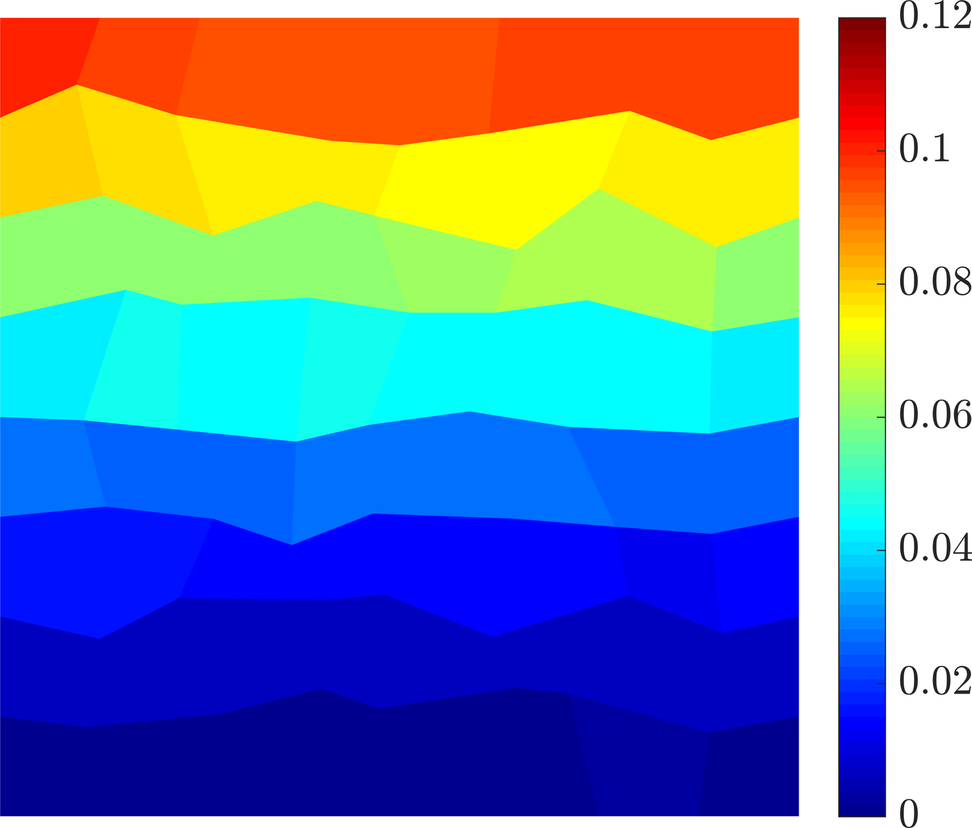}} \hfill
	\subfloat[Quad. M5D \label{fig:Couette_Mach_QUA_distorted_H6}]{\includegraphics[width=0.24\textwidth]{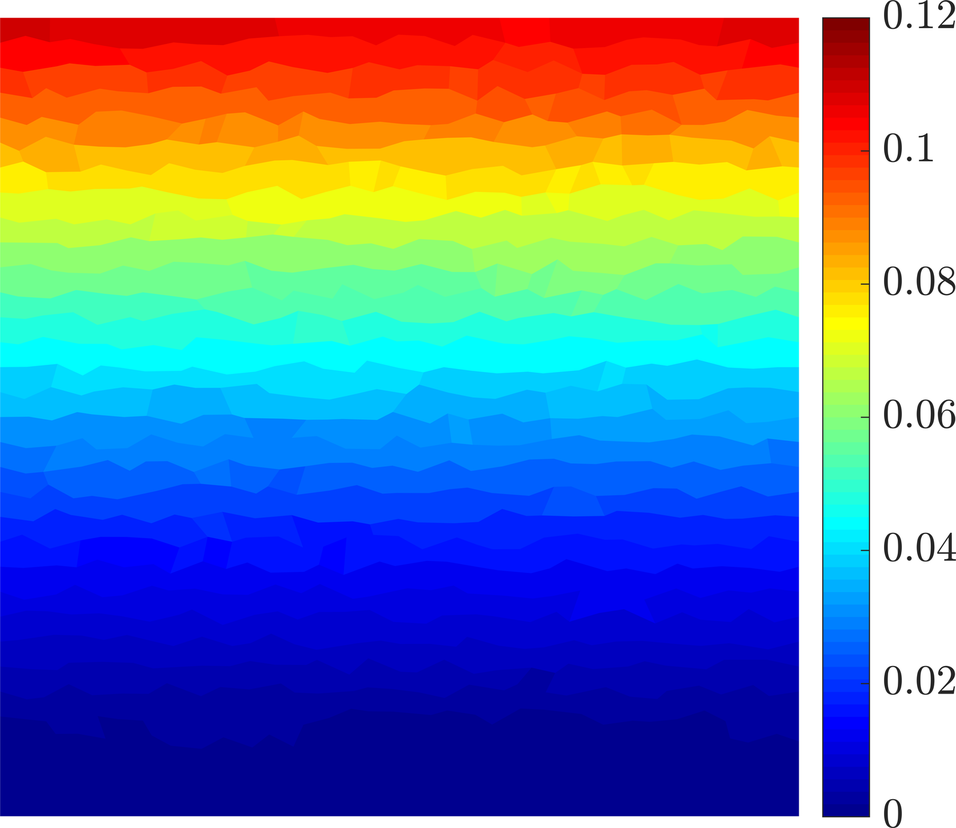}} \hfill
	\subfloat[Triangles M3D \label{fig:Couette_Mach_TRI_distorted_H4}]{\includegraphics[width=0.24\textwidth]{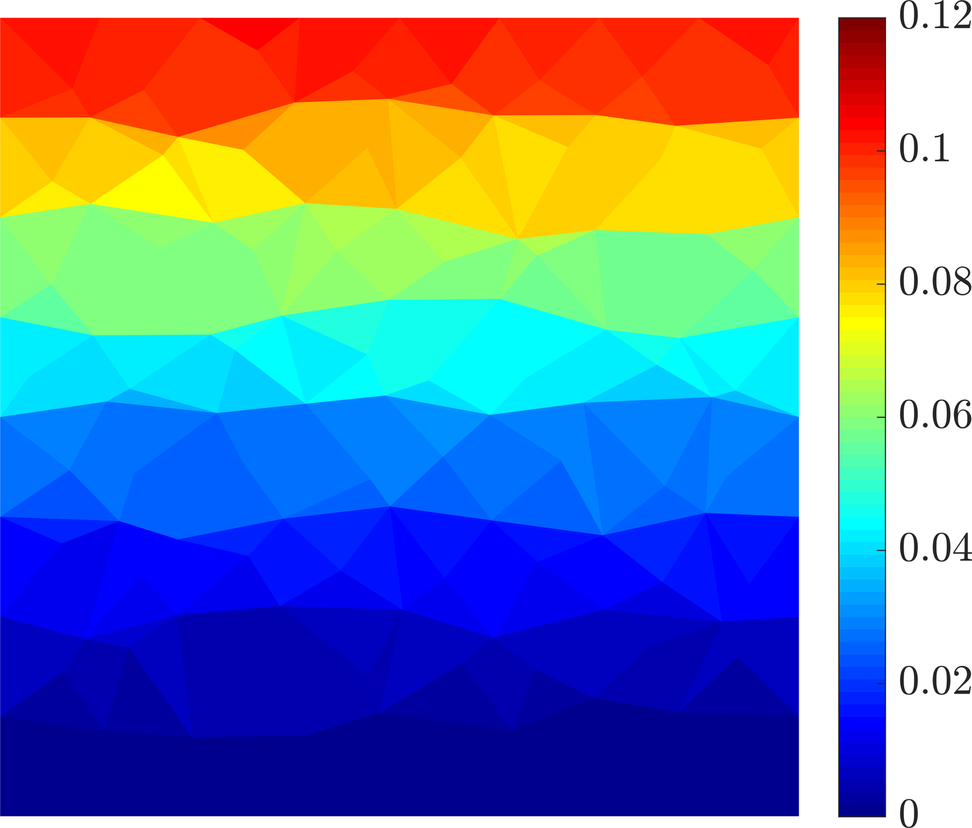}} \hfill
	\subfloat[Triangles M5D \label{fig:Couette_Mach_TRI_distorted_H6}]{\includegraphics[width=0.24\textwidth]{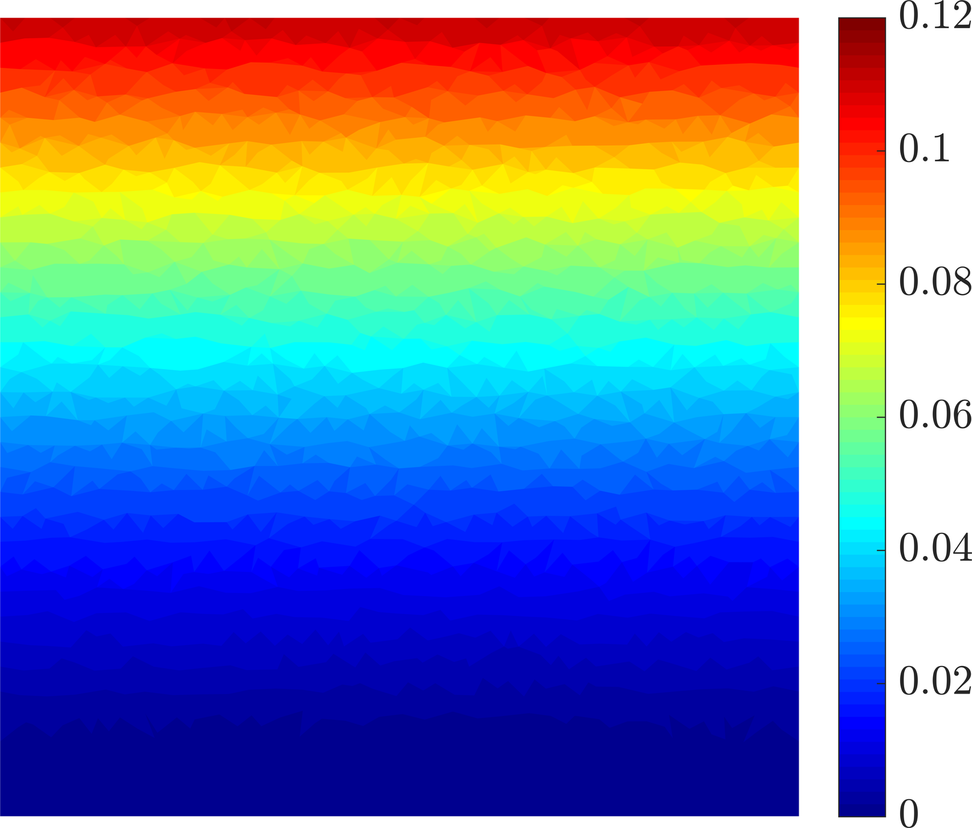}} 
	\caption{Couette flow - Mach number distribution using the distorted (a-b) quadrilateral and (c-d) triangular meshes in Figure \ref{fig:Couette_DistortedMeshes} employing the HLLEM Riemann solver.}
	\label{fig:Couette_Ma}
\end{figure}

The previously introduced Couette flow example for Reynolds number $\Rey = 100$ is also employed for the current sensitivity study. More precisely, figure~\ref{fig:Couette_Ma} shows the approximation of the Mach number distribution on the meshes of distorted cells in figure~\ref{fig:Couette_DistortedMeshes} employing the HLLEM Riemann solver.

\begin{figure}[!ht]
	\subfloat[$s=0$ \label{fig:Mesh_TRI_H3}]{\includegraphics[width=0.24\textwidth]{Ringleb_mesh_TRI_H3}} \hfill
	\subfloat[$s=10$ \label{fig:Mesh_TRI_Stretch10_H3}]{\includegraphics[width=0.24\textwidth]{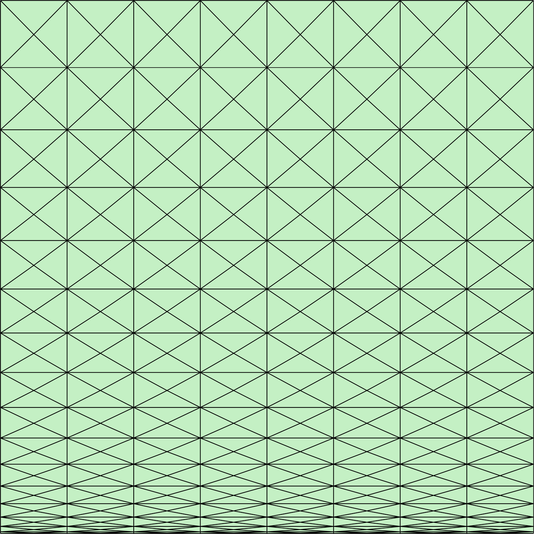}} \hfill
	\subfloat[$s=100$ \label{fig:Mesh_TRI_Stretch100_H3}]{\includegraphics[width=0.24\textwidth]{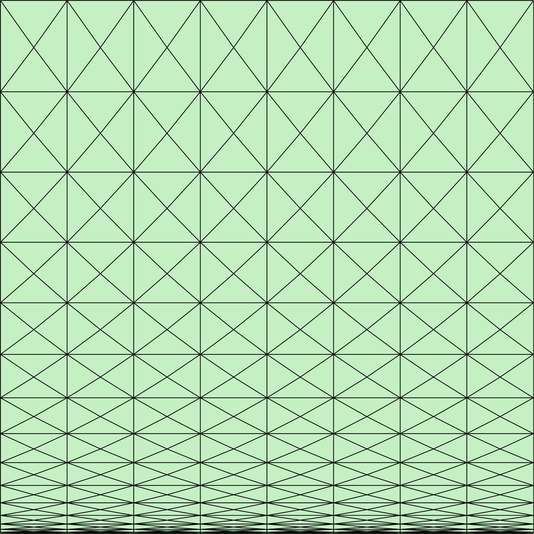}} \hfill
	\subfloat[$s=1,000$ \label{fig:Mesh_TRI_Stretch1000_H3}]{\includegraphics[width=0.24\textwidth]{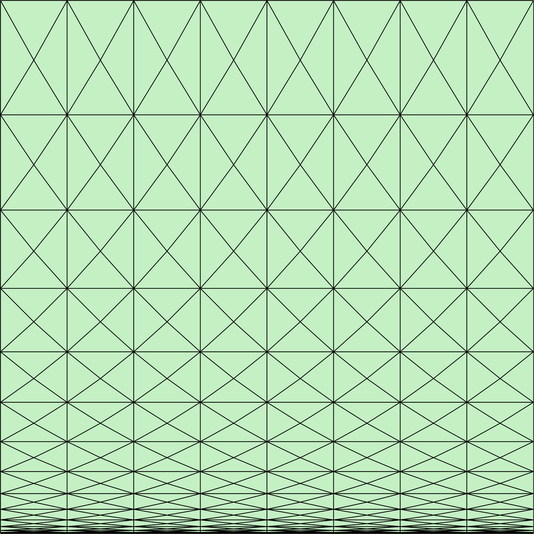}}
	\caption{Second level of refinement of the stretched meshes of $\Omega = [0,1]^2$ for different values of the stretching factor $s$.}
	\label{fig:Couette_StretchedMeshes}
\end{figure}

Second, a set of meshes with stretching near the bottom boundary is produced. For its construction in 2D, the vertical coordinate of the first mesh layer is fixed at the desired stretching factor $s$. Then, the vertical coordinate of the subsequent layers is defined as
\begin{equation}
y_k = y_{k-1} + \frac{h}{s} \beta^{k-1}, \qquad \text{for } k = 2,\dotsc,N_y + 1
\end{equation}
where $h$ is the maximum edge length of the corresponding regular mesh, $N_y$ is the number of cells in the vertical direction and the growth rate factor $\beta$ is computed by imposing that the vertical coordinate of the last layer is one, that is by finding the roots of
\begin{equation}
\frac{h}{s}\beta^{N_y} - \beta + 1 - \frac{h}{s} = 0.
\end{equation}
Figure~\ref{fig:Couette_StretchedMeshes} reports the second level of refinement of a set of triangular meshes for different levels of stretching $s$.

\begin{figure}[!ht]
	\centering
	\subfloat[Effect of distortion on mesh convergence \label{fig:Couette_convergenceDistortion}]{\includegraphics[width=0.48\textwidth]{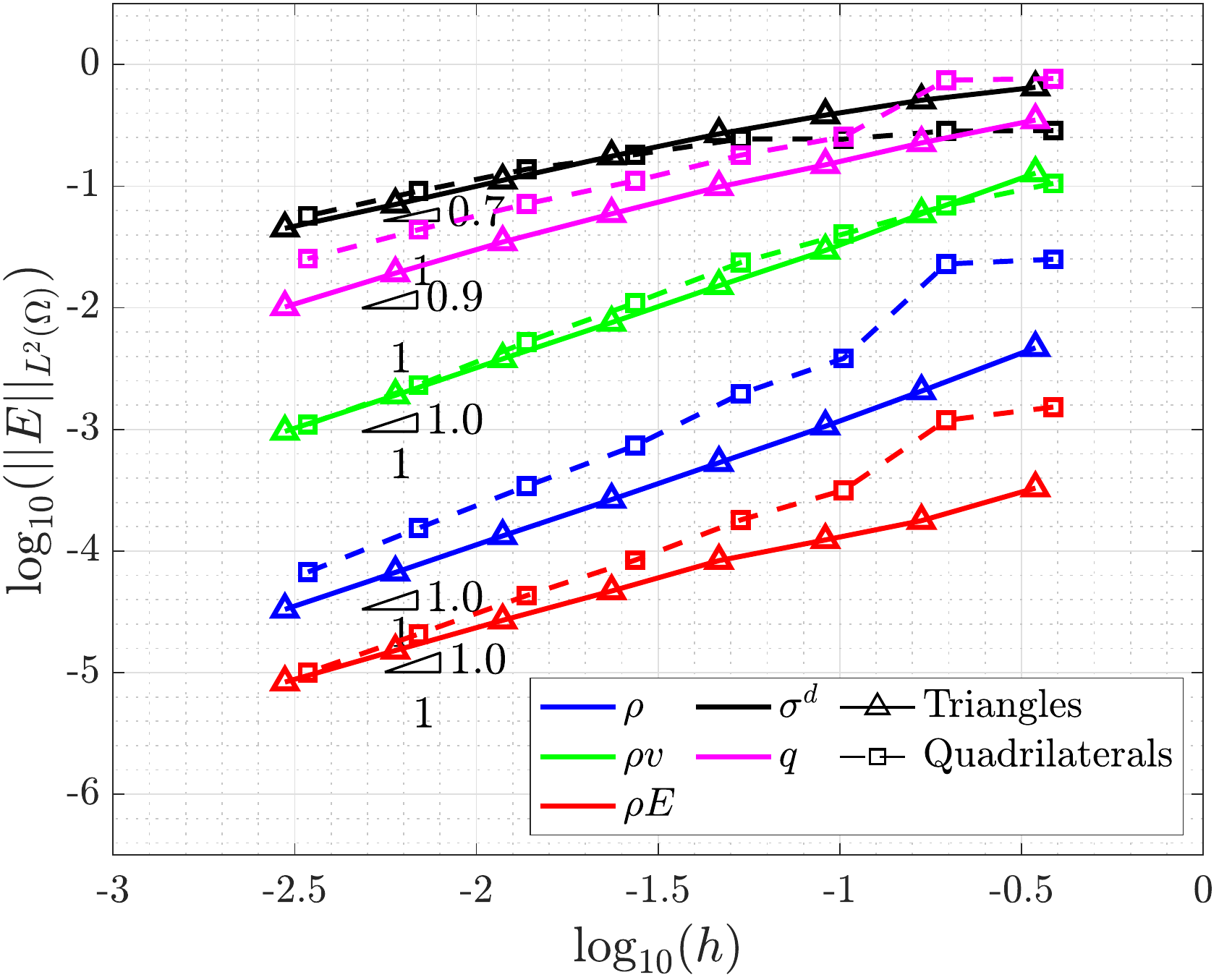}} \hfill
	\subfloat[Effect of stretching on mesh convergence \label{fig:Couette_convergenceStretch}]{\includegraphics[width=0.48\textwidth]{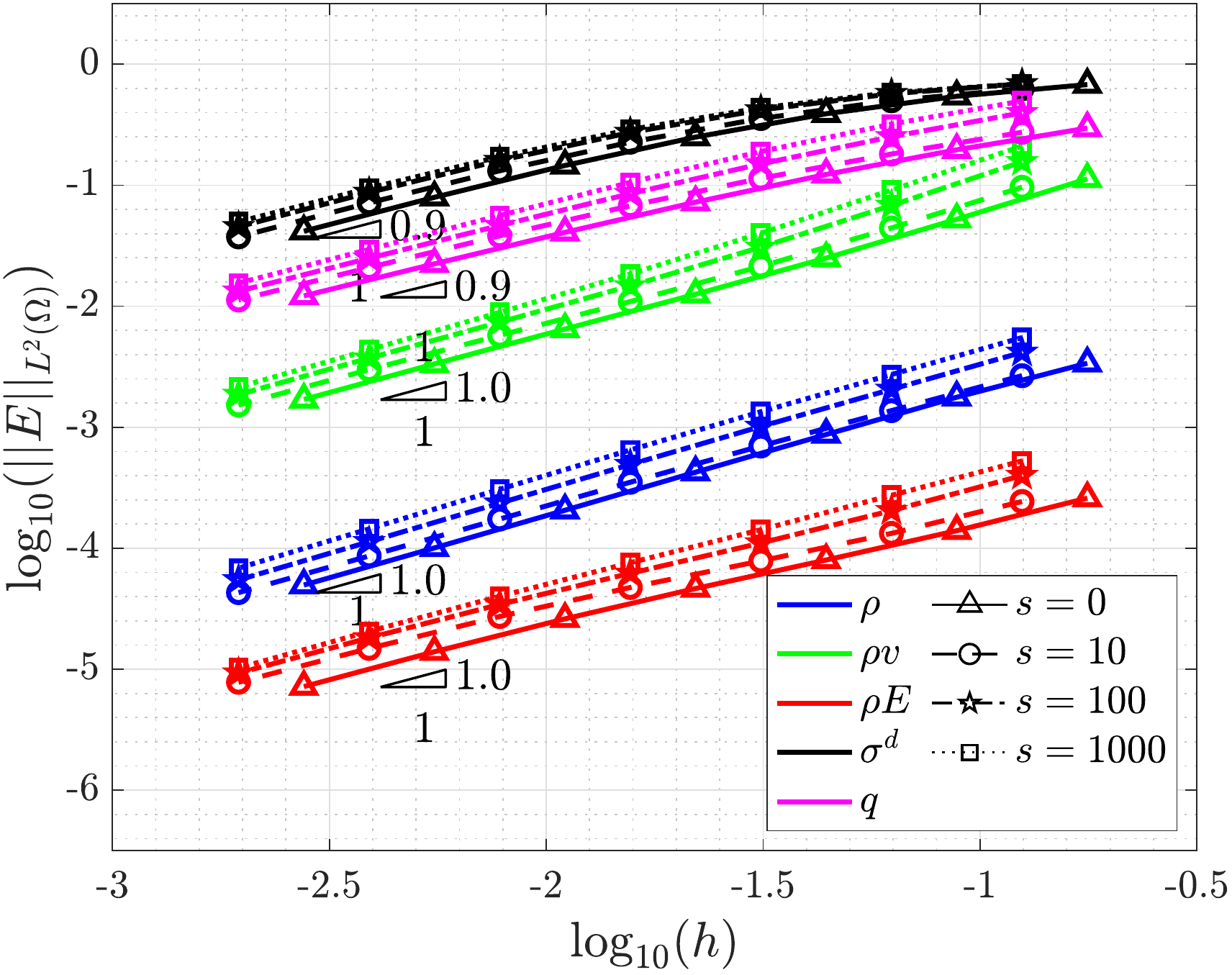}} 
	\caption{Couette flow - $h$-convergence of the error in the $\eltwo(\Omega)$ norm of density, momentum, energy, viscous stress tensor and heat flux in (a) distorted and (b) stretched meshes, using the HLLEM Riemann solver and for Reynolds number $\Rey = 100$.}
	\label{fig:Couette_ErrorStretchingDistortion}
\end{figure}

A quantitative evaluation of the influence of cell distortion and stretching on the accuracy of the FCFV approximation is performed via an $h$-convergence study of the error, measured in the $\eltwo(\Omega)$ norm, using the HLLEM Riemann solver. The results, reported in figures~\ref{fig:Couette_convergenceDistortion} and~\ref{fig:Couette_convergenceStretch}, respectively, show that optimal convergence of order $1$ is achieved for all the variables, independently of the distortion or of the stretching factor of its cells. In addition, the precision of the numerical approximation also results unaffected by the loss of orthogonality and loss of isotropy of the mesh. Indeed, by comparing the results of figure~\ref{fig:Couette_ErrorStretchingDistortion} with the ones in figure~\ref{fig:Couette_ErrorRe}, almost identical levels of accuracy are obtained in the $\eltwo(\Omega)$ error of the approximate solutions using meshes with uniform, distorted or stretched cells.

\section{Numerical benchmarks}
\label{sc:Benchmarks}

In this section, a set of numerical examples is presented to show the capabilities of the proposed FCFV method to simulate inviscid and viscous compressible flows at different regimes.

\subsection{Inviscid transonic flow over a NACA 0012 profile}

The first test case considers the inviscid transonic flow over a NACA 0012 aerofoil at free-stream Mach number $\Minf = 0.8$ and angle of attack $\alpha = 1.25\degree$. This classical benchmark for inviscid compressible flows~\cite{Sevilla-SHM:2013,Kroll2009} is proposed to evaluate the ability of the FCFV solver to capture flow solutions involving shock waves. More precisely, this benchmark is used to demonstrate the importance of the choice of the Riemann solver in the accuracy and stability of the FCFV approximate solution.

\begin{figure}[!ht]
	\centering
	
	\newsavebox{\bigimage}
	\sbox{\bigimage}{
		\subfloat[Mesh of the computational domain \label{fig:NACA0012_inviscid_mesh_domain}]{\includegraphics[width=0.48\textwidth]{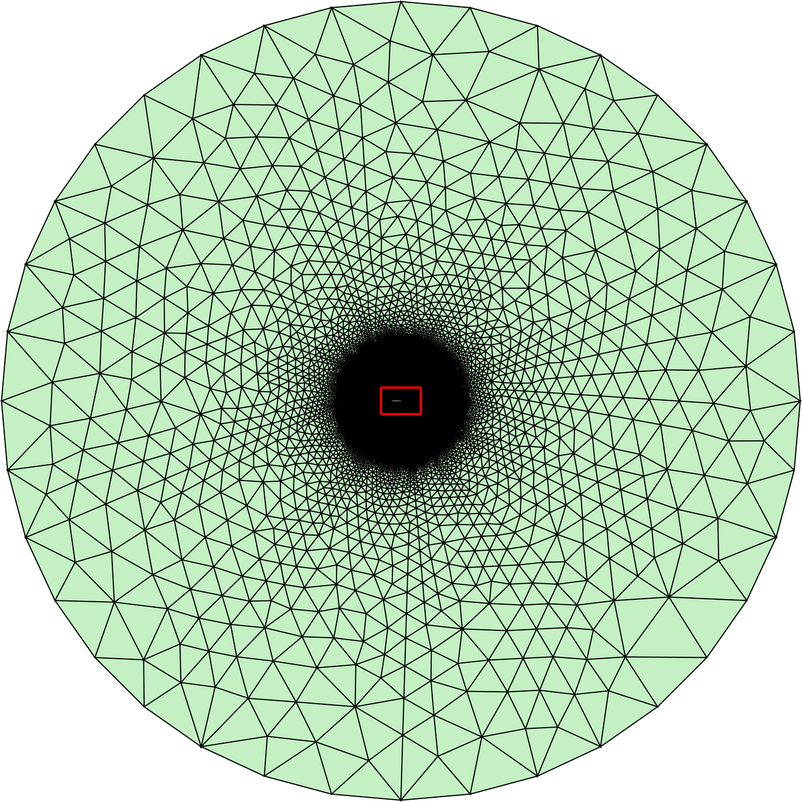}}
		\vspace{0pt}
	}
	
	\usebox{\bigimage} 
	\begin{minipage}[b][\ht\bigimage][s]{.48\textwidth}
		\centering
		\subfloat[Coarse mesh \label{fig:NACA0012_inviscid_mesh2}]{\includegraphics[width=0.66\textwidth]{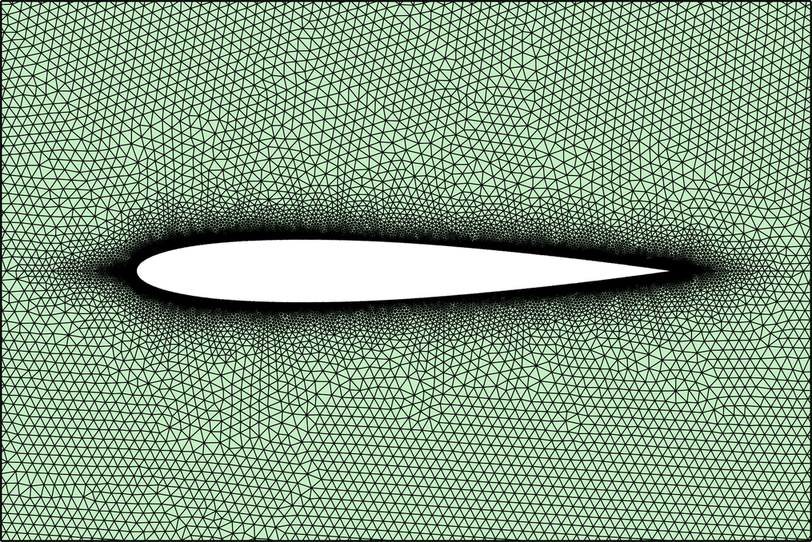}} \\
		\vfill
		\subfloat[Fine mesh \label{fig:NACA0012_inviscid_mesh4}]{\includegraphics[width=0.66\textwidth]{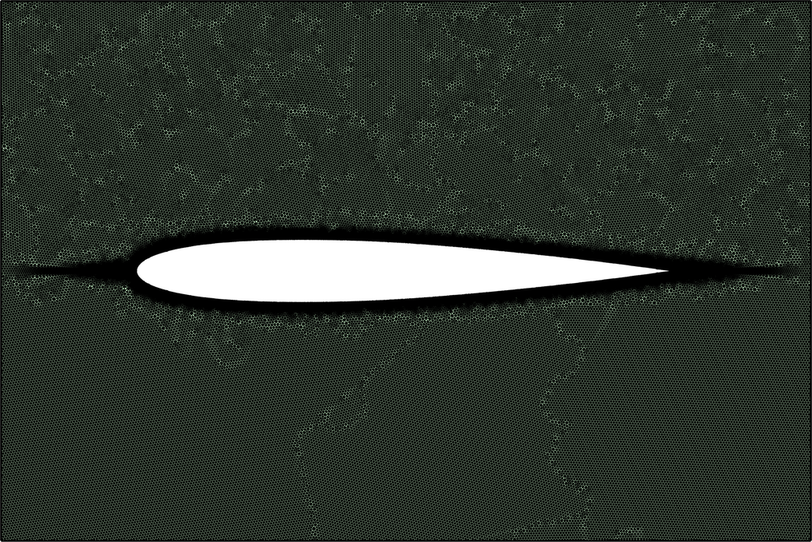}}
	\end{minipage}
	
	\caption{Mesh refinement for the inviscid transonic flow over a NACA 0012 profile.}
	\label{fig:NACA0012_inviscid_meshes}
\end{figure}

Unstructured meshes of triangular cells with non-uniform refinements on the surface of the aerofoil and at the leading and trailing edges are used for the simulation. Figure~\ref{fig:NACA0012_inviscid_meshes} reports the details of a coarse and a fine mesh featuring 89,250 and 712,164 triangular cells, respectively. The far-field boundary is located at 15 chord units away from the profile and the aerofoil surface is defined as an inviscid wall.

\begin{figure}[!ht]
	\subfloat[Mach \label{fig:NACA0012_inviscid_Ma_HLL}]{\includegraphics[width=0.48\textwidth]{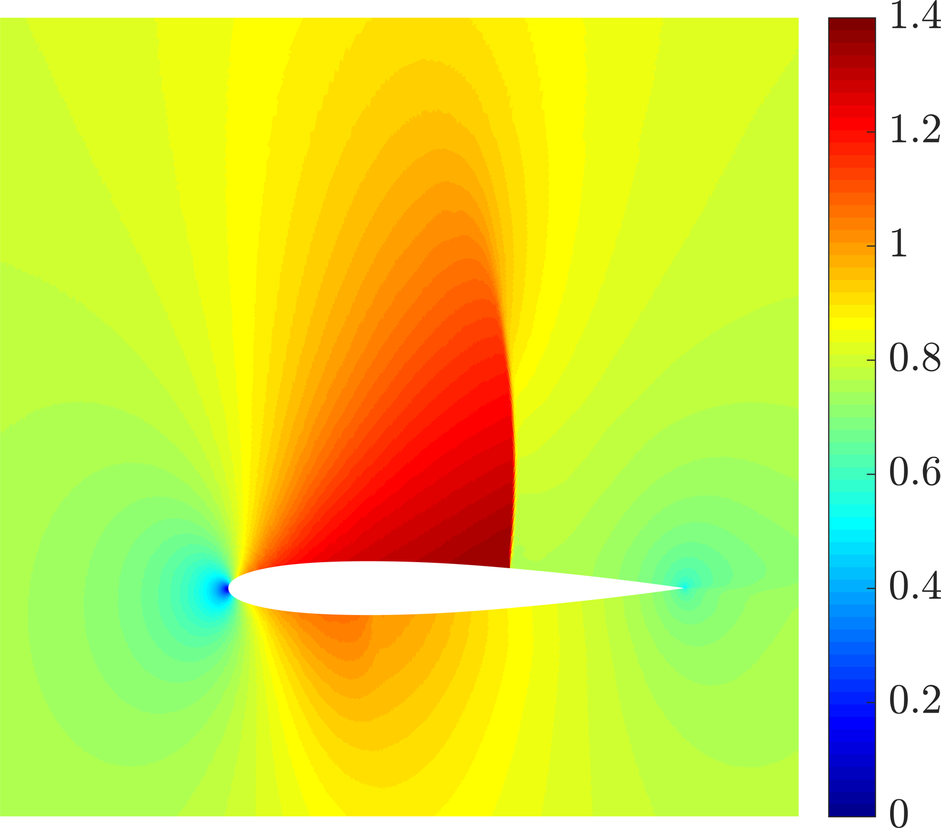}} \hfill
	\subfloat[Pressure, $p/p_\infty$ \label{fig:NACA0012_inviscid_pressure_HLL}]{\includegraphics[width=0.48\textwidth]{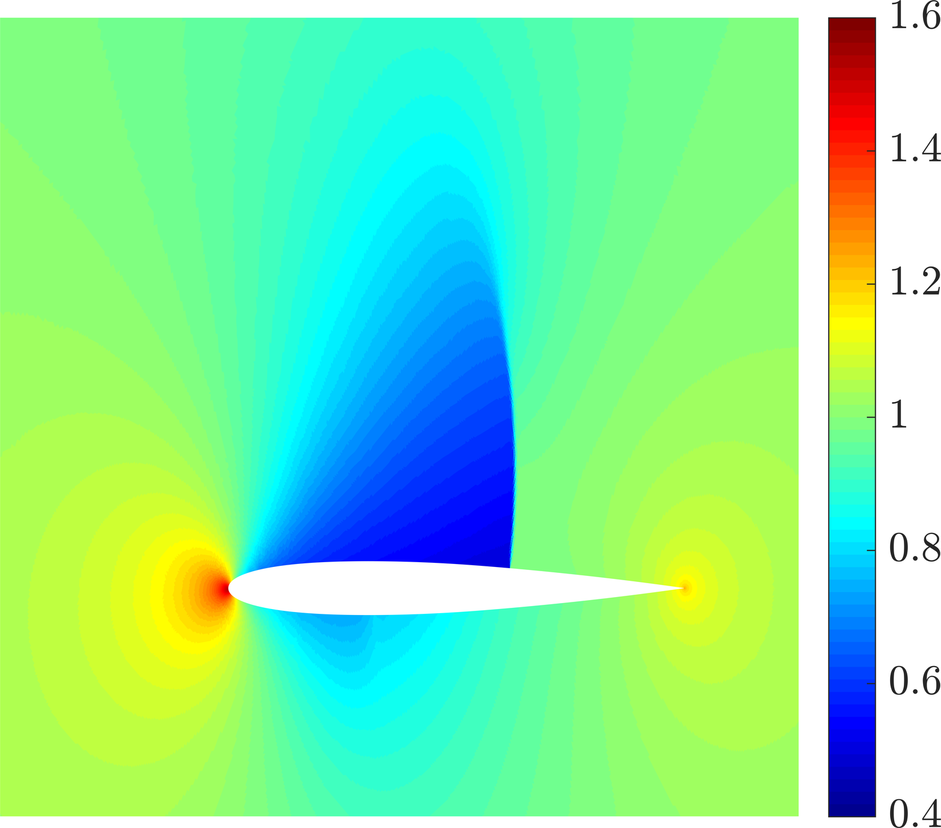}}
	\caption{Inviscid transonic flow over a NACA 0012 profile - (a) Mach number and (b) pressure distributions around the aerofoil computed on the fine mesh using the HLL Riemann solver.}
	\label{fig:NACA0012_inviscid_HLL}
\end{figure}

\begin{figure}[!ht]
	\subfloat[$y/c=0.50$ \label{fig:NACA0012_inviscid_P_section4}]{\includegraphics[width=0.48\textwidth]{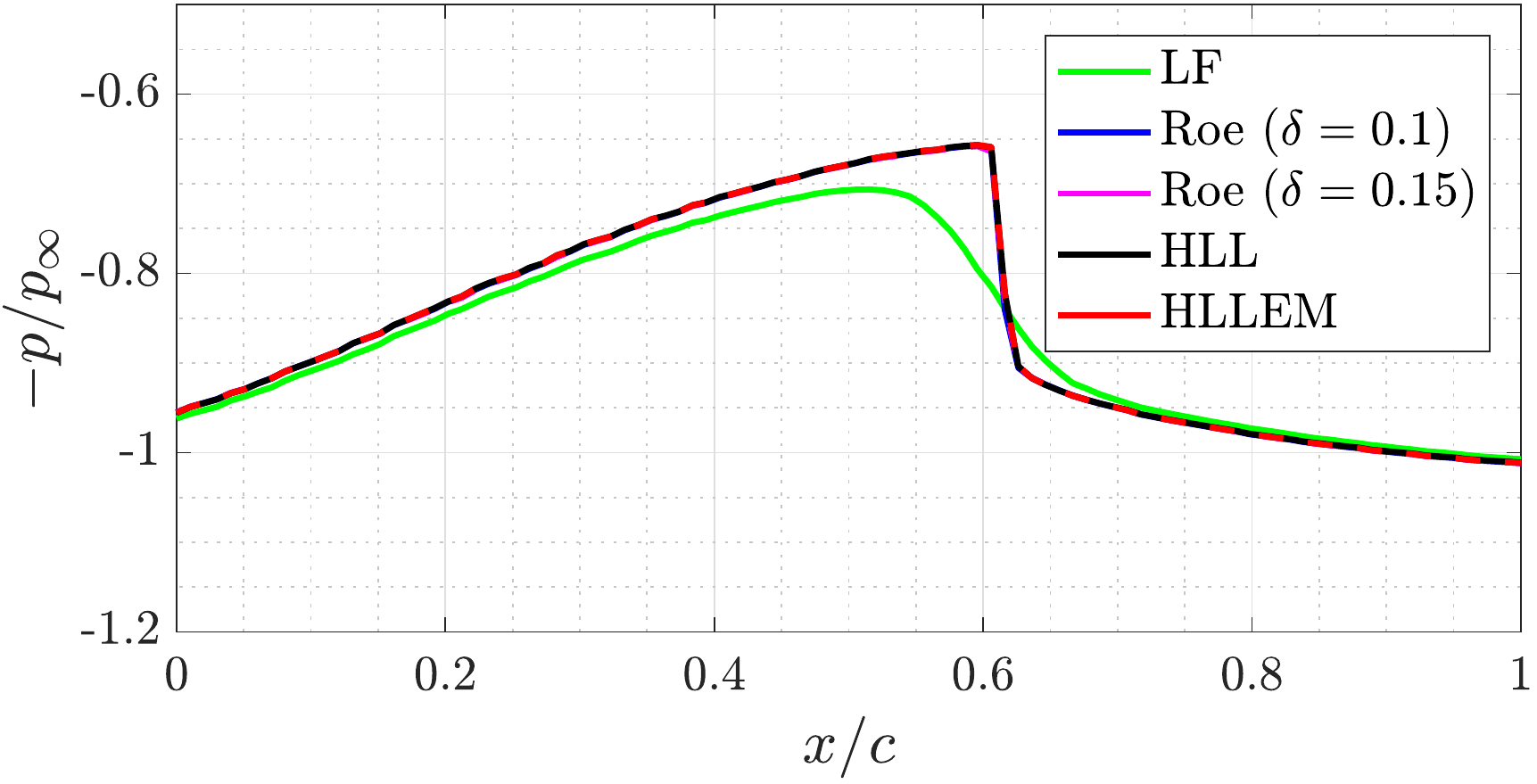}} \hfill
	\subfloat[$y/c=0.25$ \label{fig:NACA0012_inviscid_P_section1}]{\includegraphics[width=0.48\textwidth]{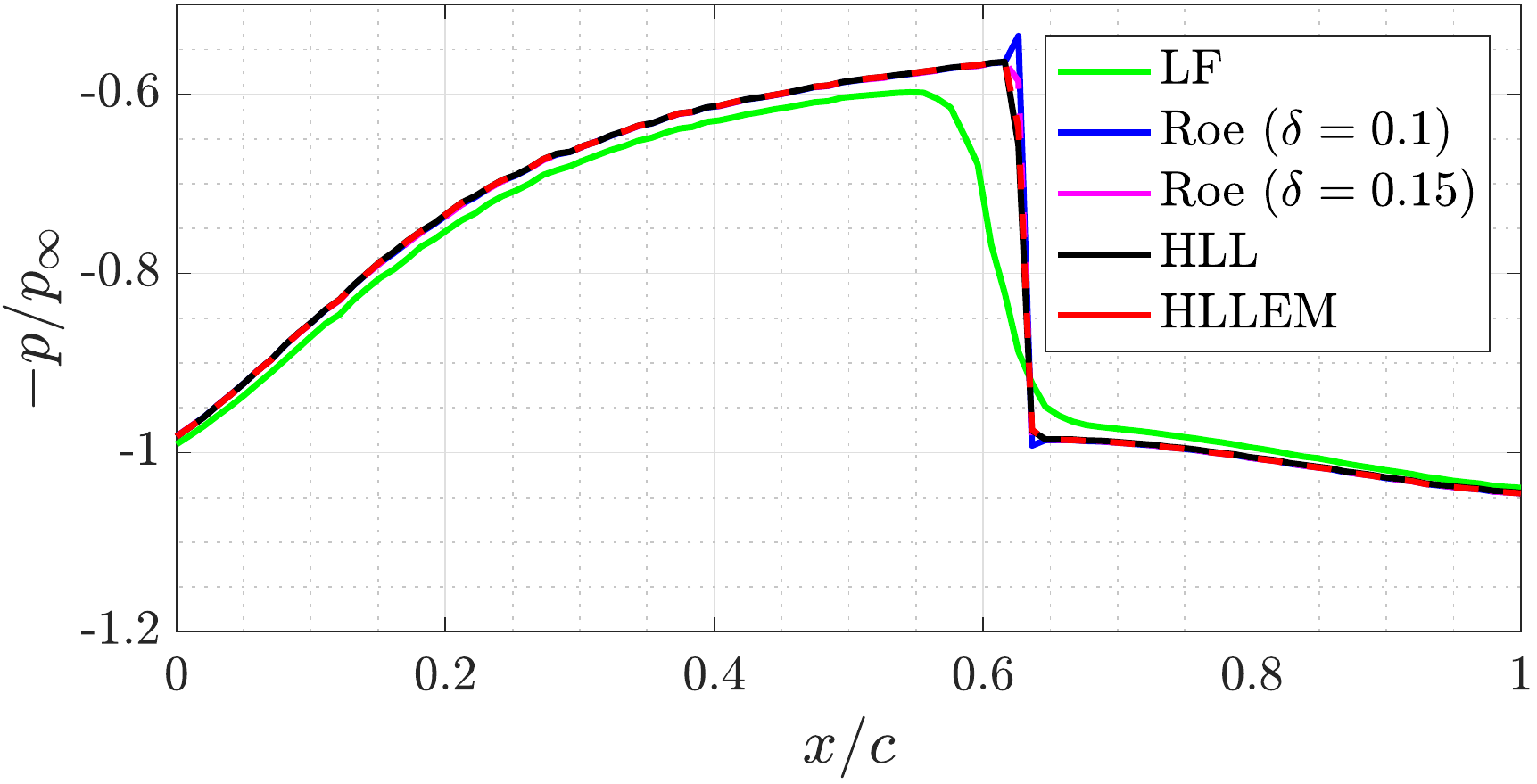}} \\
	\subfloat[$y/c=0.10$ \label{fig:NACA0012_inviscid_P_section2}]{\includegraphics[width=0.48\textwidth]{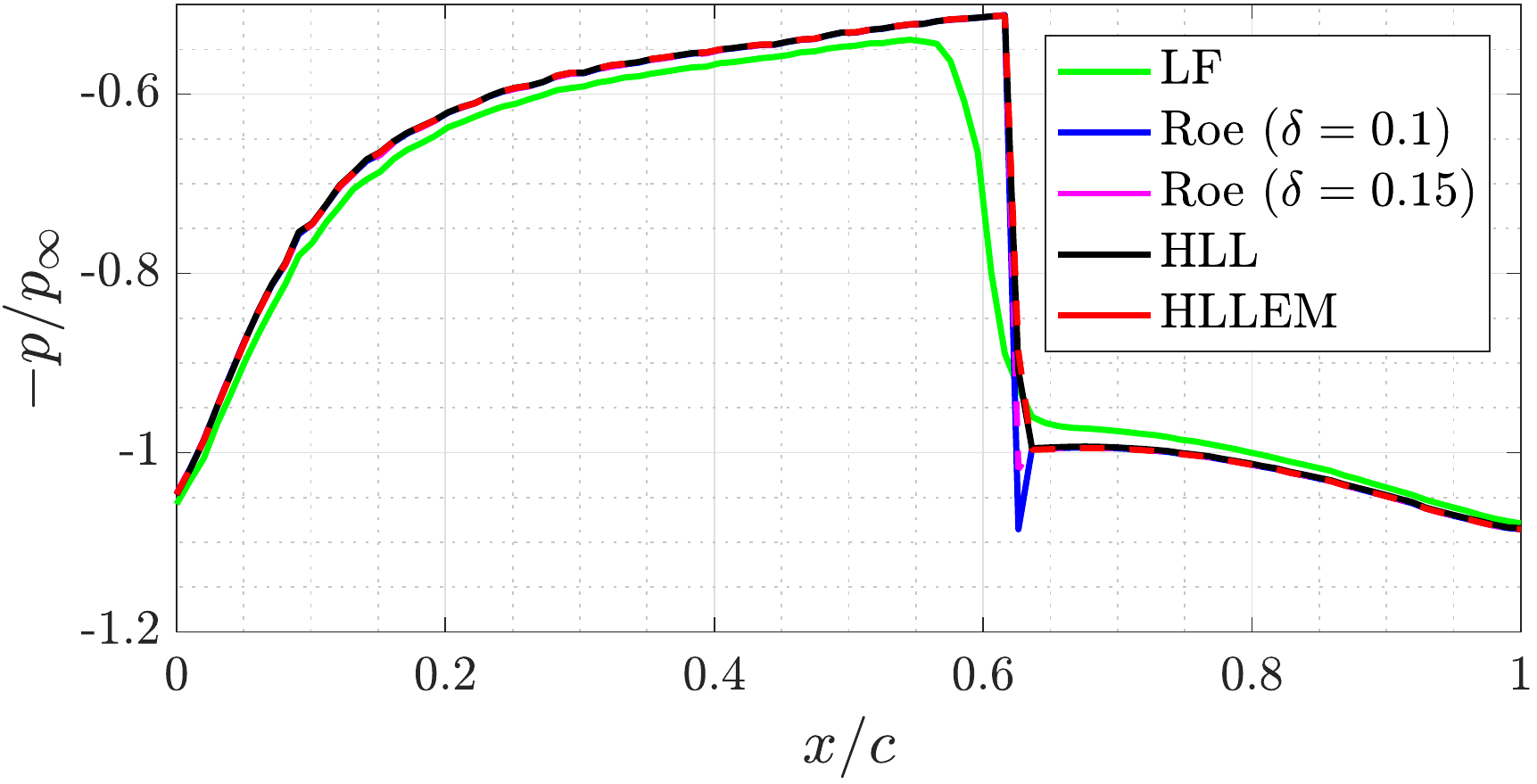}} \hfill
	\subfloat[$y/c=-0.075$ \label{fig:NACA0012_inviscid_P_section3}]{\includegraphics[width=0.48\textwidth]{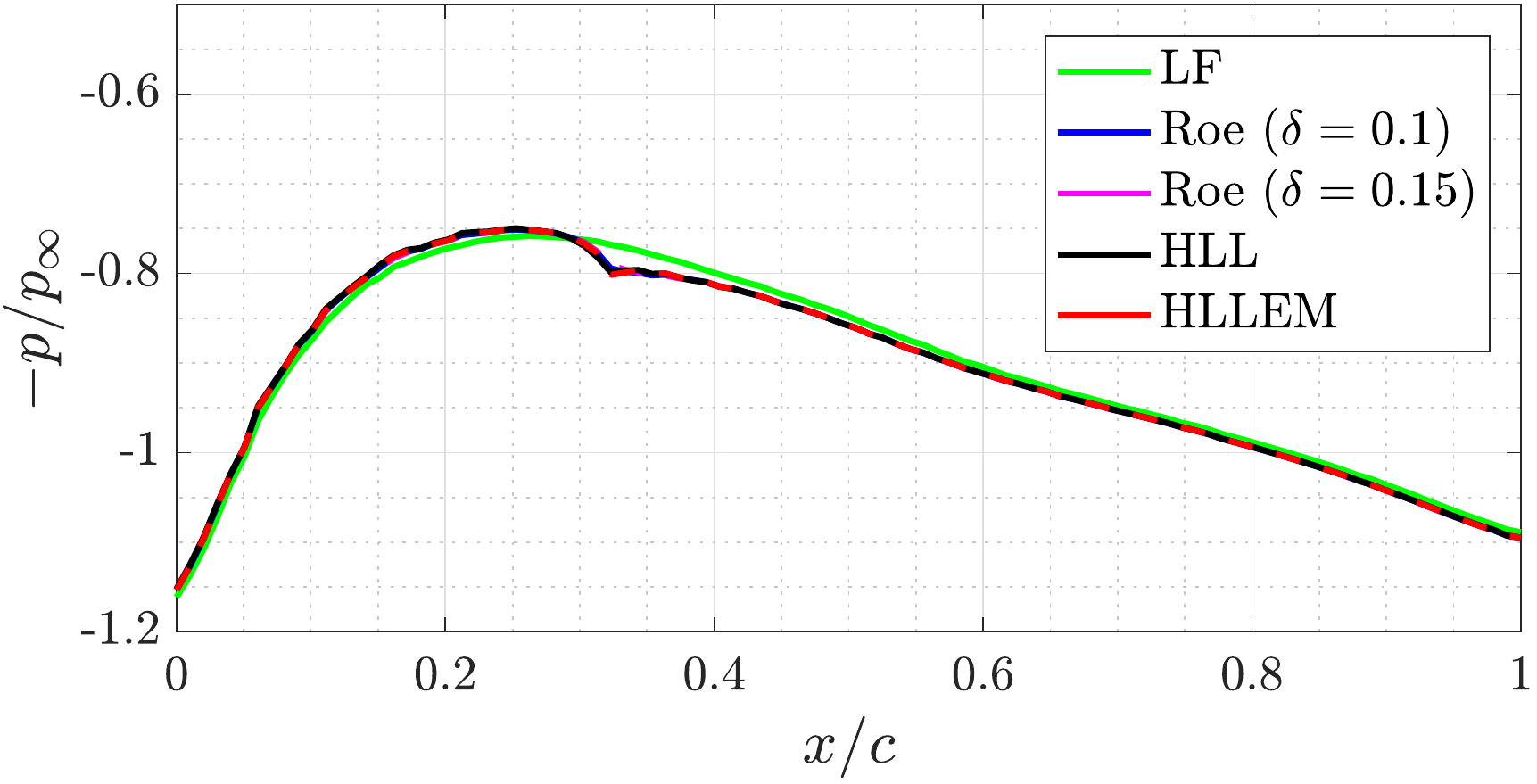}}
	\caption{Inviscid transonic flow over a NACA 0012 profile - Pressure distribution at different sections parallel to the aerofoil chord using different Riemann solvers.}
	\label{fig:NACA0012_inviscid_P_sections}
\end{figure}

The Mach number and the pressure distributions computed on the fine mesh using the HLL Riemann solver are displayed in figure~\ref{fig:NACA0012_inviscid_HLL}. Both the strong shock wave on the upper surface and the weaker shock in the lower part of the aerofoil are accurately represented.
It is worth noticing that the FCFV method provides non-oscillatory solutions in presence of abrupt variations without the need of any shock capturing or limiting mechanism. This property \hl{follows from the result} on the monotonicity of first-order schemes in Godunov's theorem~\cite{Godunov1959}. Hence, the FCFV method with the HLL numerical flux ensures the positivity properties of the approximate solution. In addition, the choice of the Riemann solver also controls the amount of numerical diffusion introduced by the FCFV method, influencing the overall accuracy of the computed solution. A qualitative comparison of the different Riemann solvers is performed in figure~\ref{fig:NACA0012_inviscid_P_sections} by illustrating the pressure distribution in the fine mesh at different sections along the vertical body axis.
The results display that the Lax-Friedrichs numerical flux provides non-oscillatory solutions. Nonetheless, it introduces excessive numerical dissipation leading to a smeared representation of the shock wave. The Roe Riemann solver is equipped with a Harten-Hyman entropy fix: without this correction, the method fails to converge and nonphysical solutions with localised overshoots appear. On the one hand, using an entropy fix with threshold parameter $\delta= 0.1$, the Roe solver shows insufficient numerical dissipation producing an approximation with oscillations in the vicinity of the shock wave. On the other hand, a threshold value $\delta= 0.15$ for the Roe solver leads to a physically-admissible and accurate solution. It is worth noticing that the parameter $\delta$, which is problem-dependent, needs to be appropriately tuned \emph{a priori} by the user. Finally, HLL-type Riemann solvers exhibit their ability to produce positivity-preserving and accurate solutions in presence of shocks without the need of any user-defined parameter, thus remedying the aforementioned issue of the Roe solver.

\begin{figure}[!ht]
	\centering
	\includegraphics*[width=0.6\textwidth]{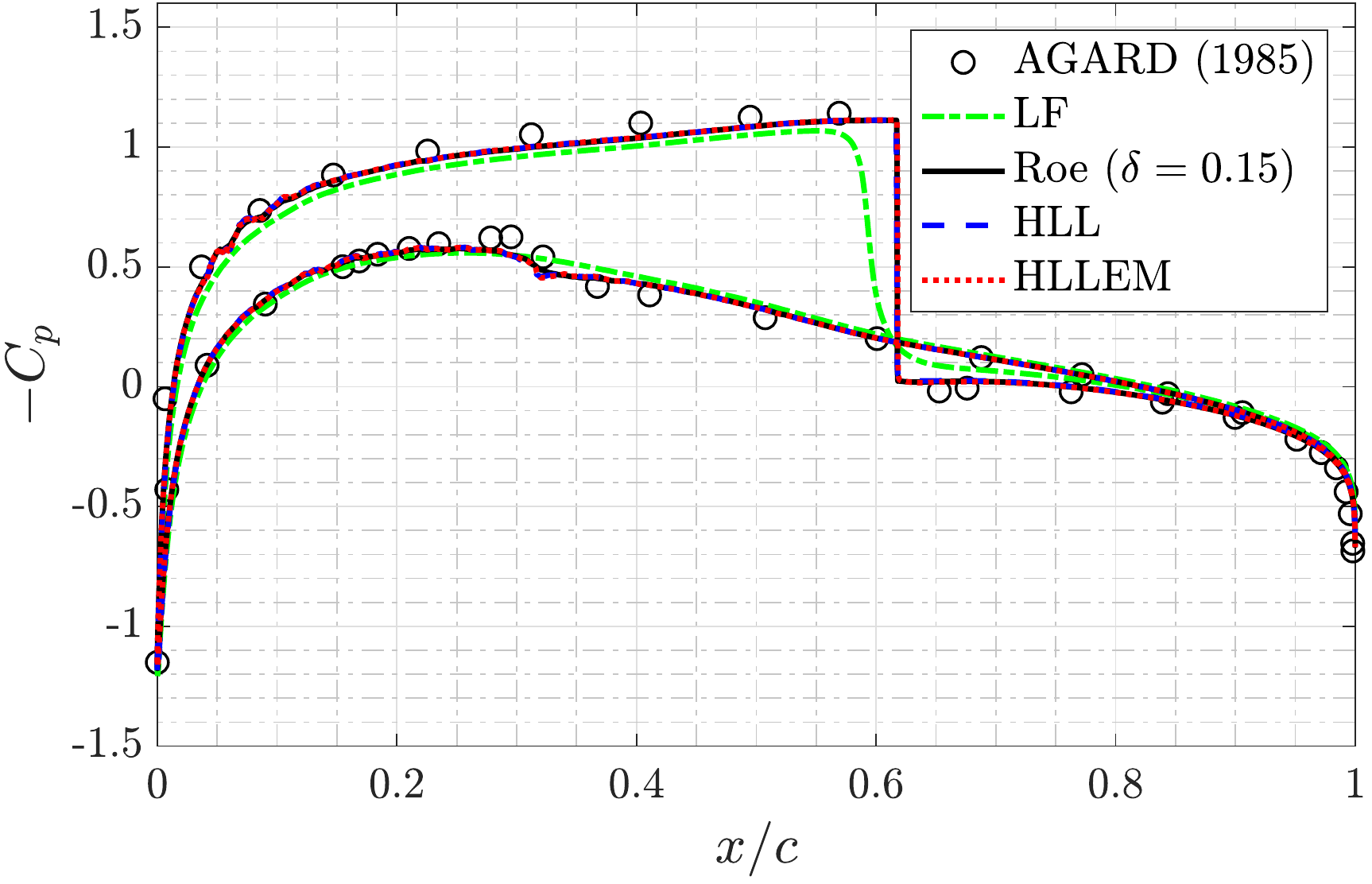}
	\caption{Inviscid transonic flow over a NACA 0012 profile - Pressure coefficient around the aerofoil surface computed on the fine mesh using Lax-Friedrichs (LF), Roe, HLL and HLLEM Riemann solvers.}
	\label{fig:NACA0012_inviscid_Cp}
\end{figure}

To further analyse the accuracy of the Riemann solvers for the FCFV method, the numerical computation of the pressure coefficient over the aerofoil surface is compared with experimental data~\cite{AGARD:1985}. Figure~\ref{fig:NACA0012_inviscid_Cp} confirms the overdissipative nature of the Lax-Friedrichs solution which shows a smeared representation of the shock wave. The HLL, HLLEM and the Roe solvers (the latter with an entropy fix parameter $\delta = 0.15$) produce nearly identical solutions with a sharp representation of the shock wave showing excellent agreement with the experimental data.

Finally, a quantitative comparison is performed by computing the lift and drag coefficients, reported in table~\ref{tb:NACA0012_inviscid_LiftDrag}. According to experimental data~\cite{Thibert-TGO:1979}, acceptable values lie within the range $[0.342,0.352]$ for the lift and $[0.0217,0.0227]$ for the drag coefficient, accounting for a tolerance of 5 lift and drag counts.
\begin{table} [!ht]
	\centering
	\caption{Inviscid transonic flow over a NACA 0012 aerofoil - Lift, $C_l$, and drag, $C_d$, coefficients computed on the fine mesh using different Riemann solvers.}
	\begin{tabular}{ L{1cm} C{3cm} C{3cm} C{2.25cm} C{2.25cm}  }
		\toprule
		&Lax-Friedrichs & Roe ($\delta = 0.15$) & HLL  & HLLEM \\
		\midrule
		$C_l$  & $0.274$ & $0.312$ & $0.314$ & $0.313$ \\
		\midrule
		$C_d$ & $0.0279$ & $0.0222$ & $0.0236$ & $0.0223$ \\
		\bottomrule
	\end{tabular}
	\label{tb:NACA0012_inviscid_LiftDrag}
\end{table}
The reported values for the drag coefficient employing the Roe and HLLEM Riemann solvers lie within the specified reference intervals, whereas the value obtained with the HLL solution is at 9 drag counts. Regarding the lift coefficient, the obtained results show an underestimation of this quantity, regardless of the employed Riemann solver. It is worth noticing that the proximity of the far-field boundary has a strong influence on the precision of the computed quantities, as reported in~\cite{YanoDarmofal2012,Wang2013}. Indeed, the presented results are in quantitative agreement with references employing a similar domain, see e.g.~\cite{Sevilla-SHM:2013} where the far-field boundary is located at 20 chord units from the aerofoil. In~\cite{Sevilla-SHM:2013}, the value of the lift coefficient computed using a first-order stabilised finite element approximation is $0.308$, differing between 4 and 6 lift counts from the FCFV solution provided by Roe, HLL and HLLEM Riemann solvers.

\subsection{Viscous laminar transonic flow over a NACA 0012 aerofoil}

The next example consists of the viscous laminar transonic flow over a NACA 0012 profile at free-stream Mach number $\Minf = 0.8$ and angle of attack $\alpha = 10\degree$. The Reynolds number, based on the chord length of the aerofoil, is $\Rey = 500$. This benchmark is presented to establish the capability of the FCFV method to concurrently capture abrupt variations due to shock waves and viscous effects in boundary layers~\cite{Bristeau1987,Sevilla-SHM:2013,Nogueira2009}.

\begin{figure}[!ht]
	\centering
	\subfloat[Mesh \label{fig:NACA0012_viscous_mesh2}]{\includegraphics[width=0.4\textwidth]{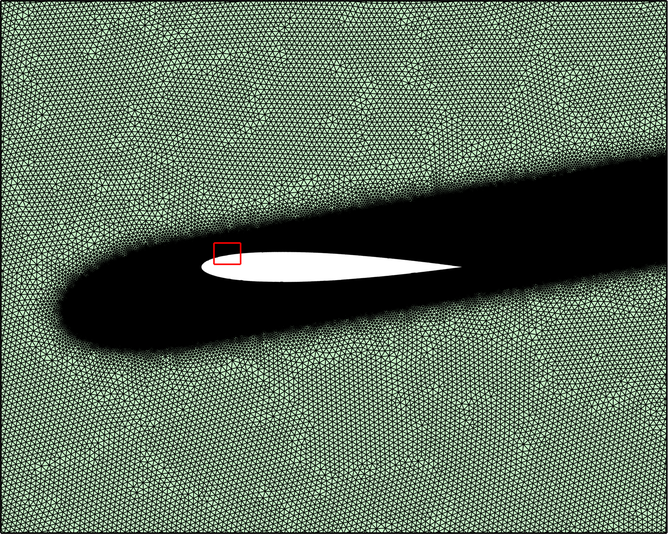}} \qquad
	\subfloat[Refinement on the surface \label{fig:NACA0012_viscous_mesh2_detail}]{\includegraphics[width=0.4\textwidth]{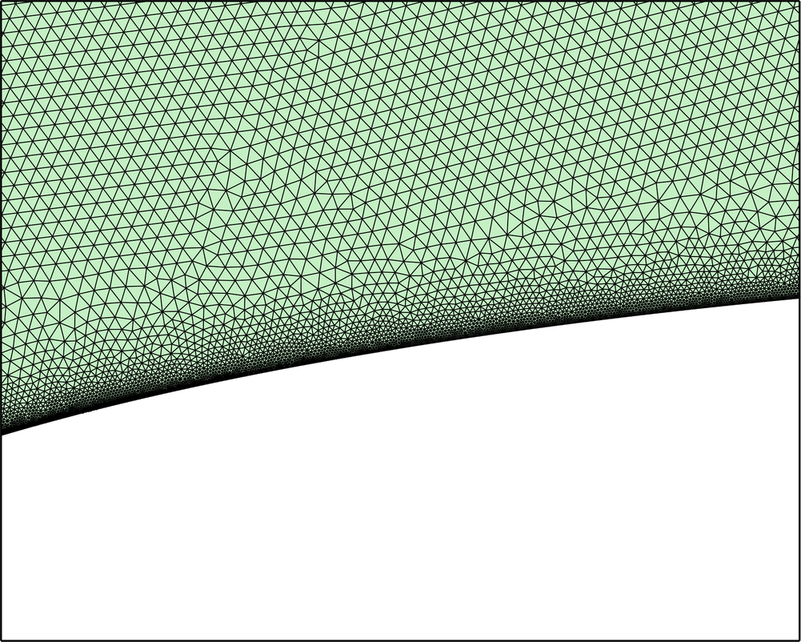}}
	\caption{Mesh for the viscous laminar transonic flow over a NACA 0012 profile.}
	\label{fig:NACA0012_viscous_meshes}
\end{figure}

As for the meshes utilised in the previous section, a non-uniform refinement is performed near the aerofoil surface. In addition, exploiting the information on the angle of attack of the free-stream, \emph{a priori} mesh refinement is introduced in a region surrounding the aerofoil, tilted $10\degree$ from its mean chord line, in order to accurately capture the viscous effects of the flow in the wake of the profile. 
An unstructured mesh of 1,005,199 triangular cells is displayed in figure~\ref{fig:NACA0012_viscous_mesh2} and a detail of its refinement on the surface of the aerofoil is reported in figure~\ref{fig:NACA0012_viscous_mesh2_detail}. The far-field boundary is located at 15 chord units from the profile and the aerofoil surface is considered adiabatic.
%
%

\begin{figure}[!ht]
	\subfloat[Mach \label{fig:NACA0012_viscous_Ma_HLLEM}]{\includegraphics[width=0.48\textwidth]{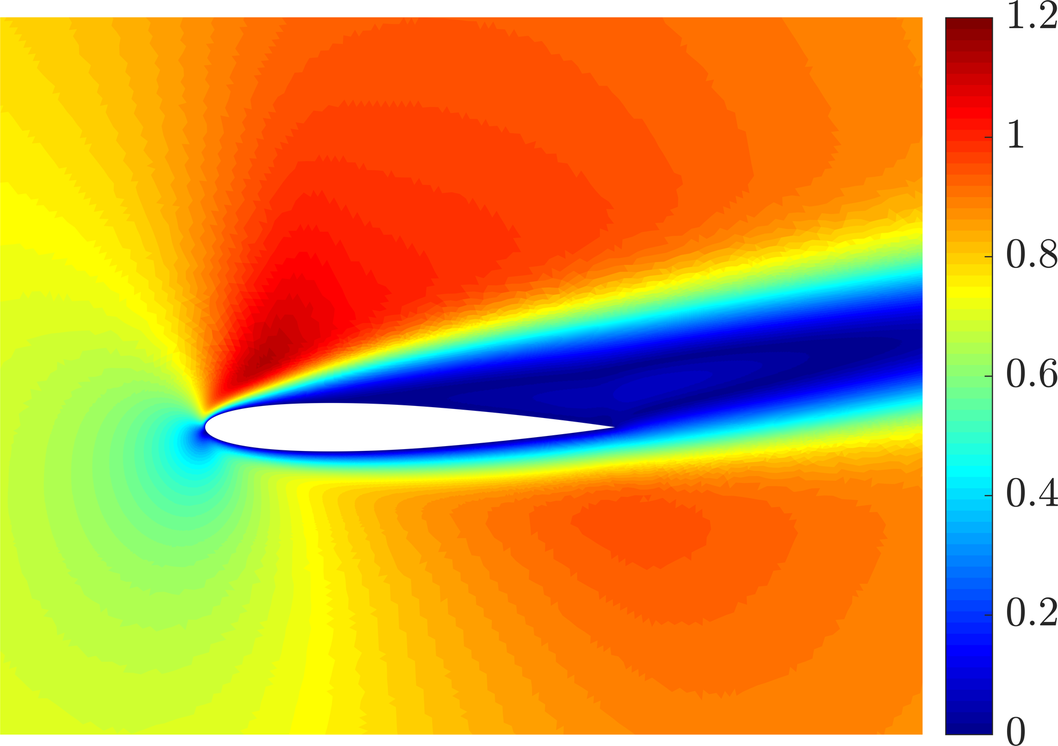}} \hfill
	\subfloat[Pressure, $p/p_\infty$ \label{fig:NACA0012_viscous_pressure_HLLEM}]{\includegraphics[width=0.48\textwidth]{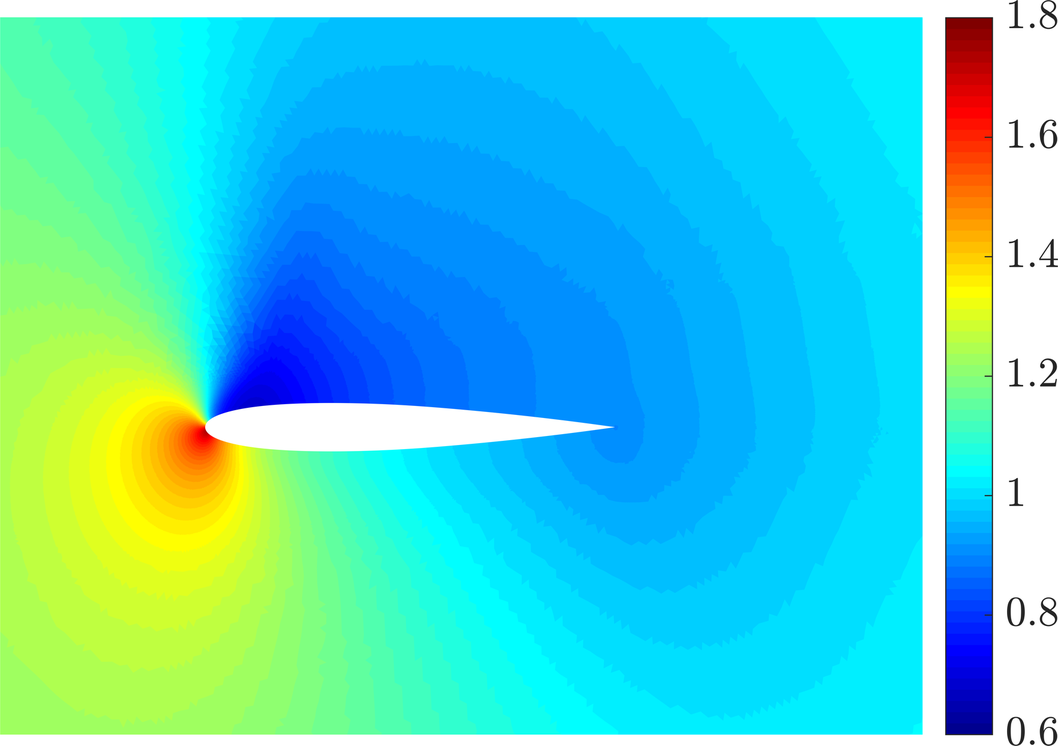}}
	\caption{Viscous laminar transonic flow over a NACA 0012 profile - (a) Mach number and (b) pressure distributions around the aerofoil computed using the HLLEM Riemann solver.}
	\label{fig:NACA0012_viscous_HLLEM}
\end{figure}

The flowfield computed with the HLLEM Riemann solver is depicted in figure~\ref{fig:NACA0012_viscous_HLLEM}. The Mach number distribution illustrates the capacity of the method to accurately describe the detached sonic region near the leading edge as well as the appearance of a wake behind the profile.

\begin{figure}[!ht]
	\subfloat[Pressure coefficient \label{fig:NACA0012_viscous_Cp}]{\includegraphics[width=0.48\textwidth]{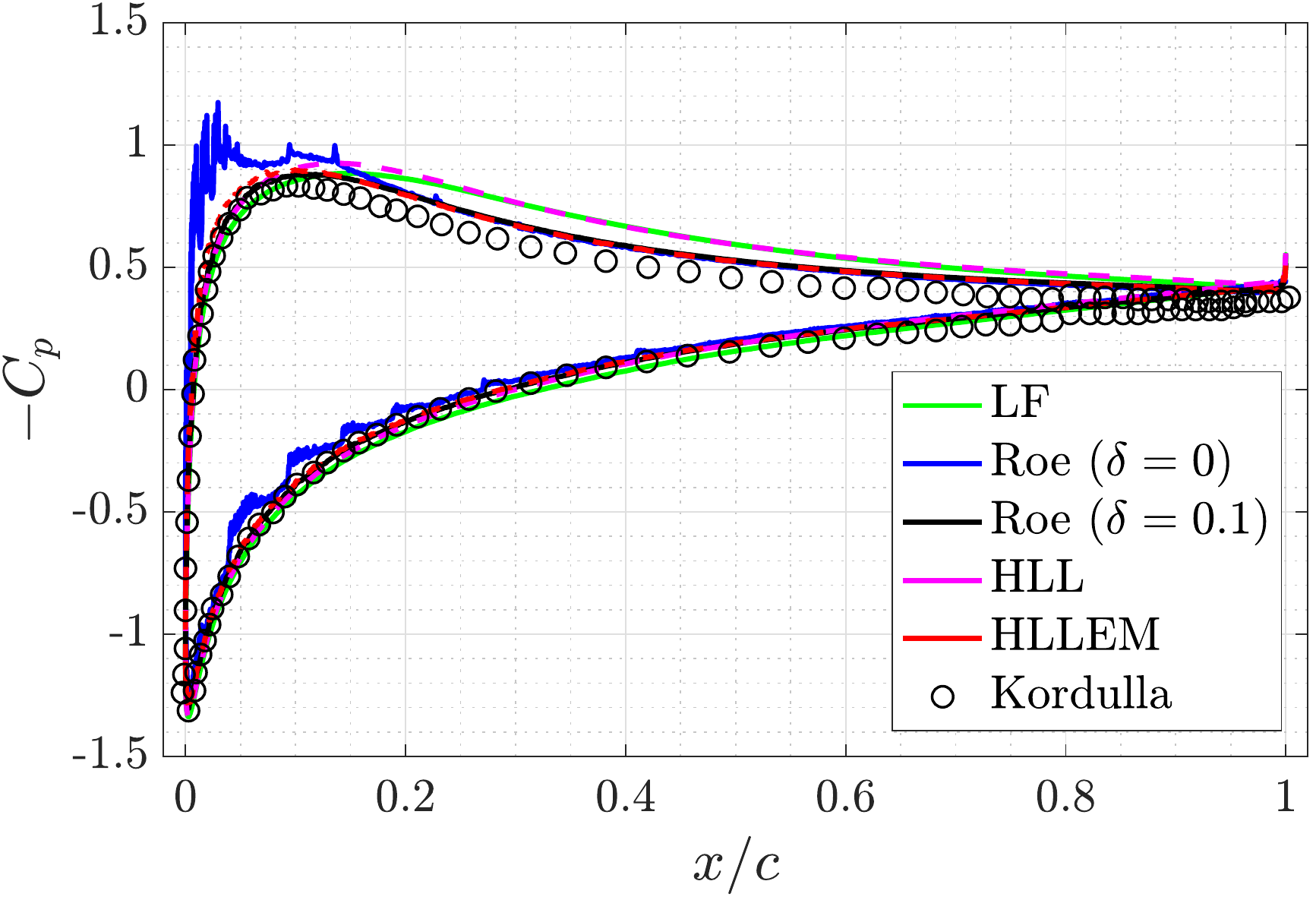}} \hfill
	\subfloat[Skin friction coefficient \label{fig:NACA0012_viscous_Cf}]{\includegraphics[width=0.48\textwidth]{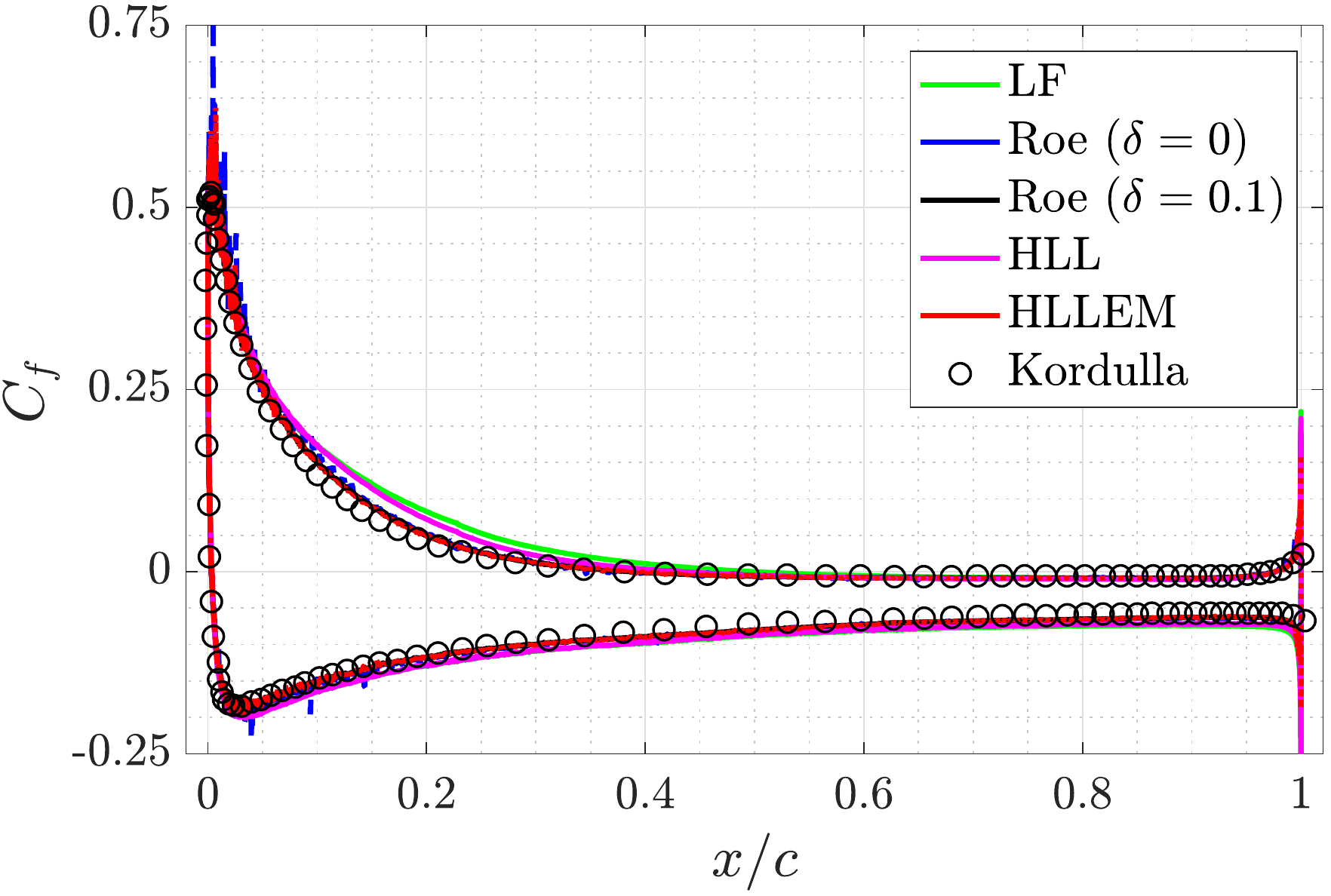}}
	\caption{Viscous laminar transonic flow over a NACA 0012 profile - (a) Pressure and (b) skin friction coefficient on the aerofoil surface computed using Lax-Friedrichs (LF), Roe, HLL and HLLEM Riemann solvers.}
	\label{fig:NACA0012_viscous_AerodynamicCoefficients}
\end{figure}

The different Riemann solvers for the FCFV method are compared for this viscous test case in figure~\ref{fig:NACA0012_viscous_AerodynamicCoefficients}. The results display the pressure and the skin friction coefficients, computed on the aerofoil surface, as well as the numerical results obtained by Kordulla in~\cite{Bristeau1987}.
Similarly to the results observed in the inviscid simulation, the Lax-Friedrichs Riemann solver displays discrepancies with respect to the reference curves for both the pressure and the skin friction coefficient. The Lax-Friedrichs results are matched by the ones provided by the HLL Riemann solver which shows an excessive numerical dissipation in the viscous boundary layer. The overdiffusive nature of the HLL numerical flux, not observed in the inviscid case, is attributed to its misrepresentation of contact and shear waves~\cite{Einfeldt1988,Einfeldt1991}. Concerning Roe numerical flux, this Riemann solver strongly depends upon the choice of the value of the entropy fix also in the viscous case. Without entropy fix ($\delta = 0$), numerical oscillations of the solution near the leading edge appear and larger values of the threshold parameter $\delta$ are required to remedy this issue. For sufficiently large values of the entropy fix, the solution computed using the Roe Riemann solver is in good agreement with the reference one. Such an accurate approximation is also achieved by the FCFV method using the HLLEM numerical flux, without the need of tuning any parameter.

Table~\ref{tb:NACA0012_viscous_LiftDrag} reports the values of the lift and drag coefficients, computed using different Riemann solvers. Reference data from several numerical studies based on various computational methods were collected in~\cite{Bristeau1987}, reporting values of the lift coefficient in the range $[0.415,0.483]$ and of the drag coefficient in the interval $[0.2430,0.2868]$.
\begin{table} [!ht]
	\centering
	\caption{Viscous laminar transonic flow over a NACA 0012 profile - Lift, $C_l$, and drag, $C_d$, coefficients computed using different Riemann solvers.}
	\begin{tabular}{ L{1cm} C{3cm} C{3cm} C{2.25cm} C{2.25cm}  }
		\toprule
		&Lax-Friedrichs & Roe ($\delta = 0.1$) & HLL  & HLLEM \\
		\midrule
		$C_l$  & $0.528$ & $0.468$ & $0.518$ & $0.466$ \\
		\midrule
		$C_d$ & $0.3215$ & $0.2845$ & $0.3135$ & $0.2832$ \\
		\bottomrule
	\end{tabular}
	\label{tb:NACA0012_viscous_LiftDrag}
\end{table}
The excessive numerical dissipation introduced by the Lax-Friedrichs and HLL numerical fluxes leads to estimate of the lift and drag coefficients with errors beyond the acceptable accuracy. On the contrary, the FCFV method equipped with the HLLEM and the Roe (with appropriate entropy fix) Riemann solvers provides values of the lift and drag coefficients liying within the ranges of published values for this benchmark, showing acceptable levels of accuracy also for the simulation of viscous laminar flows.

\subsection{Low Mach number flow over a cylinder}
\label{ssc:lowMach}

In this section, an incompressible flow over a 2D cylinder at angle of attack $\alpha = 0\degree$ is considered, both in the inviscid and viscous laminar case. The objective is to show the robustness of the proposed FCFV solver for compressible flows when low Mach number flows are considered~\cite{Peraire-WDP-01,Sevilla-SHM:2013}.

\begin{figure}[!ht]
	\centering
	\subfloat[Inviscid flow \label{fig:CylinderMesh_inviscid}]{\includegraphics[height=0.3\textwidth]{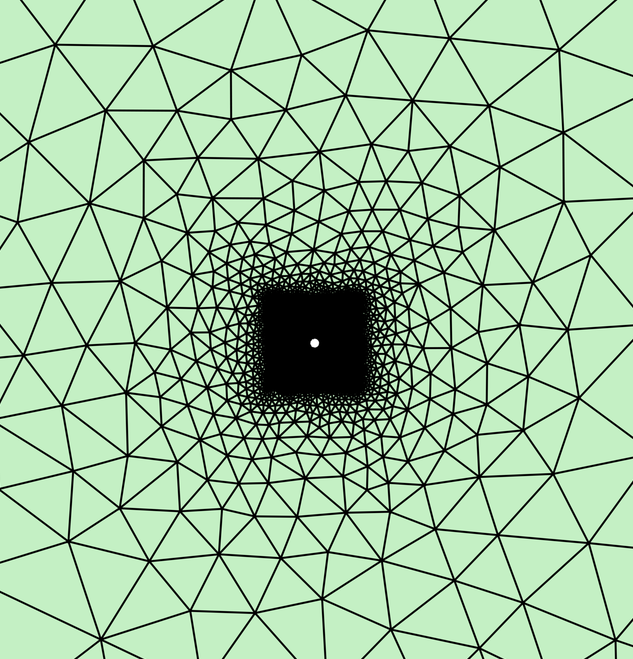}} \qquad
	\subfloat[Viscous flow \label{fig:CylinderMesh_viscous}]{\includegraphics[height=0.3\textwidth]{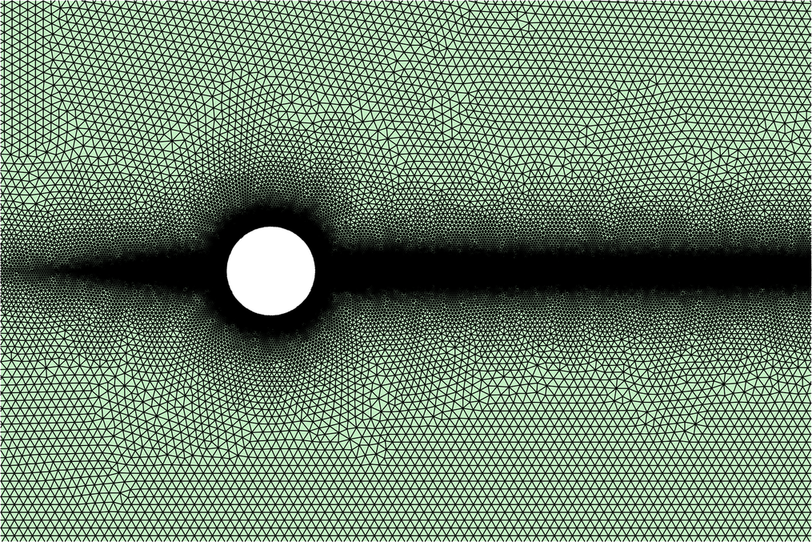}}
	\caption{Meshes for the low Mach number flows over a cylinder in the inviscid and viscous case.}
	\label{fig:CylinderMesh}
\end{figure}

Unstructured meshes of triangular cells are considered for both the inviscid and the viscous simulations. On the one hand, for the inviscid case, the mesh is isotropically refined in the vicinity of the cylinder, for a total of $359,242$ cells, as displayed in figure~\ref{fig:CylinderMesh_inviscid}. The far-field boundary is located at 50 chord units from the cylinder where inviscid wall boundary conditions are imposed. On the other hand, the refinement of the boundary layer and of the wake of the cylinder in the viscous simulation leads to a mesh of $654,194$ triangular cells, reported in figure~\ref{fig:CylinderMesh_viscous}. In this case, the far-field boundary is placed at 20 chord lengths from the obstacle and the surface of the cylinder is considered adiabatic.

\begin{figure}[!ht]
	\subfloat[$\Minf = 0.1$]{\includegraphics[width=0.48\textwidth]{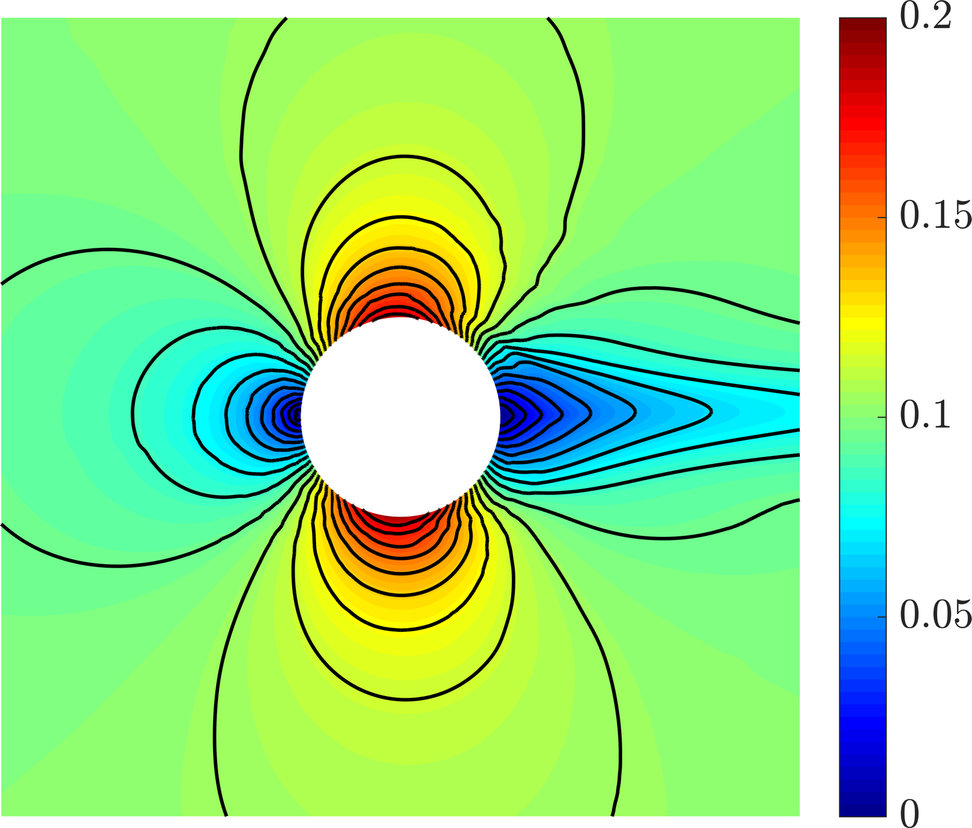}} \hfill
	\subfloat[$\Minf = 0.01$]{\includegraphics[width=0.48\textwidth]{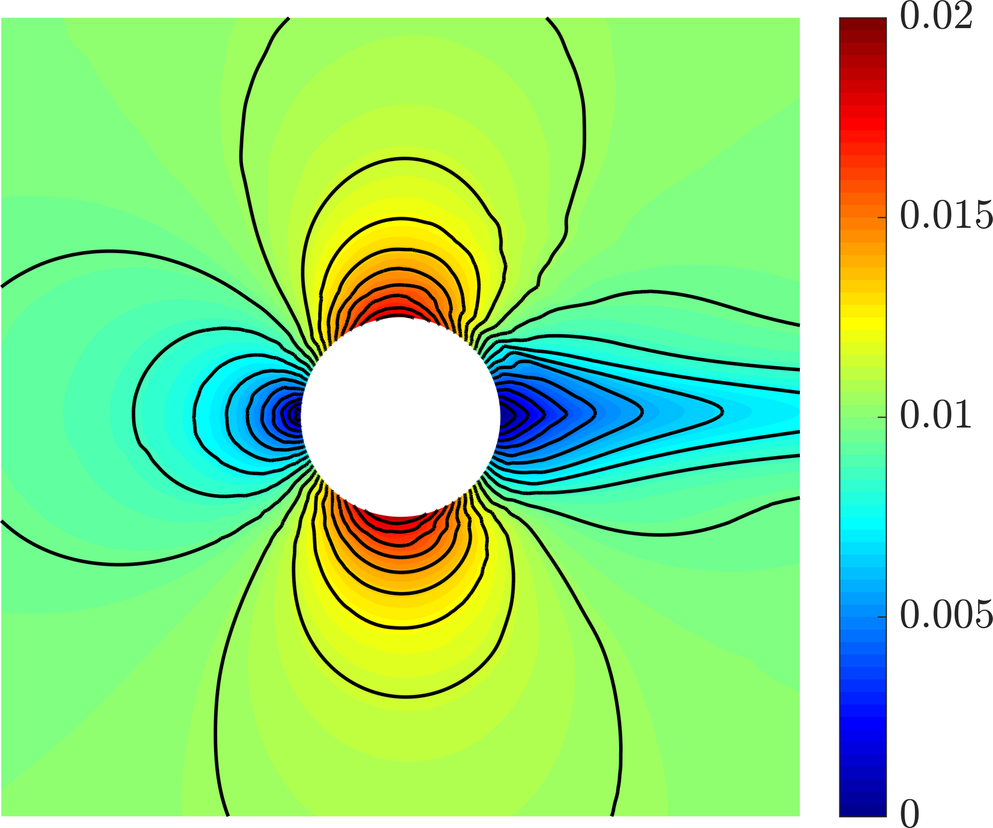}}
	\caption{Inviscid subsonic flow over a cylinder - Mach number distribution computed using the HLLEM Riemann solver for different values of the far-field condition.}
	\label{fig:Cylinder_inviscid_Mach}
\end{figure}

The isolines of the Mach number distribution computed with the HLLEM Riemann solver for the inviscid flow are reported in figure~\ref{fig:Cylinder_inviscid_Mach} for different values of the far-field condition. The results display the robustness of the FCFV solver for low Mach number simulations, highlighting the capability of the method to devise non-oscillatory solutions even in the incompressible limit. More precisely, the computed solution is not deteriorated by the decrease of the Mach number, even when it approaches zero. Note that the loss of symmetry of the solution is due to the geometric error introduced by the piecewise linear approximation of the surface of the cylinder. This well-known problem, see~\cite{Bassi-BR:97}, is related to the production of nonphysical entropy by the low-order discretisation of the curved boundary. It is worth recalling that the objective of this test is to show the robustness of the proposed method in the incompressible limit. In order to remedy the above mentioned issue, several approaches proposed in the literature can be employed within the FCFV paradigm. These include high-order approximation of the geometry~\cite{Bassi-BR:97}, appropriate modification of the wall boundary condition~\cite{Krivodonova-KB:2006} or exact treatment of the geometry via the NURBS-enhanced finite element method~\cite{Sevilla-SFH-11}.

\begin{figure}[!ht]
	\subfloat[$\Minf = 0.1$]{\includegraphics[width=0.48\textwidth]{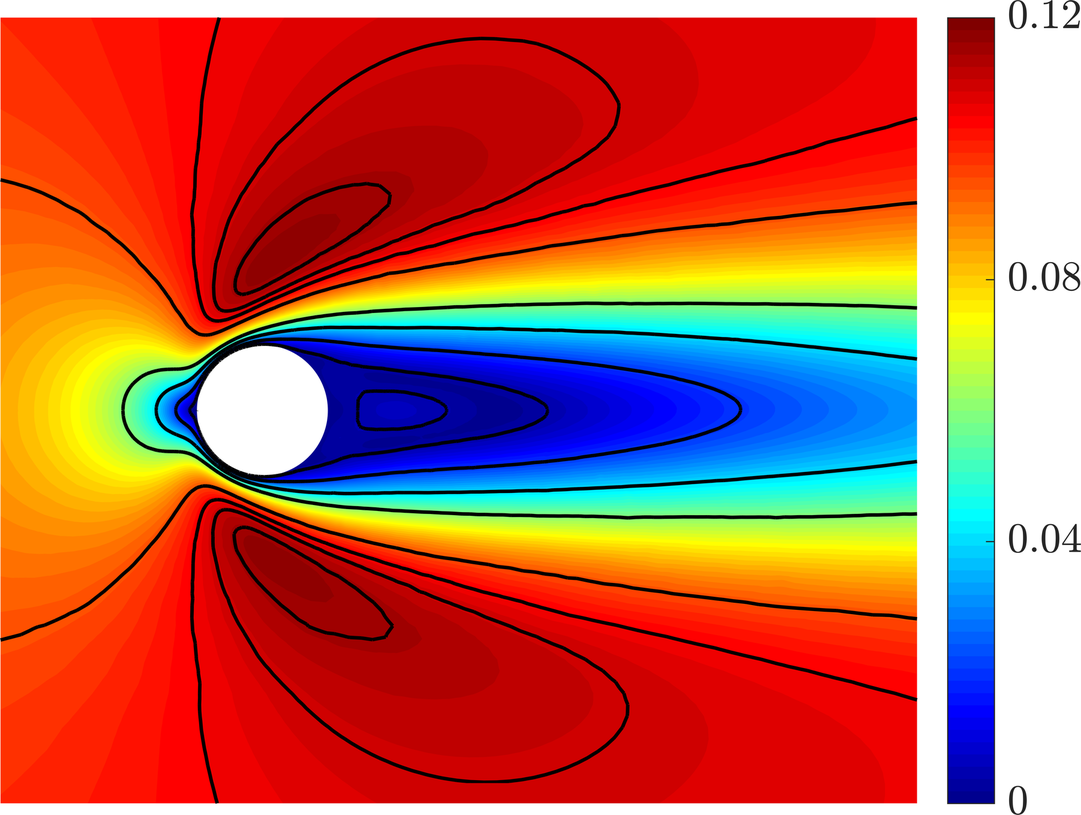}} \hfill
	\subfloat[$\Minf = 0.01$]{\includegraphics[width=0.48\textwidth]{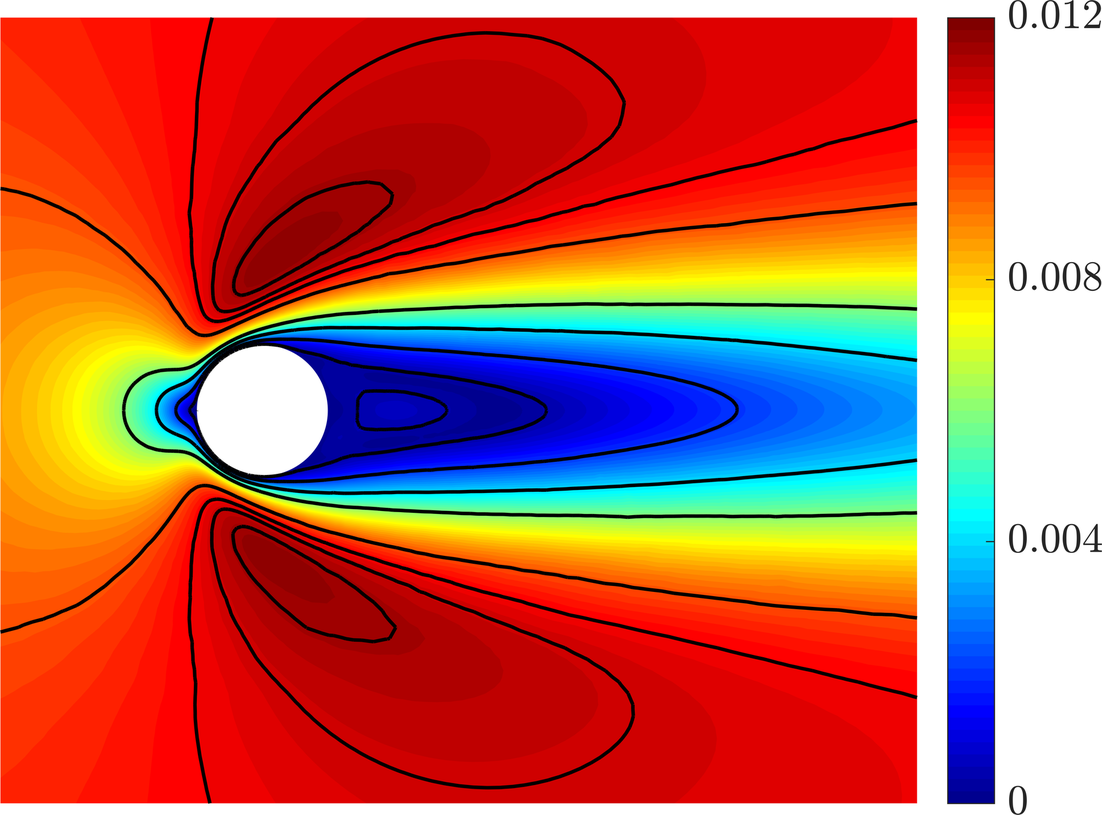}}
	\caption{Viscous laminar subsonic flow over a cylinder at $\Rey = 30$ - Mach number distribution computed using the HLLEM Riemann solver for different values of the far-field condition.}
	\label{fig:Cylinder_viscous_Mach}
\end{figure}

The robustness of the FCFV solver in the incompressible limit of the compressible Navier-Stokes equations is studied through a steady-state flow at $\Rey = 30$. Figure~\ref{fig:Cylinder_viscous_Mach} displays the isolines of the Mach number distribution for far-field conditions at $\Minf = 0.1$ and $\Minf = 0.01$. In this case, the FCFV solver is able to precisely approximate the flow in the wake of the cylinder, with no loss of accuracy when approaching the incompressible limit. It is worth noticing that in both figure~\ref{fig:Cylinder_inviscid_Mach} and figure~\ref{fig:Cylinder_viscous_Mach}, the variation of the far-field boundary condition only affects the scale of the computed Mach number and not its distribution.

\begin{figure}[!ht]
	\subfloat[Pressure coefficient]{\includegraphics[width=0.48\textwidth]{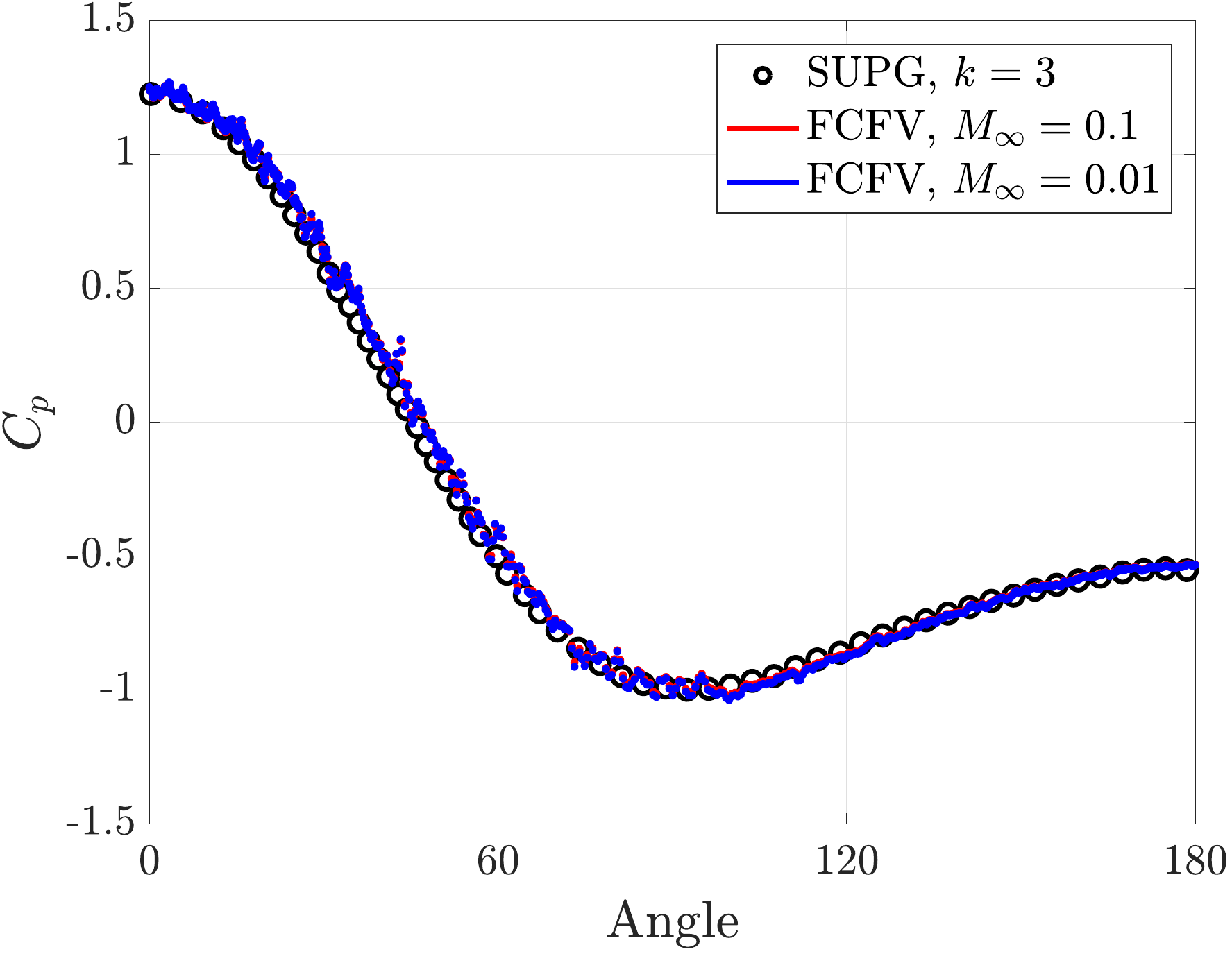}} \hfill
	\subfloat[Skin friction coefficient]{\includegraphics[width=0.48\textwidth]{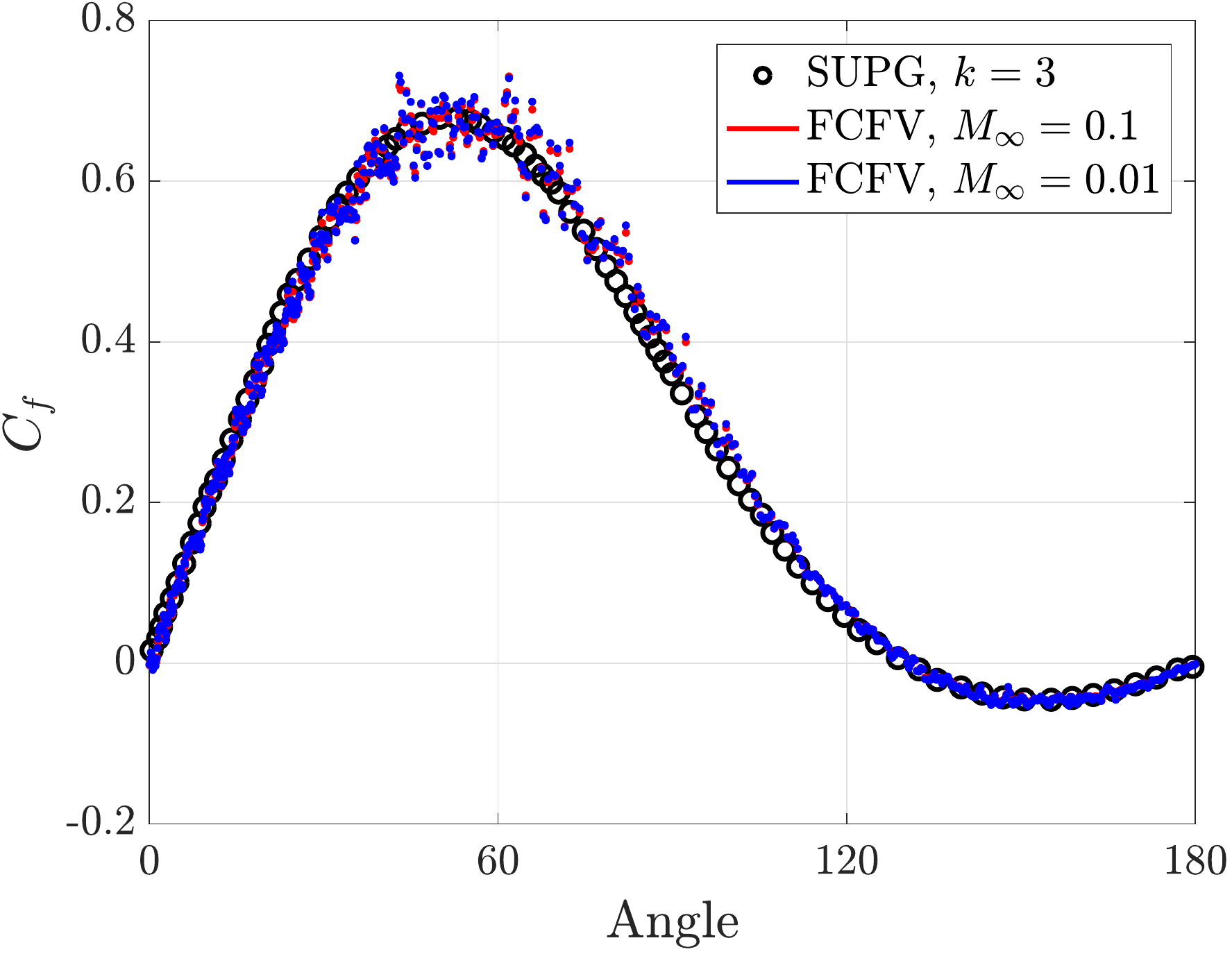}}
	\caption{Viscous laminar subsonic flow over a cylinder at $\Rey = 30$ - (a) Pressure and (b) skin friction coefficient on the object surface computed using the HLLEM Riemann.}
	\label{fig:Cylinder_viscous_AerodynamicCoefficients}
\end{figure}

Finally, the values of the pressure and skin friction coefficients computed using the HLLEM Riemann solver are compared with the results reported in~\cite{Sevilla-SHM:2013} of a stabilised finite element simulation performed with polynomial approximation of degree $k=3$. Good agreement is displayed in figure~\ref{fig:Cylinder_viscous_AerodynamicCoefficients} for both the pressure and the skin friction coefficients, confirming the capability the proposed methodology of accurately simulating viscous laminar flows also in the incompressible limit. In particular, it is worth noticing that the FCFV results computed using the two values of the far-field condition are almost identical and they are not affected by the value of the Mach number going to zero.
Hence, the FCFV method equipped with the HLLEM numerical flux provides a robust solver for steady-state compressible flows able to seamlessly handle both inviscid and viscous flows, at high and low Mach numbers.

\subsection{Inviscid transonic flow over an ONERA M6 wing}

The last example involves the 3D simulation of a steady inviscid transonic flow over an ONERA M6 wing at free-stream Mach number $\Minf = 0.84$ and angle of attack $\alpha = 3.06\degree$. This benchmark constitutes a classic CFD validation example for external flows due to its complex flow physics and the availability of experimental results at high Reynolds number~\cite{Schmitt-AGARD:1979}.

%

\begin{figure}[!ht]
	\centering
	\subfloat[Coarse mesh \label{fig:OneraM6_inviscid_mesh1}]{\includegraphics[width=0.42\textwidth]{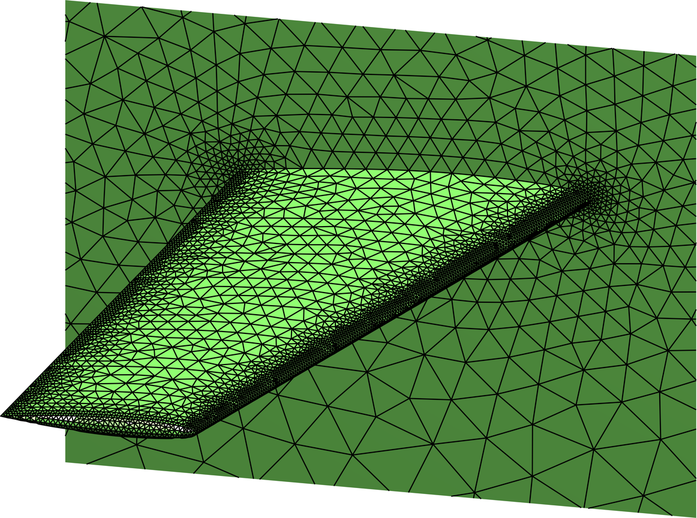}} \qquad
	\subfloat[Fine mesh \label{fig:OneraM6_inviscid_mesh3}]{\includegraphics[width=0.42\textwidth]{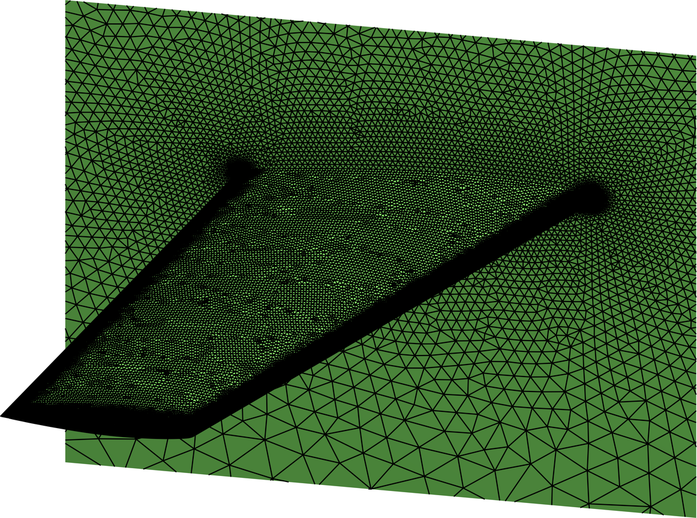}}
	\caption{Mesh refinement for the inviscid transonic flow over an ONERA M6 wing.}
	\label{fig:OneraM6_inviscid_meshes}
\end{figure}

A non-uniform mesh refinement is adopted in the vicinity of the wing surface and towards the leading and trailing edges. Figure~\ref{fig:OneraM6_inviscid_meshes} details two levels of refinement with meshes consisting of 236,682 and 5,061,252 tetrahedral cells, respectively. The far-field boundary is located at approximately 12 chord lengths from the wing and the aerofoil surface is defined as an inviscid wall.

\begin{figure}[!ht]
	\subfloat[Mach \label{fig:OneraM6_inviscid_Mach}]{\includegraphics[width=0.48\textwidth]{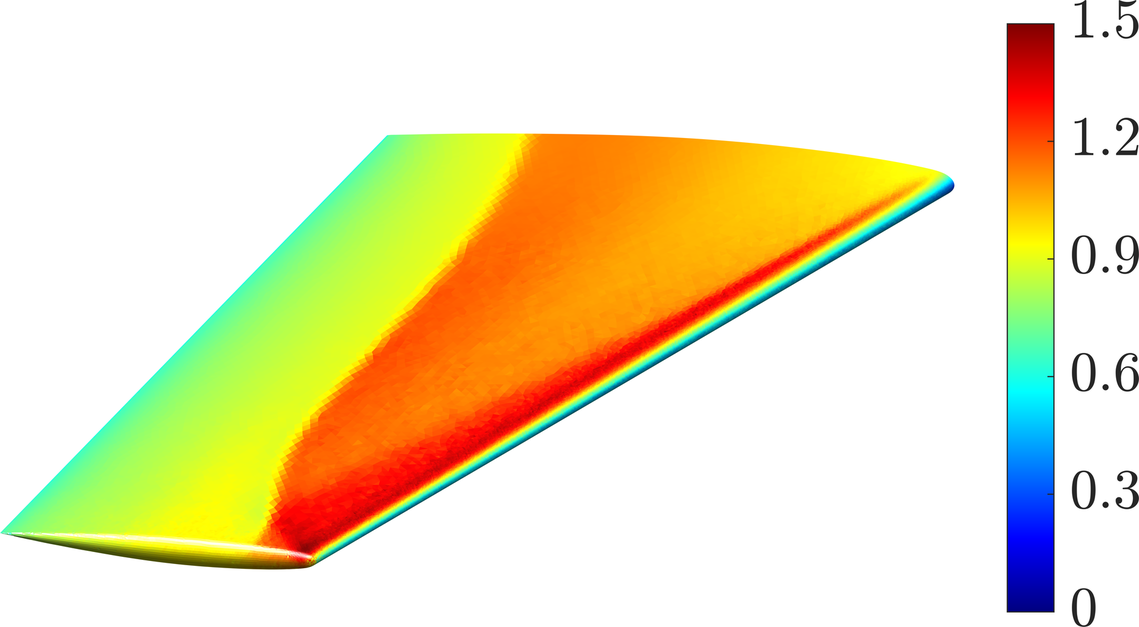}} \hfill
	\subfloat[Pressure, $p/p_\infty$ \label{fig:OneraM6_inviscid_pressure}]{\includegraphics[width=0.48\textwidth]{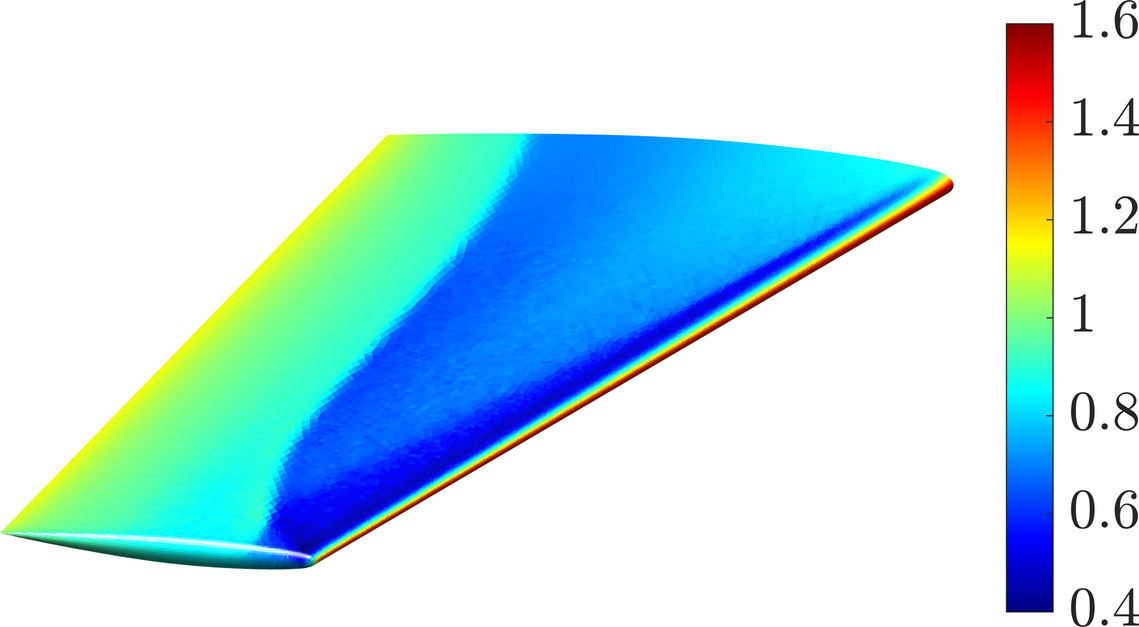}}
	\caption{Inviscid transonic flow over an ONERA M6 wing - (a) Mach number and (b) pressure distributions on the wing surface computed on the fine mesh using the HLL Riemann solver.}
	\label{fig:OneraM6_inviscid_Quantities}
\end{figure}

The FCFV simulation is performed using the HLL Riemann solver, based on its capability of producing positivity-conserving solutions and on its robustness in inviscid supersonic cases. The linear system of equations arising from the FCFV discretisation is solved using a GMRES with restarting parameter $10$ and no preconditioner. A detail of the flow computation obtained on the fine mesh is reported in figure~\ref{fig:OneraM6_inviscid_Quantities}. The Mach number and pressure distributions clearly show that the FCFV method is able of accurately capturing the characteristic \emph{lambda-shock} arising in this test benchmark.
\begin{figure}[!ht]
	\subfloat[$y/b = 0.20$ \label{fig:ONERA_inviscid_Cp_Section1}]{\includegraphics[width=0.33\textwidth]{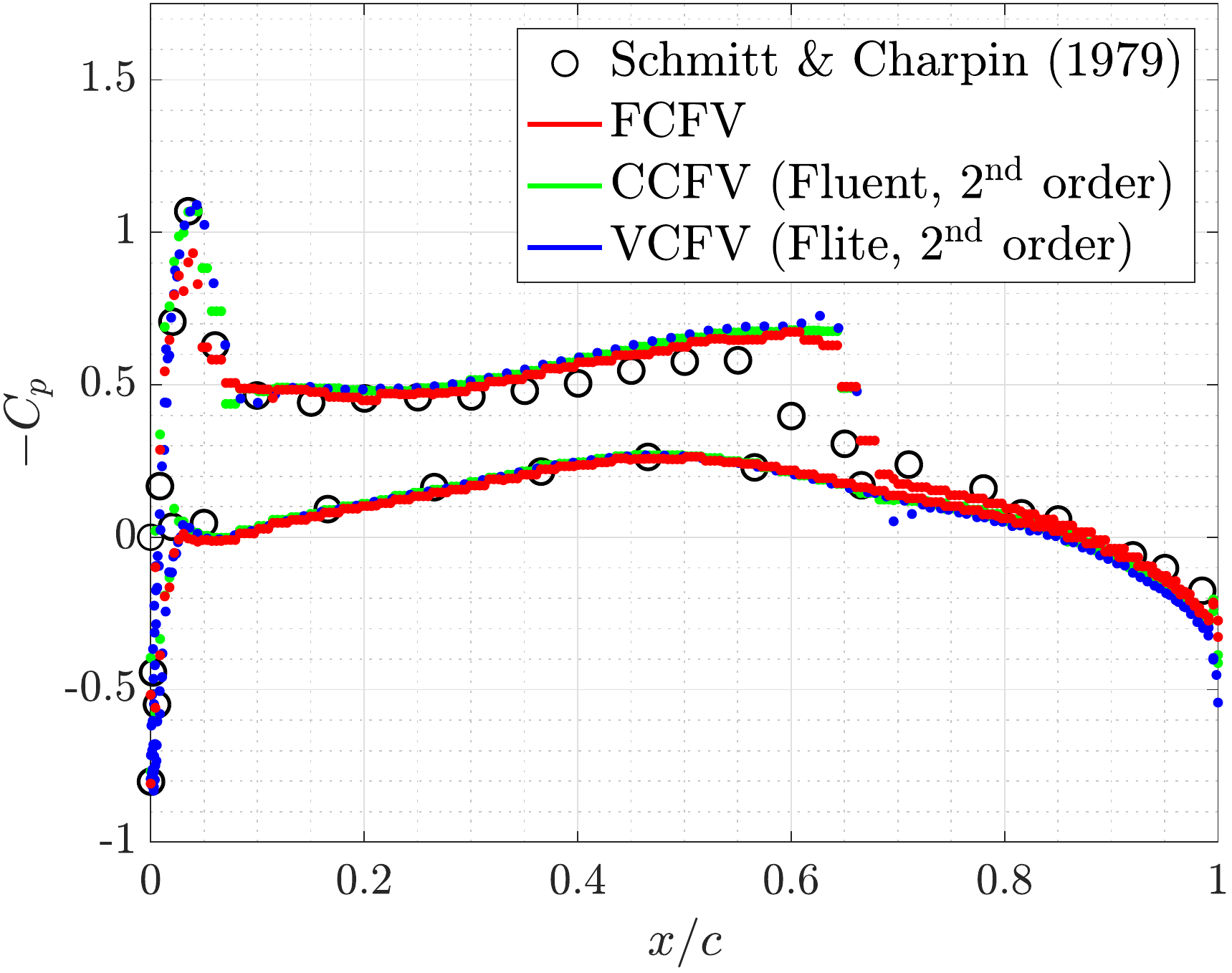}} \hfill
	\subfloat[$y/b = 0.44$ \label{fig:ONERA_inviscid_Cp_Section2}]{\includegraphics[width=0.33\textwidth]{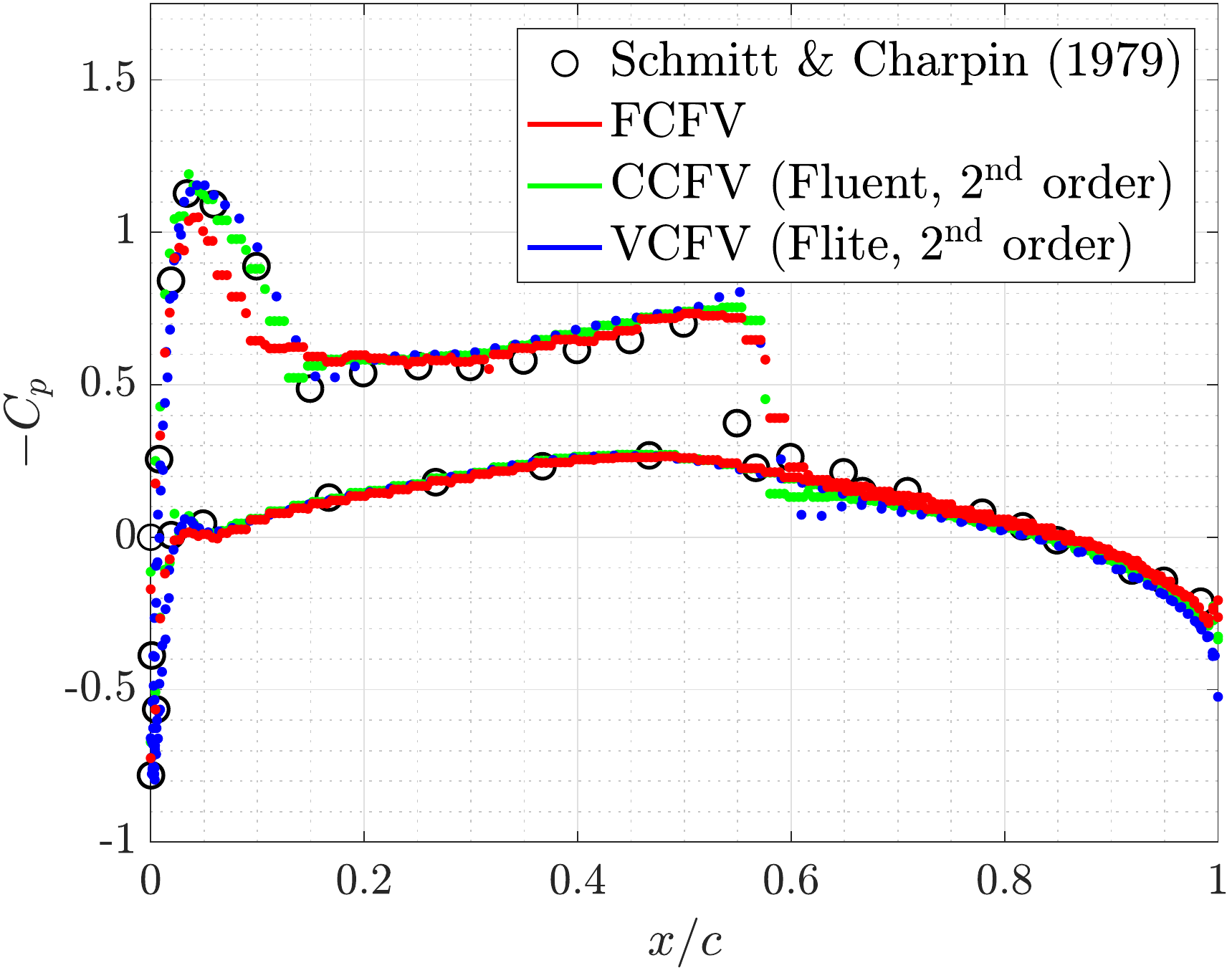}} \hfill
	\subfloat[$y/b = 0.65$ \label{fig:ONERA_inviscid_Cp_Section3}]{\includegraphics[width=0.33\textwidth]{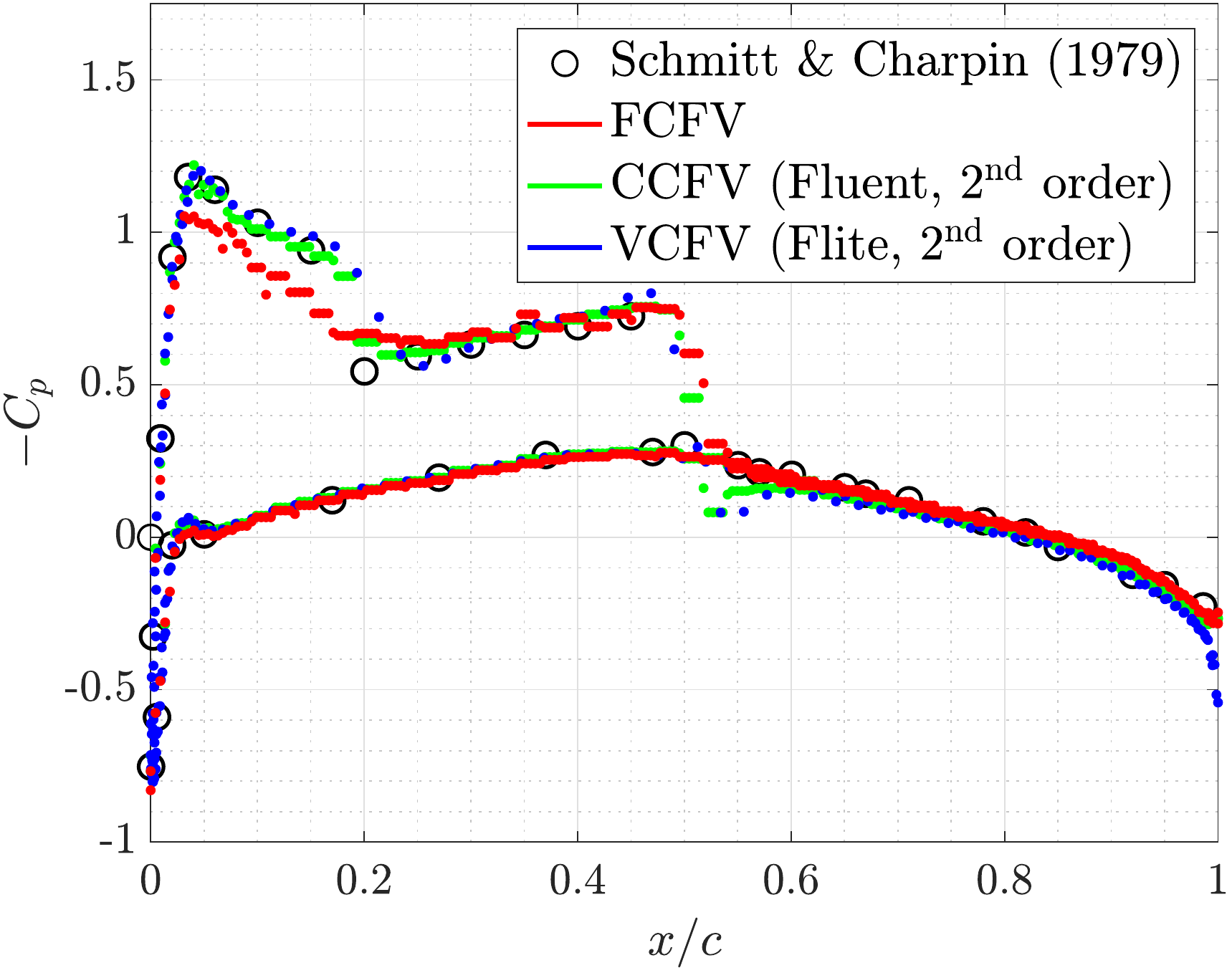}} \\
	\subfloat[$y/b = 0.80$ \label{fig:ONERA_inviscid_Cp_Section4}]{\includegraphics[width=0.33\textwidth]{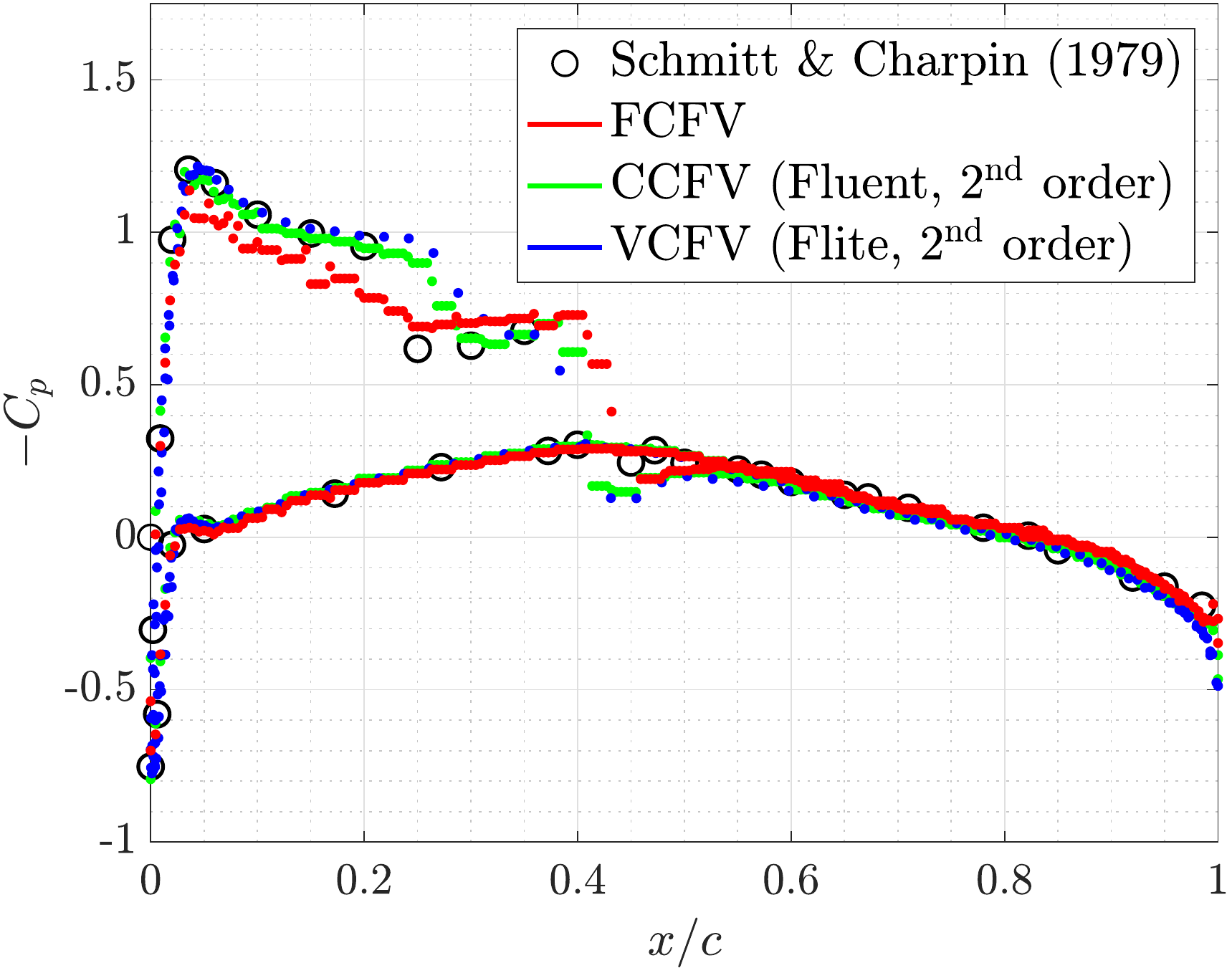}} \hfill
	\subfloat[$y/b = 0.90$ \label{fig:ONERA_inviscid_Cp_Section5}]{\includegraphics[width=0.33\textwidth]{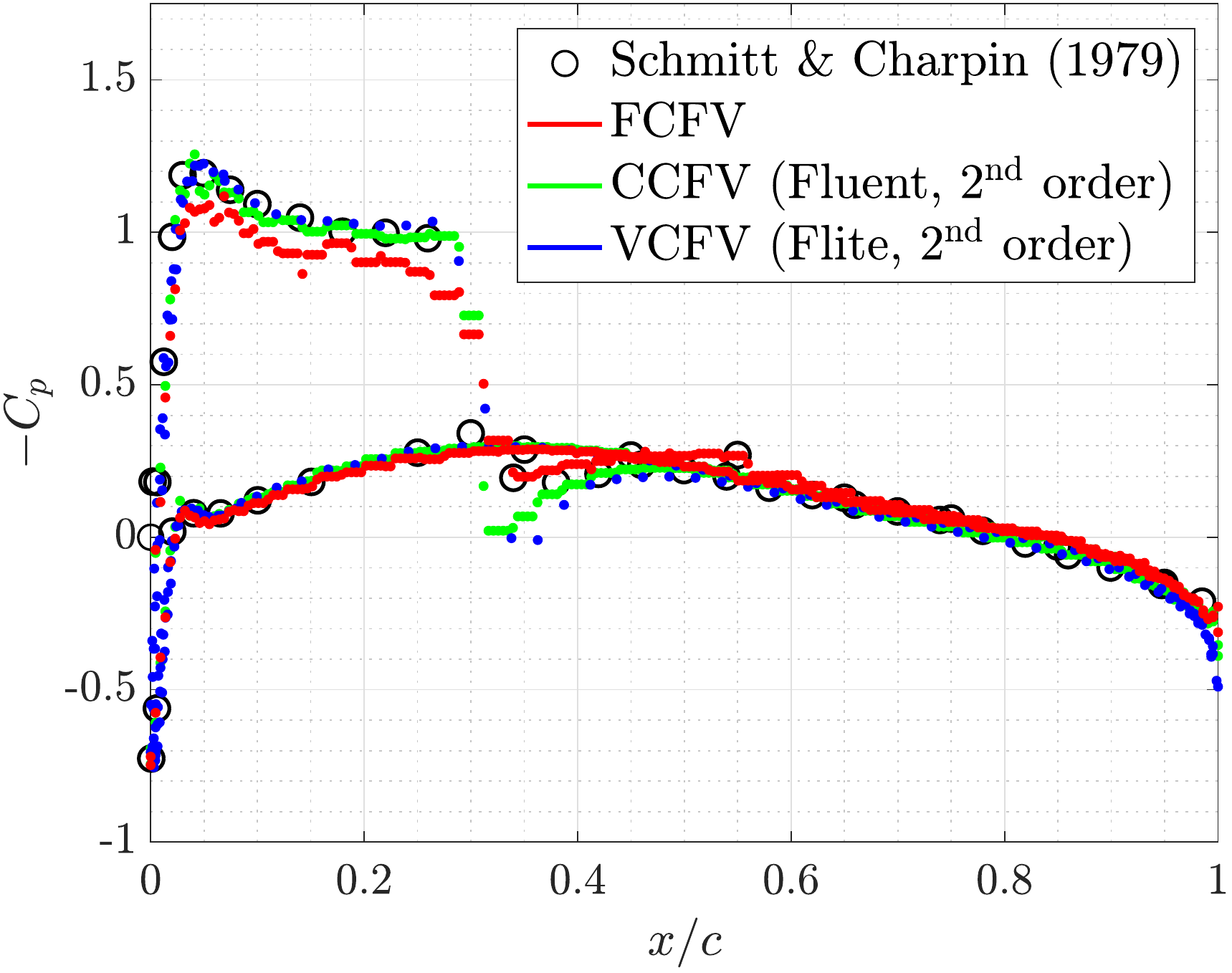}} \hfill
	\subfloat[$y/b = 0.95$ \label{fig:ONERA_inviscid_Cp_Section6}]{\includegraphics[width=0.33\textwidth]{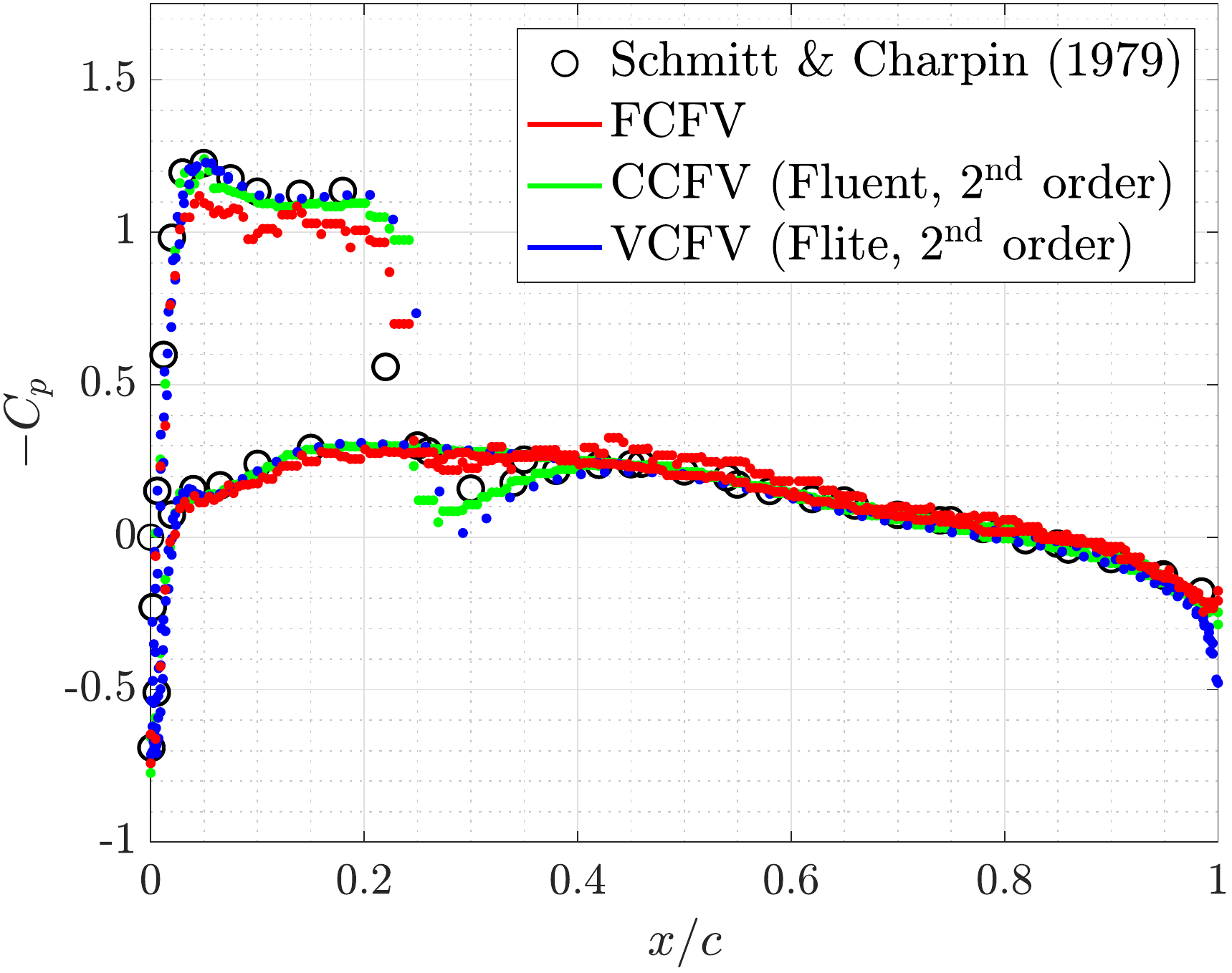}} \\
	\caption{Inviscid transonic flow over an ONERA M6 wing - Comparison of the pressure coefficient distribution at different sections along the wing span using different FV solvers.}
	\label{fig:OneraM6_inviscid_Cp}
\end{figure}

The performance of the FCFV method is evaluated by examining the pressure coefficient, computed on the fine mesh, at different sections along the wing span, see figure~\ref{fig:OneraM6_inviscid_Cp}. The obtained results are compared both to experimental data~\cite{Schmitt-AGARD:1979} and to computational simulations performed using second-order CCFV and VCFV solvers on the same mesh. More precisely, the commercial CFD software Ansys Fluent~\cite{FluentManual} is employed as CCFV solver, whereas the VCFV results are obtained using the CFD solver FLITE~\cite{sorensen2003a,sorensen2003b}. An upwind scheme is utilised for the treatment of the numerical fluxes in Fluent, whereas a Roe Riemann solver is selected for the FLITE simulation.
The numerical results obtained with the FCFV method show excellent agreement both with the experimental data and with the remaining FV schemes. It is worth noticing that both second-order FV schemes produce a small oscillation in the representation of the shock-wave located at the upper mid-chord. On the contrary, the first-order FCFV method based on the positivity-preserving HLL Riemann solver is capable of computing non-oscillatory solutions, establishing a robust framework for the simulation of 3D problems involving complex flow features.

\section{Concluding remarks}
\label{sc:Conclusions}

The face-centred finite volume paradigm was proposed for the first time for the approximation of nonlinear hyperbolic PDEs modelling compressible flows. The method is based on a mixed formulation and defines the unknowns, that is, the hybrid vector of conservative variables, at the barycentre of the faces. The unknowns in each cell, i.e. density, momentum, energy, deviatoric strain rate tensor and gradient of temperature, are eliminated via a hybridisation procedure to reduce the global number of degrees of freedom of the problem. In addition, traditional Riemann solvers, i.e. Lax-Friedrichs, Roe, HLL and HLLEM, are devised in the context of FCFV discretisations via appropriate definitions of the numerical fluxes.

\hl{The presented methodology provides first-order accuracy of the conservative quantities, as well as of the stress tensor and of the heat flux. More precisely, the first-order convergence of the stress tensor and the heat flux is achieved without the need to perform a reconstruction of the gradients as required by traditional second-order CCFV and VCFV strategies. The FCFV paradigm is thus robust on unstructured meshes and retains optimal accuracy even with highly stretched or distorted cells, avoiding well-known issues of traditional FV schemes.}

In addition, the FCFV method is able to construct non-oscillatory solutions of sharp fronts without the need of any shock capturing or limiting technique. The accurate treatment of shocks, expansion fans and shear waves is naturally handled by the Riemann solvers implicitly embedded in the FCFV numerical fluxes.

Finally, the method is robust in the incompressible limit, allowing to seamlessly simulate flows at low Mach number, without the need of introducing specific pressure corrections like the well-known SIMPLE algorithm. 

A comprehensive set of two- and three-dimensional numerical examples is employed to demonstrate the optimal convergence properties of the method and its capabilities to solve complex flow problems of aerodynamic interest across various regimes, from inviscid to viscous laminar flows, from transonic to subsonic incompressible flows. Moreover, a detailed comparison of the accuracy and robustness of different Riemann solvers is presented. The FCFV method equipped with HLL-type Riemann solvers thus provides a solution strategy suitable for all flow regimes, which outperforms the traditional Lax-Friedrichs numerical flux in terms of accuracy and the Roe solver in terms of robustness.

\hl{An extension of the proposed first-order FCFV approach to achieve second-order accuracy in the primal variable will be investigated in future works starting from the recent results obtained for the second-order FCFV method in the context of second-order elliptic PDEs~\cite{RS-VGSH:20,MG-RS-20}. This approach relies on two ingredients. On the one hand, a piecewise linear approximation is employed for the solution in the cells, while its gradient in the cell and the hybrid solution on the faces are maintained piecewise constant. On the other hand, a projection operator is introduced to propose a new definition of the numerical fluxes. In this context, the number of globally-coupled degrees of freedom of the second-order FCFV method is the same as the original FCFV method. On the contrary, to achieve second-order convergence of the primal variable, a standard HDG method~\cite{Peraire-PNC:10,Fernandez-FNP:2017,JVP_HDG-VGSH:20} employs polynomial functions of degree one for all the variables, including the hybrid one, substantially increasing the number of unknowns and the size of the resulting system of equations.}

\section*{Acknowledgements}
This work was supported by the Spanish Ministry of Economy and Competitiveness, through the Mar\'ia de Maeztu programme for units of excellence in R\&D that financed the PhD fellowship of J.V.P. (MDM-2014-0445), the Spanish Ministry of Science and Innovation and the Spanish State Research Agency MCIN/AEI/10.13039/501100011033 (PID2020-113463RB-C33 to  M.G., PID2020-113463RB-C32 to A.H., CEX2018-000797-S to A.H. and M.G.),  the Generalitat de Catalunya (2017-SGR-1278 to A.H. and M.G.) and the Engineering and Physical Sciences Research Council (EP/P033997/1 to R.S.).
M.G. also acknowledges the support of the Serra H\'unter Programme of the Generalitat de Catalunya.

\bibliographystyle{unsrt}
\bibliography{FCFV-K0_Compressible}

\begin{thebibliography}{10}

\bibitem{Versteeg2007}
H.~Versteeg and W.~Malalasekra.
\newblock {\em An Introduction to Computational Fluid Dynamics}.
\newblock Prentice Hall, 2007.

\bibitem{Leveque2013}
R.J. Leveque.
\newblock {\em Finite {V}olume {M}ethods for {H}yperbolic {P}roblems}.
\newblock Cambridge University Press, 2013.

\bibitem{FluentManual}
{ANSYS}.
\newblock {F}luent tutorial guide.
\newblock Technical report, {ANSYS Inc.}, 2017.

\bibitem{CFL3D:06}
R.E. Bartels, C.L. Rumsey, and R.T. Biedron.
\newblock {CFL3D} {V}ersion 6.4 --- {G}eneral usage and aeroelastic analysis.
\newblock Technical report, NASA/TM-2006-214301, 2006.

\bibitem{FUN3D:19}
R.T. Biedron, J.-R. Carlson, J.M. Derlaga, P.A. Gnoffo, B.~Kleb D.P.~Hammond,
  W.T.~Jones, E.M. Lee-Rausch, E.J. Nielsen, M.A. Park, C.L. Rumsey, J.L.
  Thomas, K.B. Thompson, and W.A. Wood.
\newblock {FUN3D} {M}anual: 13.6.
\newblock Technical report, NASA/TM-2019-220416, 2019.

\bibitem{Morgan1991}
K.~Morgan, J.~Peraire, J.~Peiro, and O.~Hassan.
\newblock The computation of three-dimensional flows using unstructured grids.
\newblock {\em Computer Methods in Applied Mechanics and Engineering},
  87(2-3):335--352, 1991.

\bibitem{sorensen2003a}
K.A. S{\o}rensen, O.~Hassan, K.~Morgan, and N.P. Weatherill.
\newblock A multigrid accelerated time-accurate inviscid compressible fluid
  flow solution algorithm employing mesh movement and local remeshing.
\newblock {\em International journal for numerical methods in fluids},
  43(5):517--536, 2003.

\bibitem{sorensen2003b}
K.A. S{\o}rensen, O.~Hassan, K.~Morgan, and N.P. Weatherill.
\newblock A multigrid accelerated hybrid unstructured mesh method for 3{D}
  compressible turbulent flow.
\newblock {\em Computational mechanics}, 31(1-2):101--114, 2003.

\bibitem{jasak2009openfoam}
H.~Jasak.
\newblock Open{FOAM}: open source {CFD} in research and industry.
\newblock {\em International Journal of Naval Architecture and Ocean
  Engineering}, 1(2):89--94, 2009.

\bibitem{gerhold2005overview}
T.~Gerhold.
\newblock Overview of the hybrid {RANS} code {TAU}.
\newblock In {\em MEGAFLOW-Numerical Flow Simulation for Aircraft Design},
  pages 81--92. Springer, 2005.

\bibitem{chalot2004industrial}
F.L. Chalot and P.~Perrier.
\newblock Industrial aerodynamics.
\newblock In {\em Encyclopedia of Computational Mechanics}. Wiley Online
  Library, 2004.

\bibitem{Diskin2010}
B.~Diskin, J.L. Thomas, E.J. Nielsen, H.~Nishikawa, and J.~A. White.
\newblock Comparison of node-centered and cell-centered unstructured
  finite-volume discretizations: Viscous fluxes.
\newblock {\em {AIAA} Journal}, 48(7):1326--1338, 2010.

\bibitem{Diskin2011}
B.~Diskin and J.L. Thomas.
\newblock Comparison of node-centered and cell-centered unstructured
  finite-volume discretizations: Inviscid fluxes.
\newblock {\em {AIAA} Journal}, 49(4):836--854, 2011.

\bibitem{Morton2007}
K.W. Morton and T.~Sonar.
\newblock Finite volume methods for hyperbolic conservation laws.
\newblock {\em Acta Numerica}, 16:155--238, 2007.

\bibitem{Ohlberger-BO-04}
T.~Barth and M.~Ohlberger.
\newblock Finite volume methods: Foundation and analysis.
\newblock In {\em Encyclopedia of Computational Mechanics}. John Wiley {\&}
  Sons, Ltd, 2004.

\bibitem{Eymard2000}
R.~Eymard, T.~Gallou\"{e}t, and R.~Herbin.
\newblock Finite volume methods.
\newblock In {\em Handbook of Numerical Analysis}, pages 713--1018. Elsevier,
  2000.

\bibitem{Maire2007}
P.-H. Maire, R.~Abgrall, J.~Breil, and J.~Ovadia.
\newblock A cell-centered {L}agrangian scheme for two-dimensional compressible
  flow problems.
\newblock {\em {SIAM} Journal on Scientific Computing}, 29(4):1781--1824, 2007.

\bibitem{Asouti2010}
V.G. Asouti, X.S. Trompoukis, I.C. Kampolis, and K.C. Giannakoglou.
\newblock Unsteady {CFD} computations using vertex-centered finite volumes for
  unstructured grids on {G}raphics {P}rocessing {U}nits.
\newblock {\em International Journal for Numerical Methods in Fluids},
  67(2):232--246, 2010.

\bibitem{RS-SGH:2018_FCFV1}
R.~Sevilla, M.~Giacomini, and A.~Huerta.
\newblock A face-centred finite volume method for second-order elliptic
  problems.
\newblock {\em International Journal for Numerical Methods in Engineering},
  115(8):986--1014, 2018.

\bibitem{RS-SGH:2019_FCFV2}
R.~Sevilla, M.~Giacomini, and A.~Huerta.
\newblock A locking-free face-centred finite volume ({FCFV}) method for linear
  elastostatics.
\newblock {\em Computers {\&} Structures}, 212:43--57, 2019.

\bibitem{MG-RS-20}
M.~Giacomini and R.~Sevilla.
\newblock A second-order face-centred finite volume method on general meshes
  with automatic mesh adaptation.
\newblock {\em International Journal for Numerical Methods in Engineering},
  121(23):5227--5255, 2020.

\bibitem{RS-VGSH:20}
R.~Sevilla L.~M.~Vieira, M.~Giacomini and A.~Huerta.
\newblock A second-order face-centred finite volume method for elliptic
  problems.
\newblock {\em Computer Methods in Applied Mechanics and Engineering},
  358:112655, 2020.

\bibitem{Svaerd2008}
M.~Sv\"{a}rd, J.~Gong, and J.~Nordstr\"{o}m.
\newblock An accuracy evaluation of unstructured node-centred finite volume
  methods.
\newblock {\em Applied Numerical Mathematics}, 58(8):1142--1158, 2008.

\bibitem{Toro2009}
E.F. Toro.
\newblock {\em Riemann {S}olvers and {N}umerical {M}ethods for {F}luid
  {D}ynamics}.
\newblock Springer-Verlag Berlin Heidelberg, 2009.

\bibitem{Hesthaven2017}
J.S. Hesthaven.
\newblock {\em Numerical {M}ethods for {C}onservation {L}aws}.
\newblock Society for Industrial and Applied Mathematics, 2017.

\bibitem{Cockburn-CS:1998}
B.~Cockburn and C.-W. Shu.
\newblock The {R}unge{\textendash}{K}utta {D}iscontinuous {G}alerkin {M}ethod
  for {C}onservation {L}aws {V}.
\newblock {\em Journal of Computational Physics}, 141(2):199--224, 1998.

\bibitem{Qiu-QKS:2006}
J.~Qiu, B.~Cheong Khoo, and C.-W. Shu.
\newblock A numerical study for the performance of the
  {R}unge{\textendash}{K}utta discontinuous {G}alerkin method based on
  different numerical fluxes.
\newblock {\em Journal of Computational Physics}, 212(2):540--565, 2006.

\bibitem{VonNeumann1950}
J.~{Von Neumann} and R.~D. Richtmyer.
\newblock A method for the numerical calculation of hydrodynamic shocks.
\newblock {\em Journal of Applied Physics}, 21(3):232--237, 1950.

\bibitem{Donea2003}
J.~Donea and A.~Huerta.
\newblock {\em Finite Element Methods for Flow Problems}.
\newblock John Wiley \& Sons, Ltd., 2003.

\bibitem{Persson-PP:2006}
P.-O. Persson and J.~Peraire.
\newblock Sub-{C}ell {S}hock {C}apturing for {D}iscontinuous {G}alerkin
  {M}ethods.
\newblock {\em AIAA Paper}, 0112, 2006.

\bibitem{Fernandez-FNP:2018}
P.~Fern{\'a}ndez, C.~Nguyen, and J.~Peraire.
\newblock A physics-based shock capturing method for unsteady laminar and
  turbulent flows.
\newblock {\em AIAA Paper}, 2018.

\bibitem{vanLeer1979}
Bram {van Leer}.
\newblock Towards the ultimate conservative difference scheme. {V}. a
  second-order sequel to {G}odunov's method.
\newblock {\em Journal of Computational Physics}, 32(1):101--136, 1979.

\bibitem{Sweby1984}
P.~K. Sweby.
\newblock High resolution schemes using flux limiters for hyperbolic
  conservation laws.
\newblock {\em {SIAM} Journal on Numerical Analysis}, 21(5):995--1011, 1984.

\bibitem{Toro2000}
E.~Toro.
\newblock Centred {TVD} schemes for hyperbolic conservation laws.
\newblock {\em {IMA} Journal of Numerical Analysis}, 20(1):47--79, 2000.

\bibitem{Krivodonova2007}
Lilia Krivodonova.
\newblock Limiters for high-order discontinuous {G}alerkin methods.
\newblock {\em Journal of Computational Physics}, 226(1):879--896, 2007.

\bibitem{Cockburn-CKS:2000}
B.~Cockburn, G.E. Karniadakis, and C.-W. Shu.
\newblock The development of discontinuous {G}alerkin methods.
\newblock In {\em Discontinuous {G}alerkin {M}ethods}, pages 3--50.
  Springer-Verlag Berlin Heidelberg, Berlin, Germany, 2000.

\bibitem{Selmin1993}
V.~Selmin.
\newblock The node-centred finite volume approach: {B}ridge between finite
  differences and finite elements.
\newblock {\em Computer Methods in Applied Mechanics and Engineering},
  102(1):107--138, 1993.

\bibitem{Bank1987}
R.E. Bank and D.J. Rose.
\newblock Some error estimates for the {B}ox method.
\newblock {\em {SIAM} Journal on Numerical Analysis}, 24(4):777--787, 1987.

\bibitem{Jay-CGL:09}
B.~Cockburn, J.~Gopalakrishnan, and R.~Lazarov.
\newblock Unified hybridization of discontinuous {G}alerkin, mixed, and
  continuous {G}alerkin methods for second order elliptic problems.
\newblock {\em SIAM Journal on Numerical Analysis}, 47(2):1319--1365, 2009.

\bibitem{Cockburn2016}
B.~Cockburn.
\newblock Static condensation, hybridization, and the devising of the {HDG}
  methods.
\newblock In {\em Lecture Notes in Computational Science and Engineering},
  pages 129--177. Springer International Publishing, 2016.

\bibitem{HDGlab-GSH-20}
M.~Giacomini, R.~Sevilla, and A.~Huerta.
\newblock {HDGlab}: {A}n open-source implementation of the hybridisable
  discontinuous {G}alerkin method in {MATLAB}.
\newblock {\em Archives of Computational Methods in Engineering},
  28(3):1941--1986, 2021.

\bibitem{Nguyen-NP:2012}
N.~C. Nguyen and J.~Peraire.
\newblock Hybridizable discontinuous {G}alerkin methods for partial
  differential equations in continuum mechanics.
\newblock {\em Journal of Computational Physics}, 231(18):5955--5988, 2012.

\bibitem{Peraire-PNC:10}
J.~Peraire, N.~C. Nguyen, and B.~Cockburn.
\newblock A hybridizable {D}iscontinuous {G}alerkin {M}ethod for the
  {C}ompressible {E}uler and {N}avier-{S}tokes {E}quations.
\newblock {\em AIAA Paper}, 363, 2010.

\bibitem{Jay-CG:2009}
B.~Cockburn and J.~Gopalakrishnan.
\newblock The derivation of hybridizable discontinuous {G}alerkin methods for
  {S}tokes flow.
\newblock {\em {SIAM} Journal on Numerical Analysis}, 47(2):1092--1125, 2009.

\bibitem{Patankar-PS:72}
S.V Patankar and D.B Spalding.
\newblock A calculation procedure for heat, mass and momentum transfer in
  three-dimensional parabolic flows.
\newblock {\em International Journal of Heat and Mass Transfer}, 15(10):1787 --
  1806, 1972.

\bibitem{May-WBMS-14}
M.~Woopen, A.~Balan, G.~May, and J.~Sch\"{u}tz.
\newblock A comparison of hybridized and standard {DG} methods for target-based
  hp-adaptive simulation of compressible flow.
\newblock {\em Computers \& Fluids}, 98:3 -- 16, 2014.

\bibitem{Fernandez-FNP:2017}
P.~Fern{\'a}ndez, N.~C. Nguyen, and J.~Peraire.
\newblock The hybridized {D}iscontinuous {G}alerkin method for {I}mplicit
  {L}arge-{E}ddy {S}imulation of transitional turbulent flows.
\newblock {\em Journal of Computational Physics}, 336:308--329, 2017.

\bibitem{JVP_HDG-VGSH:20}
J.~Vila-P\'erez, M.~Giacomini, R.~Sevilla, and A.~Huerta.
\newblock Hybridisable discontinuous {G}alerkin formulation of compressible
  flows.
\newblock {\em Archives of Computational Methods in Engineering},
  28(2):753--784, 2021.

\bibitem{AdM-MFH:08}
A.~Montlaur, S.~Fern\'andez-M\'endez, and A.~Huerta.
\newblock Discontinuous {G}alerkin methods for the {S}tokes equations using
  divergence-free approximations.
\newblock {\em International Journal for Numerical Methods in Fluids},
  57(9):1071--1092, 2008.

\bibitem{Peraire-PNC:11}
J.~Peraire, C.~Nguyen, and B.~Cockburn.
\newblock An {E}mbedded {D}iscontinuous {G}alerkin {M}ethod for the
  {C}ompressible {E}uler and {N}avier-{S}tokes {E}quations.
\newblock {\em AIAA Paper}, 3228, 2011.

\bibitem{Harten-HH:1983}
A.~Harten and J.M. Hyman.
\newblock Self adjusting grid methods for one-dimensional hyperbolic
  conservation laws.
\newblock {\em Journal of Computational Physics}, 50(2):235--269, 1983.

\bibitem{Harten-HLL:1983}
A.~Harten, P.D. Lax, and B.~{van Leer}.
\newblock On {U}pstream {D}ifferencing and {G}odunov-{T}ype {S}chemes for
  {H}yperbolic {C}onservation {L}aws.
\newblock {\em {SIAM} Review}, 25(1):35--61, 1983.

\bibitem{Einfeldt1988}
B.~Einfeldt.
\newblock On {G}odunov-type methods for gas dynamics.
\newblock {\em {SIAM} Journal on Numerical Analysis}, 25(2):294--318, 1988.

\bibitem{Einfeldt1991}
B~Einfeldt, C.D Munz, P.L Roe, and B~Sj\"{o}green.
\newblock On {G}odunov-type methods near low densities.
\newblock {\em Journal of Computational Physics}, 92(2):273--295, 1991.

\bibitem{Rohde2001}
Axel Rohde.
\newblock Eigenvalues and eigenvectors of the {E}uler equations in general
  geometries.
\newblock {\em AIAA Paper}, 2001.

\bibitem{Sevilla-SGKH:2018}
R.~Sevilla, M.~Giacomini, A.~Karkoulias, and A.~Huerta.
\newblock A superconvergent hybridisable discontinuous {G}alerkin method for
  linear elasticity.
\newblock {\em International Journal for Numerical Methods in Engineering},
  116(2):91--116, 2018.

\bibitem{Giacomini-GKSH:2018}
M.~Giacomini, A.~Karkoulias, R.~Sevilla, and A.~Huerta.
\newblock A superconvergent {HDG} method for {S}tokes flow with strongly
  enforced symmetry of the stress tensor.
\newblock {\em Journal of Scientific Computing}, 77(3):1679--1702, 2018.

\bibitem{MG-GS-19}
M.~Giacomini and R.~Sevilla.
\newblock Discontinuous {G}alerkin approximations in computational mechanics:
  hybridization, exact geometry and degree adaptivity.
\newblock {\em SN Applied Sciences}, 1(9):1--15, 2019.

\bibitem{MG-SGH-20}
A.~{La Spina}, M.~Giacomini, and A.~Huerta.
\newblock Hybrid coupling of {CG} and {HDG} discretizations based on
  {N}itsche's method.
\newblock {\em Computational mechanics}, 65(2):311--330, 2020.

\bibitem{Tutorial-GSH:2020}
M.~Giacomini, R.~Sevilla, and A.~Huerta.
\newblock {\em Tutorial on Hybridizable Discontinuous {G}alerkin ({HDG})
  Formulation for Incompressible Flow Problems}, pages 163--201.
\newblock Springer International Publishing, Cham, 2020.

\bibitem{AlS-SKGWH:20}
A.~{La Spina}, M.~Kronbichler, M.~Giacomini, W.~A. Wall, and A.~Huerta.
\newblock A weakly compressible hybridizable discontinuous {G}alerkin
  formulation for fluid-structure interaction problems.
\newblock {\em Computer Methods in Applied Mechanics and Engineering},
  372:113392, 2020.

\bibitem{Jaust-JS:2014}
A.~Jaust and J.~Sch\"{u}tz.
\newblock A temporally adaptive hybridized discontinuous {G}alerkin method for
  time-dependent compressible flows.
\newblock {\em Computers {\&} Fluids}, 98:177--185, 2014.

\bibitem{Sanjay2020}
A.~Komala-Sheshachala, R.~Sevilla, and O.~Hassan.
\newblock A coupled {HDG}-{FV} scheme for the simulation of transient inviscid
  compressible flows.
\newblock {\em Computers {\&} Fluids}, 202:104495, 2020.

\bibitem{Chiocchia1985}
G.~Chiocchia.
\newblock Exact solutions to transonic and supersonic flows.
\newblock {\em AGARD Technical Report AR--211}, 1985.

\bibitem{Schutz-SWM:2012}
J.~Sch\"{u}tz, M.~Woopen, and G.~May.
\newblock A hybridized {DG}/mixed scheme for nonlinear advection-diffusion
  systems, including the compressible {N}avier-{S}tokes equations.
\newblock {\em AIAA Paper}, 2012.

\bibitem{Sevilla-SHM:2013}
R.~Sevilla, O.~Hassan, and K.~Morgan.
\newblock An analysis of the performance of a high-order stabilised finite
  element method for simulating compressible flows.
\newblock {\em Computer Methods in Applied Mechanics and Engineering},
  253:15--27, 2013.

\bibitem{Kroll2009}
N.~Kroll.
\newblock {\em ADIGMA: A European Project on the Development of Adaptive Higher
  Order Variational Methods for Aerospace Applications}.
\newblock Aerospace Sciences Meetings. American Institute of Aeronautics and
  Astronautics, 2009.

\bibitem{Godunov1959}
S.K. Godunov and I.~Bohachevsky.
\newblock Finite difference method for numerical computation of discontinuous
  solutions of the equations of fluid dynamics.
\newblock {\em Matematic\v{c}eskij sbornik}, 47(3):271--306, 1959.

\bibitem{AGARD:1985}
H.~Yoshihara and P.~Sacher.
\newblock Test cases for inviscid flow field methods.
\newblock {\em AGARD Advisory Report AR-211}, 1985.

\bibitem{Thibert-TGO:1979}
J.J. Thibert, M.~Granjacques, and L.~H. Ohman.
\newblock {NACA} 0012 {A}irfoil.
\newblock {\em AGARD Advisory Report AR--138}, A1, 1979.

\bibitem{YanoDarmofal2012}
M.~Yano and D.L. Darmofal.
\newblock Case {C}1. 3: {F}low over the {NACA} 0012 airfoil: subsonic inviscid,
  transonic inviscid, and subsonic laminar flows.
\newblock In {\em First International Workshop on High-Order CFD methods},
  2012.

\bibitem{Wang2013}
Z.J. Wang, K.~Fidkowski, R.~Abgrall, F.~Bassi, D.~Caraeni, A.~Cary,
  H.~Deconinck, R.~Hartmann, K.~Hillewaert, H.T. Huynh, N.~Kroll, G.~May, P.-O.
  Persson, B.~{van Leer}, and M.~Visbal.
\newblock High-order {CFD} methods: current status and perspective.
\newblock {\em International Journal for Numerical Methods in Fluids},
  72(8):811--845, 2013.

\bibitem{Bristeau1987}
M.O. Bristeau, R.~Glowinski, J.~Periaux, and H.~Viviand.
\newblock Numerical simulation of compressible {N}avier-{S}tokes flows: a
  {GAMM} workshop.
\newblock In {\em Notes on Numerical Fluid Mechanics}, volume~18.
  Vieweg+Teubner Verlag, 1987.

\bibitem{Nogueira2009}
X.~Nogueira, L.~Cueto-Felgueroso, I.~Colominas, H.~G{\'{o}}mez, F.~Navarrina,
  and M.~Casteleiro.
\newblock On the accuracy of finite volume and discontinuous {G}alerkin
  discretizations for compressible flow on unstructured grids.
\newblock {\em International Journal for Numerical Methods in Engineering},
  78(13):1553--1584, 2009.

\bibitem{Peraire-WDP-01}
J.S. Wong, D.L. Darmofal, and J.~Peraire.
\newblock The solution of the compressible {E}uler equations at low {M}ach
  numbers using a stabilized finite element algorithm.
\newblock {\em Computer Methods in Applied Mechanics and Engineering},
  190(43-44):5719--5737, 2001.

\bibitem{Bassi-BR:97}
F.~Bassi and S.~Rebay.
\newblock High-order accurate discontinuous finite element solution of the 2{D}
  {E}uler equations.
\newblock {\em J. Comput. Phys.}, 138(2):251--285, 1997.

\bibitem{Krivodonova-KB:2006}
L.~Krivodonova and M.~Berger.
\newblock High-order accurate implementation of solid wall boundary conditions
  in curved geometries.
\newblock {\em J. Comput. Phys.}, 211(2):492--512, 2006.

\bibitem{Sevilla-SFH-11}
R.~Sevilla, S.~Fern\'{a}ndez-M\'{e}ndez, and A.~Huerta.
\newblock {NURBS}-enhanced finite element method for {E}uler equations.
\newblock {\em International Journal for Numerical Methods in Fluids},
  57(9):1051--1069, 2008.

\bibitem{Schmitt-AGARD:1979}
V.~Schmitt and F.~Charpin.
\newblock Pressure distributions on the {ONERA-M6} wing at transonic {M}ach
  numbers.
\newblock Technical report, Report of the fluid Dynamics panel. Working group
  04, 1979.

\bibitem{Mengaldo2014}
G.~Mengaldo, D.~De Grazia, F.~Witherden, A.~Farrington, P.~Vincent, S.~Sherwin,
  and J.~Peiro.
\newblock A guide to the implementation of boundary conditions in compact
  high-order methods for compressible aerodynamics.
\newblock {\em AIAA Paper}, 2014.

\bibitem{Fish-Belytschko:2007}
J.~Fish and T.~Belytschko.
\newblock {\em A First Course in Finite Elements}.
\newblock John Wiley and Sons Ltd, 2007.

\end{thebibliography}

\appendix

\section{Boundary conditions for the FCFV formulation}
\label{app:BoundaryConditions}

In this appendix, the definitions of the boundary conditions for compressible flow problems are briefly recalled in the context of hybrid discretisation methods~\cite{JVP_HDG-VGSH:20,Nguyen-NP:2012,Mengaldo2014,Peraire-PNC:10,Fernandez-FNP:2017}. Following the HDG framework, the implementation of boundary conditions in the FCFV formulation of compressible flows rely on the exploitation of the hybrid vector $\bhu$.

Let the boundary $\partial \Omega$ be partitioned as $\partial \Omega = \Ga{\infty} \cup \Ga{\text{out}} \cup \Ga{\text{ad}} \cup \Ga{\text{iso}} \cup \Ga{\text{inv}}$, where the introduced portions are disjoint by pairs. From the modelling viewpoint, $\Ga{\infty}$ refers to a far-field boundary, $\Ga{\text{out}}$ is a subsonic outlet with imposed pressure, $\Ga{\text{ad}}$ and $\Ga{\text{iso}}$ denote adiabatic and isothermal walls, respectively, whereas $\Ga{\text{inv}}$ represents a symmetry boundary or an inviscid wall with slip conditions.

The expressions of the trace boundary operator $\bb(\bu,\bhu,\beps,\bphi)$ corresponding to the above mentioned cases are reported in table~\ref{tb:boundaryConditions}. As classic in the context of compressible flows, inlet and outlet boundaries are identified through a 1D characteristics analysis in the direction of the outward normal to the boundary. More precisely, $\bm{A}_n^\pm := (\bm{A}_n \pm \abs{\bm{A}_n})/2$ denote the positive and negative parts of the matrix $\bm{A}_n(\bhu)$ and they are defined exploiting the spectral decomposition introduced in section~\ref{ssc:RiemannSolvers}. Moreover, in table~\ref{tb:boundaryConditions}, $\bUinf$ denotes the free-stream values of the conservative variables at the far-field, whereas $p_{\text{out}}$ and $T_{\text{w}}$ are the prescribed values of the outlet pressure and the wall temperature, respectively. Finally, the stabilisation coefficient $\tau^d_{\rho E} = 1/\left[ \Rey (\gamma - 1) \Minf^2 \Pra \right]$ is obtained extracting the component associated with the energy equation from the diffusive stabilisation tensor~\eqref{eq:diffStabilisation}.
\begin{table} [htbp]
	\centering
	\caption{Definition of the boundary conditions for compressible flows in FCFV discretisations.}
	\makebox[\linewidth]{
		\begin{tabular}[t]{ L{1cm} L{12cm}}
			\midrule
			$\Ga{\infty}$ & Far-field, subsonic inlet, supersonic inlet/outlet \\
			& $\bb = \bm{A}_n^+(\bhu) (\bu - \bhu) + \bm{A}_n^-(\bhu) (\bUinf - \bhu)$ \\
			\midrule
			$\Ga{\text{out}}$ & Subsonic outlet (pressure outlet) \\
			& $\displaystyle \bb = \left\lbrace \rho - \widehat{\rho}, \, \left[\rho \bv - \widehat{\rho \bv} \right]\tras,\,  \frac{p_{\text{out}}}{\gamma - 1} + \frac{\rho \norm{\bv}^2}{2} - \widehat{\rho E} \right\rbrace\tras$ \\
			\midrule
			$\Ga{\text{ad}}$ & Adiabatic wall \\
			& $\displaystyle \bb = \left\lbrace \rho - \widehat{\rho},\,  \widehat{\rho \bv}\tras,\,  \frac{\mu}{\Rey \Pra} \bphi \bn - \tau^d_{\rho E} (\rho E - \widehat{\rho E}) \right\rbrace\tras$ \\
			\midrule
			$\Ga{\text{iso}}$ & Isothermal wall \\
			& $\displaystyle \bb = \left\lbrace \rho - \widehat{\rho},\,  \widehat{\rho \bv}\tras,\,  \frac{\rho T_{\text{w}}}{\gamma} - \widehat{\rho E} \right\rbrace\tras$ \\
			\midrule
			$\Ga{\text{inv}}$ & Symmetry surface or inviscid wall \\
			& $ \bb = \lbrace \rho - \widehat{\rho},\,  \left[ (\Id{\nsd} - \bn\otimes \bn) \rho \bv - \widehat{\rho \bv} \right]\tras,\,  \rho E - \widehat{\rho E} \rbrace\tras$ \\
			\bottomrule
		\end{tabular}
	}
	\label{tb:boundaryConditions}
\end{table}

\section{Enforcing the symmetry of the mixed variable}
\label{app:Voigt}

In section~\ref{ssc:mixedVar}, the deviatoric strain rate tensor $\beps$ was introduced as mixed variable in the FCFV formulation. In this appendix, the implementation details to construct this symmetric mixed variable are provided. 

First, it is worth recalling that the symmetric second-order tensor $\beps$ is commonly represented using a matrix of dimension $\nsd \times \nsd$. Nonetheless, only $\msd = \nsd(\nsd + 1)/2$ components of this tensor are non-redundant. In order to exploit such symmetry in the discretisation, Voigt notation~\cite{Fish-Belytschko:2007} is employed. Its discrete counterpart $\bepsV$ can thus be expressed as an $\msd$-dimensional vector after a rearrangement of its non-redundant components, namely
\begin{equation} \label{eq:strainVoigt}
\bepsV :=\begin{cases}
\bigl[\epsD_{11} ,\; \epsD_{22} ,\; \epsD_{12} \bigr]^T
&\text{in 2D,} \\
\bigl[\epsD_{11} ,\; \epsD_{22} ,\; \epsD_{33} ,\; \epsD_{12} ,\; \epsD_{13} ,\; \epsD_{23} \bigr]^T
&\text{in 3D.} 
\end{cases}
\end{equation}

Following remark~\ref{rmrk:operatorVoigt}, the discrete strain rate tensor is defined as $\bepsV = \bDv \Gsym \bv$. Here, the matrices $\bDv$ and $\Gsym$ stand for the Voigt counterparts of the 
operator $\dev$ and of the symmetric part of the gradient $\GradS$, respectively, and are given by
\begin{subequations}\label{eq:VoigtDefinitions}
	\begin{equation}\label{eq:matD}
	\bDv := \begin{bmatrix} \displaystyle	 2\Id{\nsd}  - \frac{2}{3} \Jd{\nsd} & \mat{0}_{\nsd \times \nrr}  \\ \mat{0}_{\nrr \times \nsd} & \Id{\nrr} \end{bmatrix} 
	\end{equation}
	and
	\begin{equation} \label{eq:symmGrad}
	\Gsym :=\begin{cases}
	\begin{bmatrix}
	\partial/\partial x_1 & 0 & \partial/\partial x_2 \\
	0 & \partial/\partial x_2 & \partial/\partial x_1
	\end{bmatrix}^T
	&\text{in 2D,} \\
	\begin{bmatrix}
	\partial/\partial x_1 & 0 & 0 & \partial/\partial x_2 & \partial/\partial x_3 & 0 \\
	0 & \partial/\partial x_2 & 0 & \partial/\partial x_1 & 0 & \partial/\partial x_3 \\
	0 & 0 & \partial/\partial x_3 & 0 & \partial/\partial x_1 & \partial/\partial x_2
	\end{bmatrix}^T
	&\text{in 3D.} 
	\end{cases}
	\end{equation}
\end{subequations}
where $\Id{m}$ and $\Jd{\ell}$ denote the $m \times m$ identity matrix and the $\ell \times \ell$ matrix with all components equal to 1, respectively, and $\nrr = \nsd(\nsd - 1)/2$ stands for the number of rigid body rotations, that is, $\nrr = 1$ in 2D and $\nrr = 3$ in 3D.

By employing Voigt notation and exploiting the definitions in~\eqref{eq:VoigtDefinitions}, the second term of equation~\eqref{eq:FCFVLocal_eps} in the FCFV local problem is implemented as
\begin{equation}\label{eq:localTermVoigt}
\left( \dev\bhv \otimes \bn \right)_{\texttt{V}} = \bDv \bNv \bhv ,
\end{equation}
the matrix $\bNv$, accounting for the normal to a surface, being defined as
\begin{equation} \label{eq:normalVoigt}
\bNv :=\begin{cases}
\begin{bmatrix}
n_1 & 0 & n_2 \\
0 & n_2 & n_1
\end{bmatrix}^T
&\text{in 2D,} \\
\begin{bmatrix}
n_1 & 0 & 0 & n_2 & n_3 & 0\\
0 & n_2 & 0 & n_1 & 0 & n_3 \\
0 & 0 & n_3 & 0 & n_1 & n_2
\end{bmatrix}^T
&\text{in 3D,} 
\end{cases}
\end{equation}
where $n_k$ denotes the $k$-th component of the unit normal vector $\bn$.

\section{Hybridisation of the FCFV solver}
\label{app:FCFVsolver}

This appendix describes the implementation details for the hybridisation procedure and the final system obtained for the hybrid vector $\hu$. 

It is worth recalling that the FCFV method features two stages, see section~\ref{ssc:IntegralForm}. First, the $\numel$ local problems~\eqref{eq:FCFVLocal} are devised, in order to eliminate the unknowns $(\buE,\bepsE,\bphiE)$ within each cell by expressing them as functions of the hybrid vector $\bhu$. Stemming from equation~\eqref{eq:LocalSemiDiscreteCompact}, the FCFV local problem for $\Omega_e, \ e=1,\dotsc,\numel$ is given by
\begin{subequations} \label{eq:FCFVdiscretisationLocal}
	\begin{align}
	\AmatE{Q}{Q} \qE &= \AmatE{Q}{\HU} \hu + \FvectE{Q}, \\
	\AmatE{U}{U} \uE + \AmatE{U}{Q} \qE &= \AmatE{U}{\HU} \hu+ \FvectE{U},
	\end{align}
\end{subequations}
where the matrices and vectors above arise from the Newton-Raphson linearisation of nonlinear system of equations~\eqref{eq:FCFVLocal}.

Following from the constant degree of approximation utilised to approximate $\uE$ and $\qE$ at the centroid of each cell and $\hu$ at the barycentre of each face and from the quadrature rule employing a single integration point on cell and faces, the primal, $\uE$, and mixed, $\qE$, variables are expressed as functions of the hybrid unknown $\bhu$ in a decoupled manner, namely
\begin{subequations} \label{eq:FCFVlocalSolution}
	\begin{align}
	\qE &= \left[ \AmatE{Q}{Q} \right]^{-1} \AmatE{Q}{\HU} \hu + \left[ \AmatE{Q}{Q} \right]^{-1} \FvectE{Q}, \\
	\uE &= \left[ \AmatE{U}{U} \right]^{-1} \left( \AmatE{U}{\HU}  - \AmatE{U}{Q} \left[ \AmatE{Q}{Q} \right]^{-1} \AmatE{Q}{\HU} \right) \hu + \left[ \AmatE{U}{U} \right]^{-1} \left( \FvectE{U}  - \AmatE{U}{Q} \left[\AmatE{Q}{Q} \right]^{-1} \FvectE{Q} \right).
	\end{align}
\end{subequations}
It is worth noticing that the computations in equation~\eqref{eq:FCFVlocalSolution} are independent cell-by-cell and only involve the inverses of matrices $\AmatE{U}{U}$ and $\AmatE{Q}{Q}$. The former is a matrix of dimension $(\nsd+2) \times (\nsd+2)$, that is, $4 \times 4$ in 2D and $5 \times 5$ in 3D. The latter is the identity matrix of dimension $(\msd+\nsd) \times (\msd+\nsd)$ (i.e., $5 \times 5$ in 2D and $9 \times 9$ in 3D) scaled by the volume of the cell $\Omega_e$. Hence, this step requires a reduced computational effort and can be easily performed in parallel.

Similarly, upon linearisation via the Newton-Raphson method, the global problem~\eqref{eq:FCFVGlobal} is expressed as
\begin{equation} \label{eq:FCFVdiscretisationGlobal}
\sum_{e = 1}^{\numel} \left\lbrace \AmatE{\HU}{\HU}  \hu  + 
\begin{bmatrix} \AmatE{\HU}{U}  & \AmatE{\HU}{Q} \end{bmatrix} \begin{Bmatrix} \uE \\ \qE \end{Bmatrix}  -  \FvectE{\HU} \right\rbrace = \vect{0}.
\end{equation}
By plugging the expressions obtained from equation~\eqref{eq:FCFVlocalSolution} into equation~\eqref{eq:FCFVdiscretisationGlobal}, the number of unknowns is reduced by eliminating the local unknowns $\uE$ and $\qE$ from the global problem. Hence, at each Newton-Raphson iteration, the linear system
\begin{equation}\label{eq:finalLinSys}
\Kmat \Dhu  = \Fvect ,
\end{equation}
is solved, where the matrix $\Kmat$ and the vector $\Fvect$ are obtained from the assembly of the contributions from each cell, namely
\begin{subequations} \label{eq:GlobalSystemTerms}
	\begin{align}
	\Kmat^e &= \AmatE{\HU}{\HU}  + \begin{bmatrix} \AmatE{\HU}{U}  & \AmatE{\HU}{Q} \end{bmatrix}
	\begin{bmatrix} 
	\left[ \AmatE{U}{U} \right]^{-1} \left( \AmatE{U}{\HU}  - \AmatE{U}{Q} \left[ \AmatE{Q}{Q} \right]^{-1} \AmatE{Q}{\HU} \right) \\ 
	\left[ \AmatE{Q}{Q} \right]^{-1} \AmatE{Q}{\HU}  
	\end{bmatrix},  \\
	\Fvect^e &= \FvectE{\HU} - \begin{bmatrix} \AmatE{\HU}{U}  & \AmatE{\HU}{Q} \end{bmatrix} 
	\begin{Bmatrix} 
	\left[ \AmatE{U}{U} \right]^{-1} \left( \FvectE{U}  - \AmatE{U}{Q} \left[\AmatE{Q}{Q} \right]^{-1} \FvectE{Q} \right) \\ 
	\left[ \AmatE{Q}{Q} \right]^{-1} \FvectE{Q}
	\end{Bmatrix} .
	\end{align}
\end{subequations}

\end{document}